\documentclass{jfm}

\usepackage{subfigure} 
\usepackage{natbib}
\usepackage{amssymb,amsmath}
\usepackage{gensymb} % for \celsius command

\usepackage{graphicx}
\usepackage[usenames,dvipsnames,svgnames,table]{xcolor} % for colors
\usepackage{graphicx}
\usepackage{hyperref}
\hypersetup{
	colorlinks = true,
	allcolors = {blue}
}
\usepackage{lipsum}  % for blank text
\usepackage{subfigure}
\usepackage[titletoc]{appendix} % for appendix in toc
\usepackage{mathtools}

\usepackage{rotating}
\usepackage{booktabs}
\usepackage{multirow}
\usepackage{bm}
\usepackage{tabularx} 
\usepackage{ragged2e}  % for '\RaggedRight' macro (allows hyphenation)
\newcolumntype{Y}{>{\RaggedRight\arraybackslash}X} 
\usepackage{tikz}
\usepackage{comment}
\usepackage{mathrsfs}
\usepackage{ upgreek }

\newcommand{\vect}[1]{\ensuremath{{\bm{#1}}}}

\newcommand{\partiald}[2]{\ensuremath{\frac{\partial#1}{\partial#2}}}

\usepackage{setspace}

\title{Droplet settling on solids coated\\ with a soft layer}
\author{St\'ephane Poulain \and Andreas Carlson
	\corresp{\email{acarlson@math.uio.no}}
}
\shortauthor{S. Poulain and A. Carlson} %for header on even pages
\shorttitle{Droplet settling on solids coated with a soft layer}

\affiliation{Department of Mathematics, Mechanics Division, University of Oslo, N-0851 Oslo, Norway}

\begin{document}

\maketitle

\begin{abstract}
Gravitational settling of a droplet in air onto a soft substrate is a ubiquitous event relevant to many natural processes and engineering applications. We study this phenomenon by developing a three-phase lubrication model of droplet settling onto a solid substrate coated by a thin soft layer represented by a viscous film, an elastic compressible layer and an elastic sheet supported by a viscous film. By combining scaling analysis, analytical methods, and numerical simulations we elucidate how the resulting droplet dynamics is affected by the nature of the soft layer. We show that these soft layers can significantly affect the droplet shape during gravitational settling.
When there is a linear response of the deformations of the soft layer, the air layer takes longer to drain as compared to the case of a droplet settling onto a rigid substrate. Our results provide new insight into the coupled interactions between droplets and solids coated by a thin film of a soft material. \\
\end{abstract}

%elastohydrodynamics

%!TEX root = version_JFM.tex

\section{Introduction}
Impact of drops onto solid substrates or liquids are ubiquitous in many natural and industrial processes. They include inkjet printing, spray coating, forensic analysis, air-sea transfer, and epidemiology of foliar diseases to name but a few examples. 
The broad relevance of droplet impact has made it a widely studied topic, for which the effects of the interfacial and bulk properties of the drop as well as the substrate properties have been characterized \citep{Rein1993,Neitzel2002,Yarin2006,Kavehpour2015,Josserand2016,Ajaev2021}.
The drop interface dynamics has been elucidated through asymptotic analysis, scaling laws, and simulations when the impact speed is very slow \citep{Yiantsios1990,Duchemin2020}, a situation we refer to as settling.
When the drop approaches the solid it must drain the air layer separating the two interfaces, which leads to a build-up of pressure in the air film.
As a consequence, the drop interface deforms and takes the shape of a dimple before making direct contact with the solid.
The dimple-shaped interface has a maximum thickness at the axis of symmetry of the drop and a minimum near its outer edge where a neck radius can be defined.
The formation of a dimple-shaped interface is not only observed for settling droplets but also when a bubble slowly approaches a rigid surface \citep{Chan2011}, as well as for inertial drop impacts where it significantly affects the %flow dynamics
dynamics  \citep{Thorodssen2005,Xu2005,Mani2010,Hendrix2016}.

\begin{figure}
	\centering
	\includegraphics[width=0.95\textwidth]{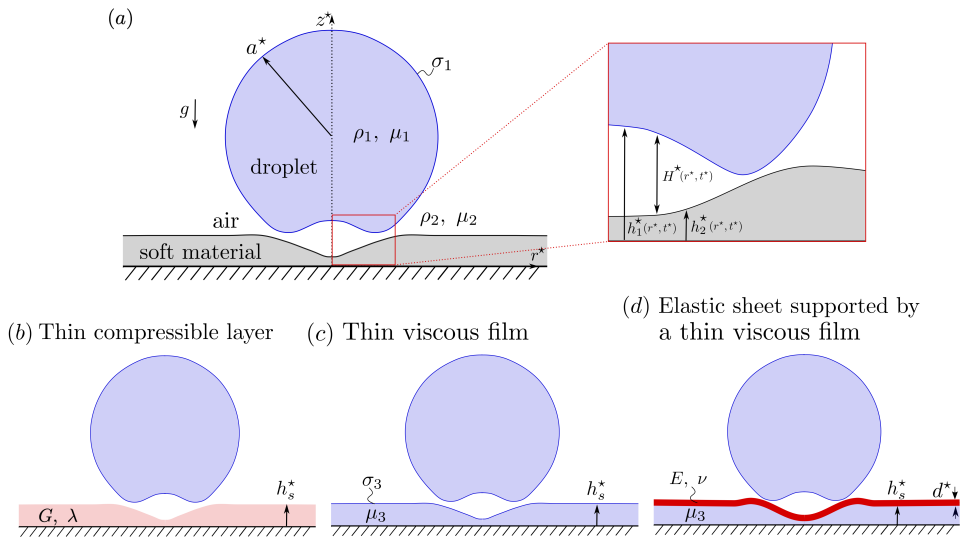}
	\caption{
	$(a)$ Schematic of a droplet of radius $a^\star$, density $\rho_1$, viscosity $\mu_1$, surface tension coefficient $\sigma_1$, suspended in a fluid at rest (typically air) of density $\rho_2$, viscosity $\mu_2$ and settling on a soft material  under the influence of the gravity field $-g\vect{e}_z$.
	Deformations are not to scale. The inset shows the definition of the droplet profile $h_1^\star(r^\star,t^\star)$, the profile of the soft material $h_2^\star(r^\star,t^\star)$, and the  thickness of the air layer $H^\star(r^\star,t^\star)=h_1^\star(r^\star,t^\star)-h_2^\star(r^\star,t^\star)$ between the two.
	We consider three different soft substrates: $(b)$ a compressible Hookean solid characterized by its Lam\'e coefficients $G$ and $\lambda$; $(c)$ a thin viscous film with surface tension coefficient $\sigma_3$; $(d)$ an elastic sheet with thickness $d^\star$, Young's Modulus $E$, and Poisson's ratio $\nu$, giving a bending stiffness  $B=Ed^{\star3}/12(1-\nu^2)$.
	In all cases the height of the undeformed soft layer is $h_s^\star$, and for $(c)$ and $(d)$ the liquid film has a viscosity $\mu_3$.
	The dimensional heights denoted by stars ()$^\star$ are nondimensionalized with the initial air layer thickness at $r^\star=0$, $H_0^\star$, whilst the radial coordinate $r^\star$ is nondimensionalized with $(H_0^\star a^\star)^{1/2}$. The dimensional time $t^\star$ is nondimensionlized with $t_0^\star=\mu_2/\Delta\rho g a^\star$.
	\label{fig:setup}}
\end{figure}

A situation as generic as a drop impact onto a rigid solid arises when it instead impacts onto a compliant soft substrate (see figure \ref{fig:setup}).
 This soft substrate can for instance be a viscous liquid film, a soft elastic layer, or an elastic sheet supported by a viscous film.
These are the three cases we describe herein.
 \citet{Yiantsios1990} described the settling of a droplet onto a bath of liquid and derived the long-term asymptotics of the quantities defining the dimple.
 They also considered the effects of slip at the droplet interface by coupling the lubrication equations with boundary integral equations to also account for the flow inside the droplet. 
 More recently, \citet{Duchemin2020} presented numerical simulations and scaling analysis using  similar lubrication equations to rationalize the settling of a large drop onto a thin liquid film. They notably showed numerically that slip at the film interface considerably accelerates the settling process.

Over the last years there has been an emerging interest in how soft materials influence capillary flows \citep{Bico2018,Andreotti2020}.
Problems involving elastohydrodynamic lubrication, or soft lubrication \citep{Skotheim2005,Essink2021}, are also critical for a wide variety of systems ranging from stereolithography to biological adhesion \citep{Wang2017,Chan2019,Wang2020}.
In this context, experiments have demonstrated that the dynamics of drop impacts can be controlled by the softness of the solid \citep{Pepper2008, Chen2010, Chen2016, Howland2016,Langley2020}.
The deformations of the substrate can affect the dynamics after contact, either by absorbing some of its energy or through the contact line motion \citep{Andreotti2020,Dervaux2020}. The high lubrication pressure can also deform the substrate before contact occurs, as recently observed in experiments \citep{Langley2020} as well as in numerical and theoretical work \citep{Pegg2018,Henman2021}.
How the compliance of the substrate affects the air drainage and dimple formation during settling of droplets has so far been overlooked, but could produce  significant effects for the resulting interfacial flow. Recent experimental advances \citep{Pack2017,Lo2017,Zhang2021,Lakshman2021} now also allow us to probe the influence of these surface deformations for drop impacts on thin liquid films down to the nanoscale.

To describe the settling of a droplet onto a soft surface, represented by a thin compliant layer, we study a minimal model considering a very viscous flow (very small Reynolds numbers) and droplets small enough for the interface deformations due to the flow to be localized near the substrate and to not affect the overall drop shape; this requires a small Bond number, i.e., capillary effects to dominate over gravity. These assumptions allow us to describe the settling dynamics based on the lubrication approximation, and we also consider a regime where deformations of the drop and the interface only appear once the lubrication assumptions hold.
Our analysis builds on the work from \citet{Yiantsios1990} and \citet{Duchemin2020}, but focuses on how the droplet settling dynamics is affected by a soft surface, represented by either a compressible elastic layer, a thin viscous liquid film, or an elastic sheet supported by a viscous film, as outlined in figure \ref{fig:setup}.

\section{Problem setup and droplet settling onto a rigid substrate}
\label{sec:solidcase}

\renewcommand{\arraystretch}{1.3}
\begin{table}
	\centering
	\begin{tabular}{lll}
	Substrate                                                 & Description                       & Expression                                                                  \\ \hline
	\multicolumn{1}{c}{-} & Aspect ratio & $\varepsilon=(H_0^\star/a^\star)^{1/2}$ \\
	\multicolumn{1}{c}{-}                                                       & Gravity / Capillarity & $\delta=\frac1{\varepsilon^2}\frac{\Delta\rho g a^{\star 2}}{\sigma_1} $                \\
	Compressible layer (\S\ref{sec:compressible_elastic})                				& Gravity / Elasticity              & $\eta=\frac{\Delta \rho g a^{\star2} h_s^\star}{H_0^{\star2} (2G+\lambda)}$ \\
	Viscous film, capillary interface (\S\ref{sec:capillary}) 				& Gravity / Capillarity & $\xi =\frac1{\varepsilon^2} \frac{\Delta\rho ga^{\star 2}}{\sigma_3}$                               \\
	Viscous film, capillary interface (\S\ref{sec:capillary}) 				& Gravity / Capillarity   & $\beta_{\rm cap} = \frac{4\pi\xi}{3h_s}$                               \\
	Viscous film (\S\ref{sec:capillary} and \ref{sec:EHD})                				& Film thickness                   & $h_s=h_s^\star/H_0^\star$                                                \\
	Viscous film (\S\ref{sec:capillary} and \ref{sec:EHD})                				& Film viscosity                   & $\lambda=\mu_2/\mu_3$                                                    \\
	Viscous film, elastic interface (\S\ref{sec:EHD})   				& Gravity / Elasticity & $\alpha=\frac{\Delta\rho g a^{\star 4}}B$ \\
	Viscous film, elastic interface (\S\ref{sec:EHD}) 				& Gravity / (Capillarity $\times$ Elasticity)  & $\beta_{\rm el} =\frac{8\pi\alpha\delta}{9h_s}$                                                                    
	\end{tabular}
	\caption{Dimensionless numbers characterizing the settling of a droplet on soft layers.
		\label{table:dimensionless}}
\end{table}

\subsection{Lubrication equations}
\label{subsec:governing}
Figure \ref{fig:setup}$(a)$ gives a generic illustration of the settling of a droplet towards a solid substrate coated with a thin  soft layer. 
The initially spherical droplet of radius $a^\star$, viscosity $\mu_1$, density $\rho_1$, is suspended in a fluid (typically air) with density  $\rho_2<\rho_1$, viscosity $\mu_2$, and surface tension coefficient $\sigma_1$. The droplet settles towards the soft surface by gravity, characterized by the gravitational acceleration $g$.
To describe this interface dynamics, we assume an axisymmetric flow and start with the lubrication theory derived by \citet{Yiantsios1990} for a rigid substrate, before moving on to a description of how the dynamics is altered when the substrate is coated with a soft layer.

The governing equations describing the flow in the air layer are made dimensionless; we denote dimensional lengths, times, and pressures with a star $(~)^\star$.
Since the droplet is initially spherical and the substrate is undeformed, the air layer thickness is well approximated by a parabolic profile $H^\star(r^\star,0)=H_0^\star+ r^{\star2}/2a^\star$ near the axis of symmetry ($r^\star=0$) at time $t^\star=0$.
Therefore the characteristic vertical length scale is the initial thickness of the film at $r^\star=0$, $H_0^\star$, whilst the characteristic radial length scale is $r_0^\star=(H_0^\star a^\star)^{1/2}$.
The droplet motion is driven by its weight, giving a characteristic pressure in the air film $p_0^\star=\Delta\rho g a^{\star2}/H_0^\star$ where $\Delta\rho=\rho_1-\rho_2$.
The time scale of the process is $t_0^\star=\mu_2/\Delta\rho ga^\star$.
We assume a small aspect ratio, $\varepsilon= H_0^\star/r_0^\star=\left(H_0^\star/a^\star\right)^{1/2}\ll 1$, as well as small Reynolds numbers both in the droplet and in the air film.
This allows us to take advantage of the lubrication approximation, i.e. we neglect any inertial effects and consider the equations at leading order in $\varepsilon$.
There are also additional assumptions we make in order to simplify the interfacial boundary conditions.
First, we only consider small interfacial slopes.
Second, we assume  no-slip at the two surfaces for the air film, an approximation valid as long as $\mu_2/\mu_1 \ll \varepsilon$.
This approximation becomes inaccurate  when the droplet is very close to the substrate, but since slip at one of the interfaces only changes the prefactor of the governing equations, the dynamics is expected to be similar.
Third, we assume that the Bond number is small enough, namely $\delta=\mathrm{Bo}/\varepsilon^2 \ll 1$ with $\mathrm{Bo}=\Delta\rho g a^{\star2}/\sigma_1$, to considerably simplify the normal stress balance. 
Considering a typical fluid with a capillary length of approximately 3 mm, and setting $\varepsilon=0.1$, this last condition amounts to considering a droplet of radius $a^\star\ll 300{\rm \mu m}$. 

By combining these assumptions we obtain Reynolds' thin film equation for the evolution of the air film thickness $H(r,t)$. The pressure $p_2(r,t)$ in the air is the capillary pressure, measured relative to the undeformed droplet, which supports the weight of the droplet at  quasi-steady state.
The following dimensionless system of equations is then satisfied \citep{Yiantsios1990}:
\begin{subequations}
\begin{align}
  \partiald{H}{t}(r,t)&=\frac{1}{12}\frac{1}{r} \partiald{}{r}\left(rH^3(r,t)\partiald{p_2}{r}(r,t)\right) 
  \label{eq:thinfilm_H}, \\
  \delta p_2(r,t) &= 2-\frac1r\partiald{}{r}\left(r\partiald{h_1}{r}(r,t)\right)
     \label{eq:pressure_h1}, \\
  \int_0^{+\infty} p_2(r,t) r~{\rm d}r &= \frac{2}{3},
     \label{eq:p_forcebalance}
\end{align}
    \label{eq:systemglobal}
\end{subequations}
where $H(r,t)=h_1(r,t)-h_2(r,t)$ is the air film thickness and $h_1(r,t)$ represents the droplet interface  (figure \ref{fig:setup}a). 
This system must be coupled with a governing equation for the height of the soft substrate $h_2(r,t)$.
We recall that the heights $H(r,t),~h_1(r,t),~h_2(r,t)$ are scaled with $H_0^\star$,
the radial coordinate $r$ with $r_0^\star=(H_0^\star a^\star)^{1/2}$,
the time $t$ with $t_0^\star=\mu_2/\Delta\rho g a^\star$,
and the pressure $p_2(r,t)$ with $p_0^\star = \Delta\rho g a^{\star 2}/H_0^\star$.
Table \ref{table:dimensionless} summarizes the definitions of the dimensionless numbers characterizing the settling dynamics, along with those relevant for describing settling onto soft layers.

\subsection{Numerical procedure}
The systems of equations we study, \eqref{eq:systemglobal} supplemented with equations for $h_2(r,t)$, are  solved numerically using the finite element method  implemented in the FEniCS code \citep{Alnaes2015} with quartic polynomial elements and an implicit time integration procedure.
For all cases studied, the initial condition for the air layer thickness
is fixed, presenting an initially small deviation to the spherical droplet following \citet{Yiantsios1990}: $H(r,0)=1+r^2/2 + \delta\left(5/18 - \ln(1+r^2/2)/3\right)$.
The integral condition \eqref{eq:p_forcebalance} is implemented as a boundary condition using \eqref{eq:pressure_h1}: $(\partial h_1/\partial r) (r,t) \overset{r\rightarrow\infty}{\sim} r- 2\delta/3r$.

\subsection{Droplet settling onto a rigid substrate}
\label{subsec:pheno}
\begin{figure}
  \centering
  \begin{tikzpicture}
    \draw (0, 0) node[inner sep=0] (fig) {\includegraphics[width=0.43\textwidth]{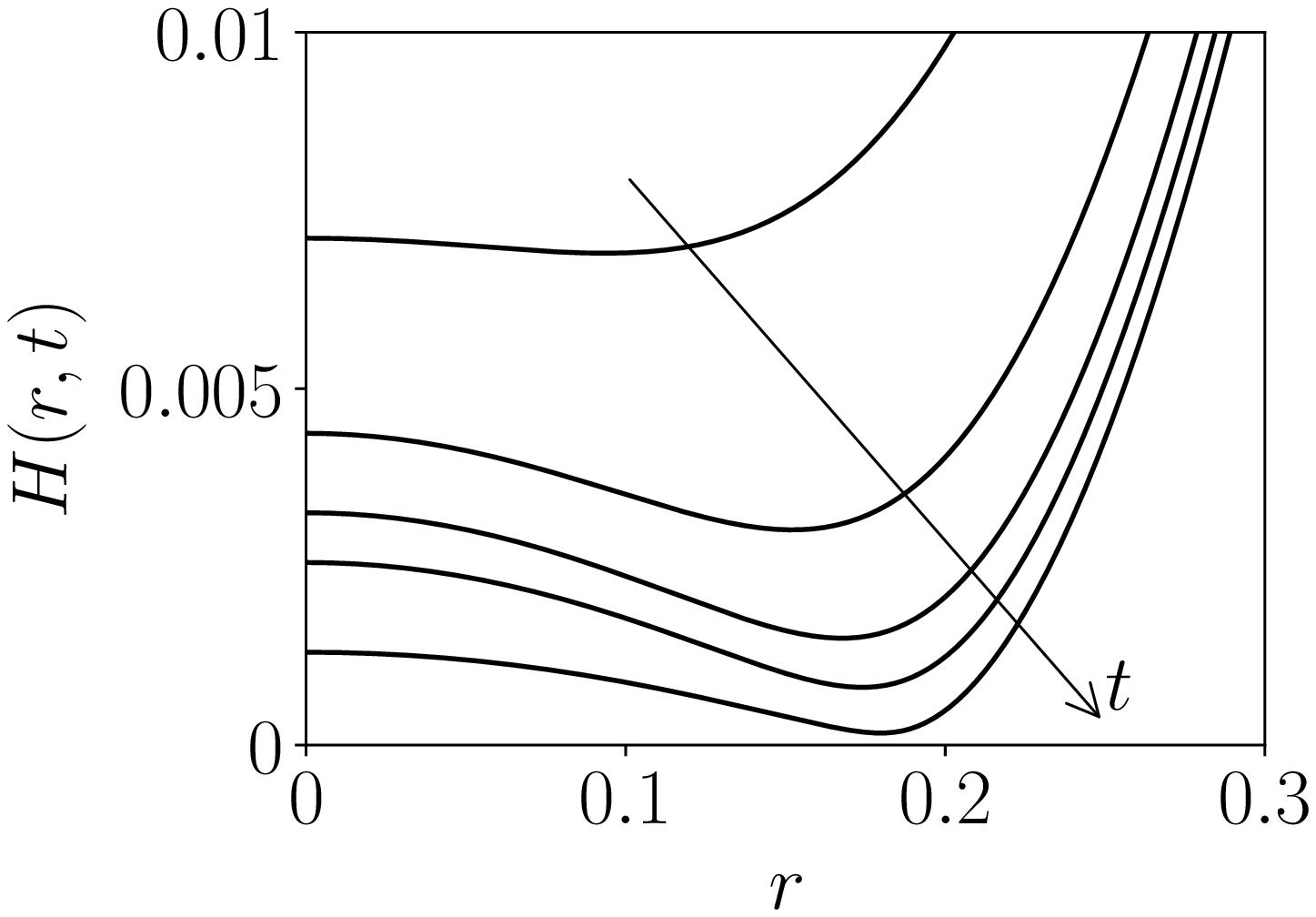}};
    \node[below right] at (fig.north west) {$(a)$};
  \end{tikzpicture}
  \begin{tikzpicture}
    \draw (0, 0) node[inner sep=0] (fig) {\includegraphics[width=0.43\textwidth]{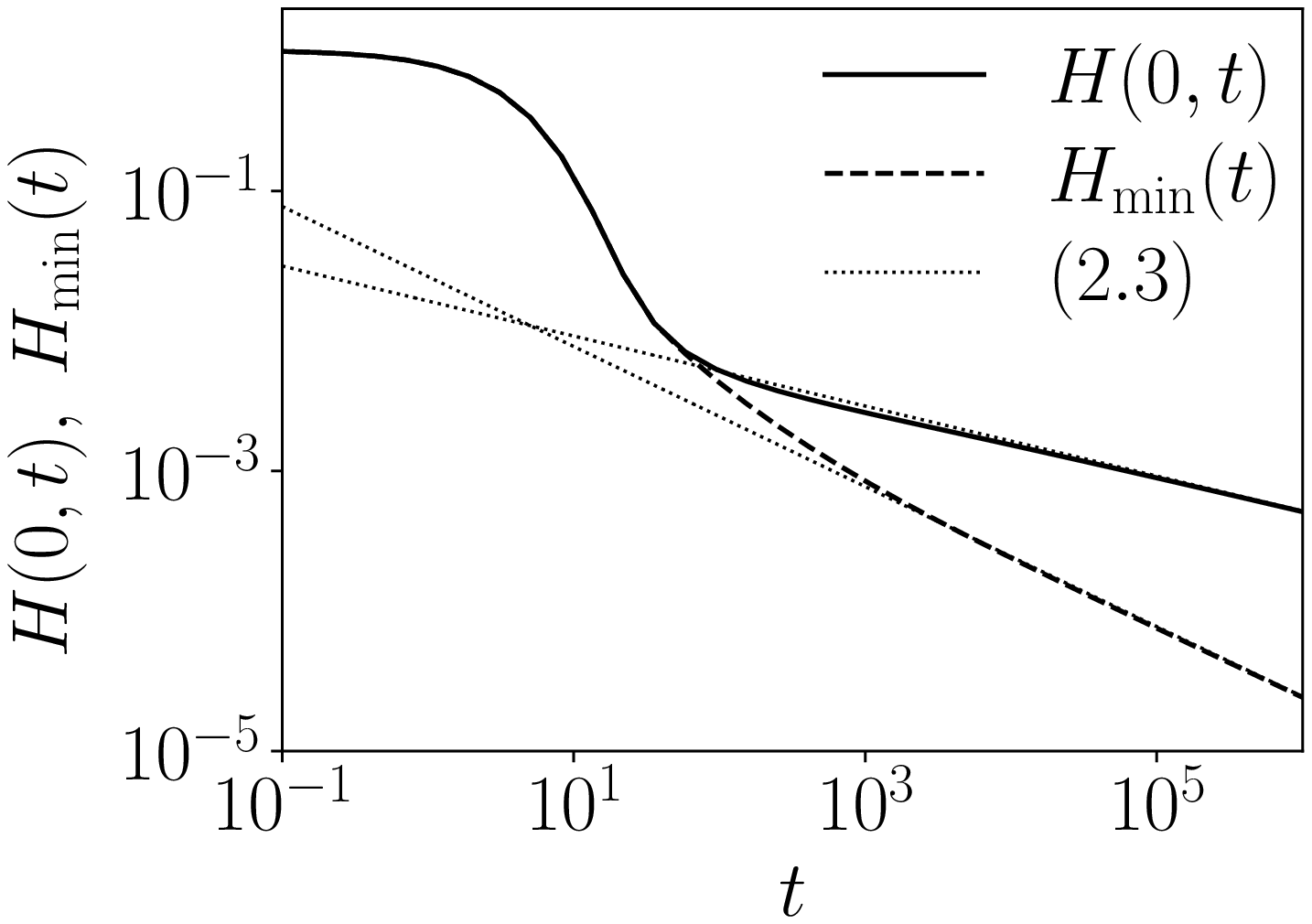}};
    \node[below right] at (fig.north west) {$(b)$};
  \end{tikzpicture}
	\caption{\label{fig:rigid}
	Results from a simulation of \eqref{eq:systemglobal} for droplet settling on a rigid substrate, with $\delta=0.05$.
	$(a)$ Droplet profile at $t=50$, 150, 400, 1000 and 20000 in the dimple.
	The expected dimple radius is $r_d=(2\delta/3)^{1/2}\simeq0.18$.
	$(b)$ Evolution of the height at the axis of symmetry, $H(0,t)$, and of the minimum height, $H_{\rm min}(t)$.
	}
\end{figure}

For a droplet settling onto a rigid substrate the system \eqref{eq:systemglobal} is closed since  $h_2(r,t)$ is a constant, taken here as $0$ without loss of generality, and $h_1(r,t)=H(r,t)$.  
A remarkable feature of the air film dynamics is that the droplet interface evolves into a dimple, a small region centered around the axis of symmetry where its deformations are localized.
Assuming that the droplet becomes almost flat in this dimple region, the pressure there is then almost uniform and given by the Laplace pressure of the droplet: $p_2(r<r_d,t) \simeq 2/\delta$. The force balance  \eqref{eq:p_forcebalance} then gives the dimple radius as $r_d\simeq(2\delta/3)^{1/2}$  \citep{Derjaguin1939,Frankel1962}.
This radius corresponds to the neck of the dimple  where the air layer thickness $H(r,t)$ is minimum.
Outside the dimple, for $r>r_d$, the droplet is nearly undeformed and spherical.
The dimple geometry  is  illustrated by numerical simulations of \eqref{eq:systemglobal} in figure \ref{fig:rigid}$(a)$.

Further insight into the interfacial dynamics may be gained through scaling analysis \citep{Duchemin2020}. Following the above discussion
we assume a dimple profile to be formed with $r_d\sim \delta^{1/2}$ and a uniform pressure $p_2(r<r_d,t)\sim\delta^{-1}$. Integrating  \eqref{eq:thinfilm_H} then gives a mass balance as:
\begin{align}
    r_d^2 \partiald{H}{t}(0,t) \sim r_d H_{\rm min}^3(t)  \partiald{p_2}{r}(r_d,t)
    \label{eq:scaling_rigidlaws}
\end{align}
where the gradient of pressure  $(\partial p_2/\partial r)(r_d,t) \sim p_2/\ell(t) \sim 1/\delta\ell(t)$  is localized within the radial extent of the dimple neck $\ell(t)$, and  the minimum thickness is $H_{\rm min}(t)=H(r_d,t)$.
Matching the curvature of the neck with the curvature of the undeformed droplet yields $H_{\rm min}(t)/\ell(t)^2 \sim 1$, whilst the slope is matched between the dimple and the neck: $H_{\rm min}(t)/\ell(t) \sim H(0,t)/r_d$. 
These two matching conditions simplify \eqref{eq:scaling_rigidlaws} to $({\rm d}H/{\rm d}t)(0,t) \sim \delta^{-4}H(0,t)^5$, which finally allows us to find the following scaling laws: $H(0,t) \sim \delta t^{-1/4}$, $H_{\rm min}(t) \sim \delta t^{-1/2}$, $\ell(t) \sim \delta^{1/2} t^{-1/4}$.
A more rigorous mathematical analysis has also been conducted by \citet{Yiantsios1990}, who derived and validated numerically the following long-time behaviors from an asymptotic expansion of the governing equations \eqref{eq:systemglobal}:
\begin{subequations}
\begin{align}
     H(0,t)&=(0.3273)\delta t^{-1/4} ,
     \label{eq:h0_rigid}\\
     H_{\rm min}(t)&=(0.4897)\delta t^{-1/2},
     \label{eq:hmin_rigid}
\end{align}
     \label{eq:Yiantsios_scalings}
\end{subequations}
where $r_d$ approaches a constant value as  $(2\delta/3)^{1/2} \left(1-0.1652 t^{-1/4}\right)$.
 These asymptotic results are in very close agreement with the numerical simulations of \eqref{eq:systemglobal} as shown in figure \ref{fig:rigid}$(b)$.
 In the next sections we describe how this physical picture changes when considering the settling of a droplet onto soft surfaces.

\section{Solid substrate coated with a thin compressible elastic layer}
\label{sec:compressible_elastic}

We move on to describe the case when the rigid substrate is coated with a thin and compressible elastic layer (figure \ref{fig:setup}$b$).
At rest, the layer has uniform thickness $h_s^\star$, such that $h_2(r,0)=h_2(r\rightarrow \infty,t)=h_s^\star/H_0^\star=h_s$.
The response of the layer to stresses follows Hooke's law: ${\mathbf{T}^\star=2G\mathbf{e}^\star+\lambda\mathrm{tr}(\mathbf{e}^\star)\mathbf{I}}$, where $\mathbf{T}^\star$ is the stress tensor, $\mathbf{I}$ is the identity matrix, and $\mathbf{e}^\star=\left(\mathbf{\nabla^\star u^\star} +(\mathbf{\nabla^\star u^\star})^\intercal \right)$ is the strain tensor for small displacements  $\mathbf{u}^\star$.
The Lam\'e coefficients $G$ and $\lambda$ are assumed to be of the same order of magnitude. 
Following \citet{Skotheim2005}, we use the lubrication approach to simplify the governing equations of the elastic medium.
In the elastic layer the vertical length scale is its thickness $h_s^\star$ whilst the vertical length scale remains $r_0^\star$.
Displacements are assumed to scale as $H_0^\star$, and the stresses scale as $p_0^\star=\Delta\rho g a^2/H_0$.
By using these scaling arguments and assuming $h_s^\star/(H_0^\star a^\star)^{1/2}=\varepsilon h_s \ll 1$, which requires $h_s$ to be at most $\mathcal{O}(\varepsilon^0)$, the governing equations of the solid reduce at leading order in $\varepsilon$ to:
\begin{subequations}
\begin{align}
    h_2(r,t)&=h_s-\eta p_2(r,t), \\
    \eta&= \frac{\Delta \rho g a^{\star2} h_s^\star}{H_0^{\star2} (2G+\lambda)}.
    \label{eq:def_eta}
\end{align}
    \label{eq:pressure_h2}
\end{subequations}
The dimensionless number $\eta$ is a softness parameter measuring the relative importance of the pressure exerted by the drop on the compressible layer compared to its Lam\'e coefficients, and which controls the importance of the layer deformations compared to its thickness \citep{Skotheim2005}.
Thus, the displacement of the elastic layer is proportional to the external pressure acting on it.
This is known as the Winkler model \citep{Dillard2018,Chandler2020}, also used to describe thin liquid-infused poroelastic layers \citep{Skotheim2005} as well as soft polymer brushes \citep{Gopinath2011,Davies2018}.

    \begin{figure}
  \centering
  \begin{tikzpicture}
    \draw (0, 0) node[inner sep=0] (fig) {\includegraphics[width=0.46\textwidth]{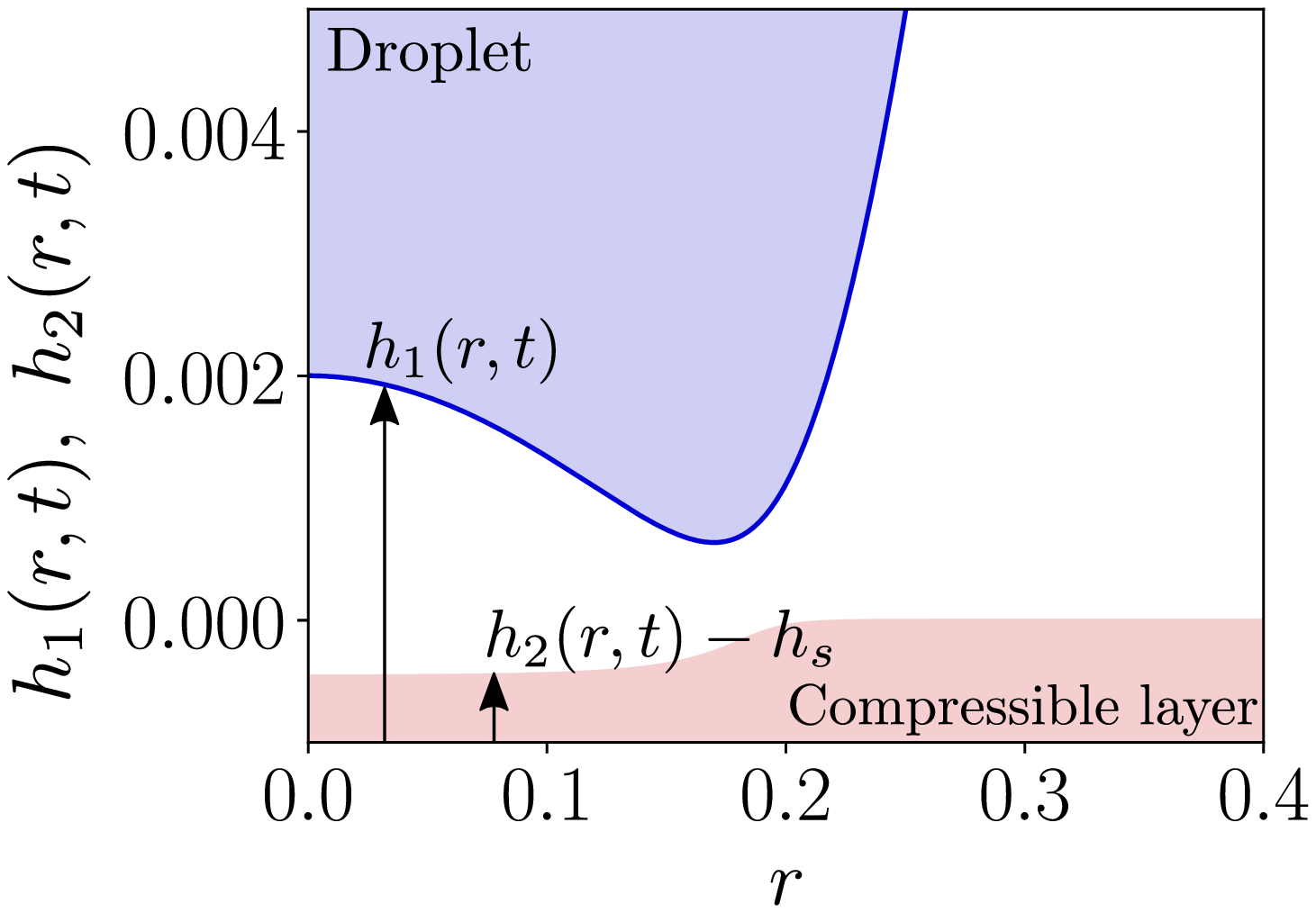}};
    \node[below right] at (fig.north west) {$(a)$};
  \end{tikzpicture}
  \begin{tikzpicture}
    \draw (0, 0) node[inner sep=0] (fig) {\includegraphics[width=0.46\textwidth]{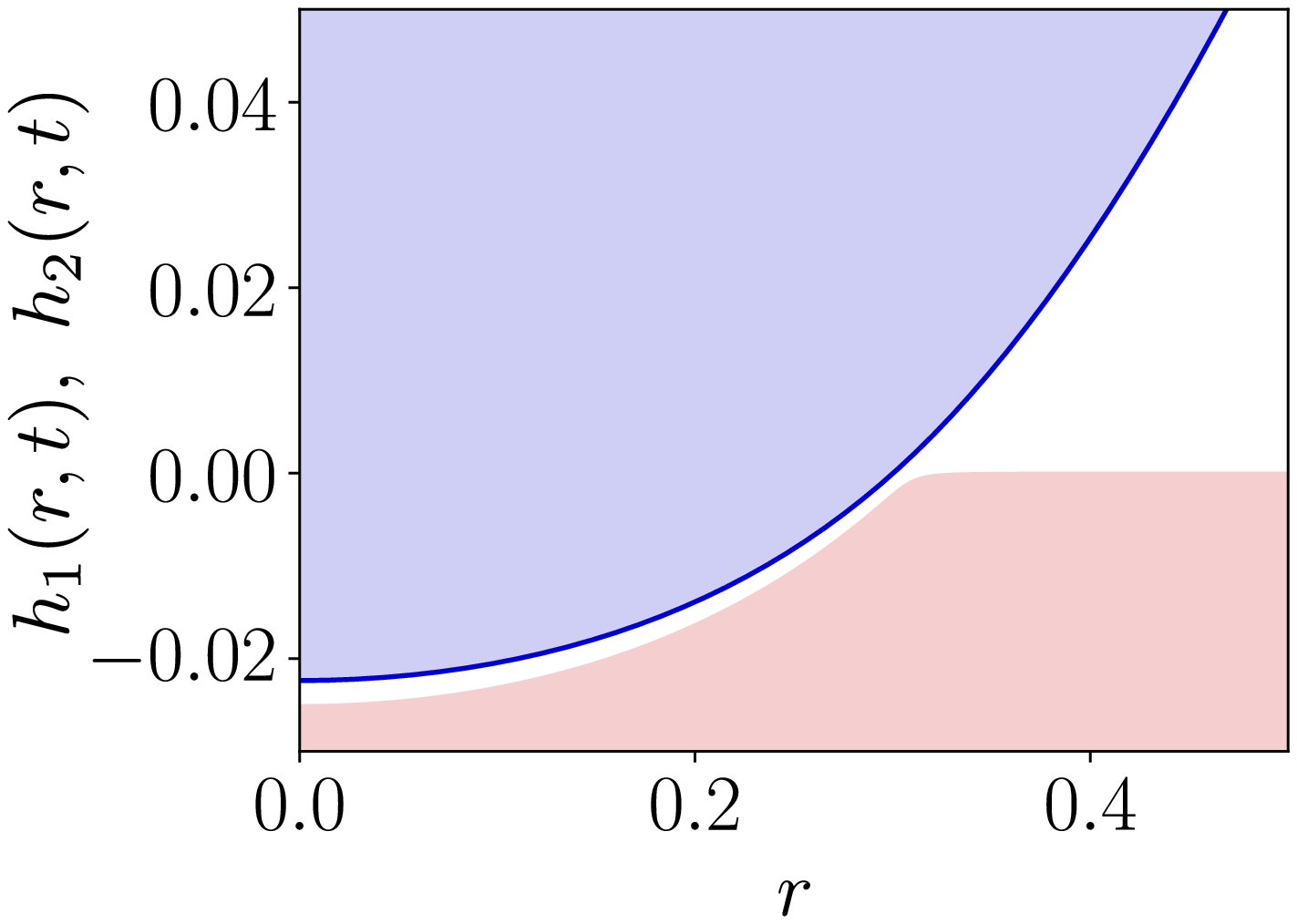}};
    \node[below right] at (fig.north west) {$(b)$};
  \end{tikzpicture}
	\caption{\label{fig:compressible1}
	Profiles of the droplet, $h_1(r,t)$, and of the compressible elastic layer $h_2(r,t)$,  at $t=1000$ and with $\delta=0.05$ and $(a)$ $\eta=10^{-5}$, $(b)$ $\eta=10^{-3}$. 
	The initial height of the elastic layer, $h_s$, is only significant in the definition of $\eta$ \eqref{eq:def_eta} and $h_2(r,t)$ is translated in the figures so that it is 0 in its undeformed state.
	}
	\end{figure}
	
To illustrate the effect of a soft layer on the droplet settling dynamics, we perform numerical simulations of \eqref{eq:systemglobal} and \eqref{eq:pressure_h2}.
 When the layer is stiff compared to the droplet, i.e. when $\eta$ is sufficiently  small, we expect the droplet settling dynamics to approach the analytical solution for a rigid wall \eqref{eq:Yiantsios_scalings}.
 Figure \ref{fig:compressible1} also shows that when $\eta$ is large, the droplet profile $h_1(r,t)$ becomes nearly undeformed while the elastic layer $h_2(r,t)$ also adopts the parabolic shape of the droplet.
 Accordingly, we start the analysis in the limit of a very soft compressible layer by assuming the following ansatz:
 \begin{equation}
    h_2(r,t)=h_2(0,t)+r_e^2/2,
    \label{eq:elastic_ansatz}
 \end{equation}
 valid up to a radius $r_e$, and an undeformed elastic layer $h_2(r,t) = h_s$ for $r>r_e$.
 From \eqref{eq:pressure_h2}, the corresponding pressure at the axis of symmetry is $\eta p_2(0,t) = r_e^2/2$.
The force balance \eqref{eq:p_forcebalance}, limiting the integration  to $0\leq r\leq r_e$, gives: $r_e^4-4\eta p_2(0,t) r_e^2 + 16\eta/3=0$.
These two relations yield:
\begin{subequations}
\begin{align}
    r_e&=\left(\frac{16}{3}\eta\right)^{1/4},
    \label{eq:rneck_compressible}\\
    p_2(r=0,t)&=\left(\frac{3}{4}\eta\right)^{-1/2}.
    \label{eq:p0_compressible}
\end{align}
\label{eq:p0andrneck_compressible}
\end{subequations}
To arrive at these results we have assumed that the droplet maintains its spherical shape and that the elastic layer follows a similar profile with the same curvature as the droplet.
This means  that the thickness of the air film is uniform: $H(r,t)\simeq H(t)$ for $r<r_e$. The evolution equation \eqref{eq:thinfilm_H}, with the parabolic pressure profile given by \eqref{eq:pressure_h2},\eqref{eq:elastic_ansatz} and \eqref{eq:p0andrneck_compressible} as $p_2(r,t)=2/\sqrt{3\eta}-r^2/2\eta$, then yields for $t \gg \eta^{-1}$ and $r<r_e$:
\begin{equation}
    H(r,t)=\left(\frac{t}{3\eta}\right)^{-1/2}.
    \label{eq:h_compressible}
\end{equation}

To find the range of applicability of the above results we can integrate \eqref{eq:pressure_h1}, which we have not yet used, with the pressure profile derived herein:
\begin{equation}
    h_1(r,t)=h_1(0,t)+\frac{r^2}{2}\left(1-\frac{1}{\sqrt{3}}\frac{\delta}{\sqrt{\eta}}\right)+\frac{1}{32}\frac{\delta}{\eta}r^4.
\end{equation}
The analysis is consistent only if  the corrections to the ansatz \eqref{eq:elastic_ansatz} are small, i.e. when $h_1(r,t)\simeq h_1(0,t) + r^2/2$, which requires $\eta\gg\delta^2$. Cancelling the quartic term up to $r=r_e=(16\eta/3)^{1/4}$ reduces to $\delta\ll 1$, consistent with our assumptions discussed in \S\ref{subsec:governing}.

\begin{figure}
    \centering
  \begin{tikzpicture}
    \draw (0, 0) node[inner sep=0] (fig) {\includegraphics[width=0.48\textwidth]{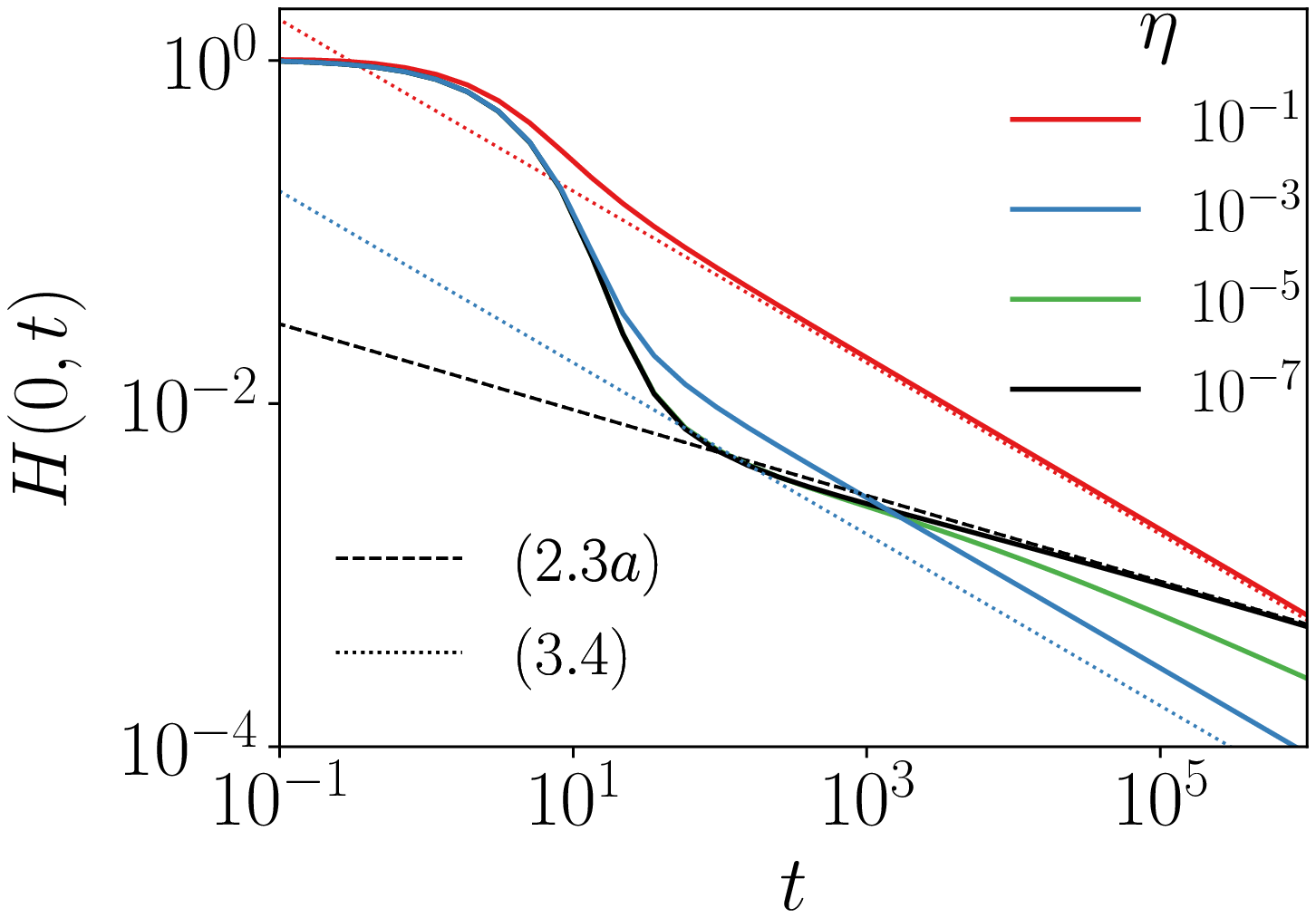}};
    \node[below right] at (fig.north west) {$(a)$};
  \end{tikzpicture}
  \begin{tikzpicture}
    \draw (0, 0) node[inner sep=0] (fig) {\includegraphics[width=0.48\textwidth]{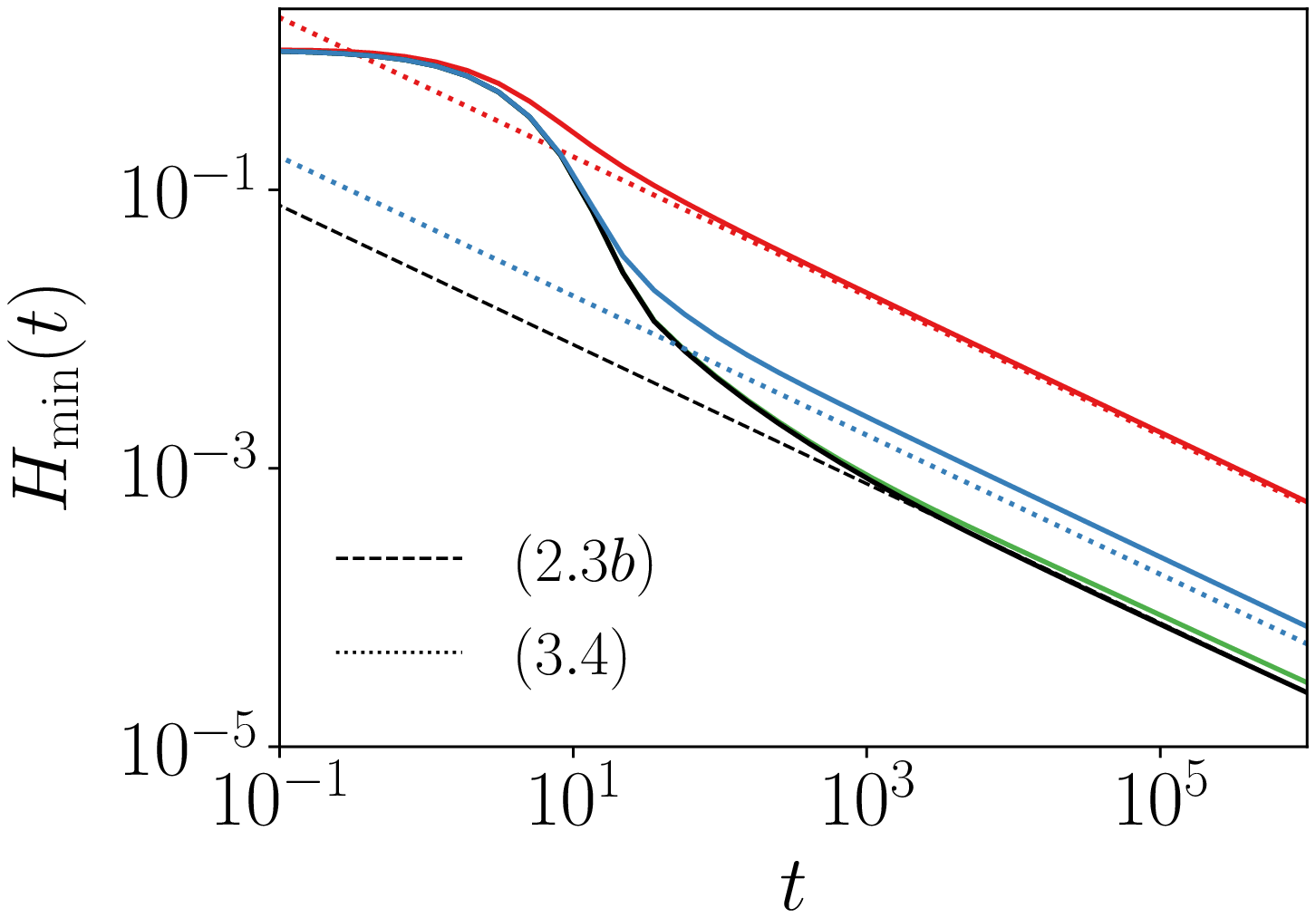}};
    \node[below right] at (fig.north west) {$(b)$};
  \end{tikzpicture}
  \begin{tikzpicture}
    \draw (0, 0) node[inner sep=0] (fig) {\includegraphics[width=0.48\textwidth]{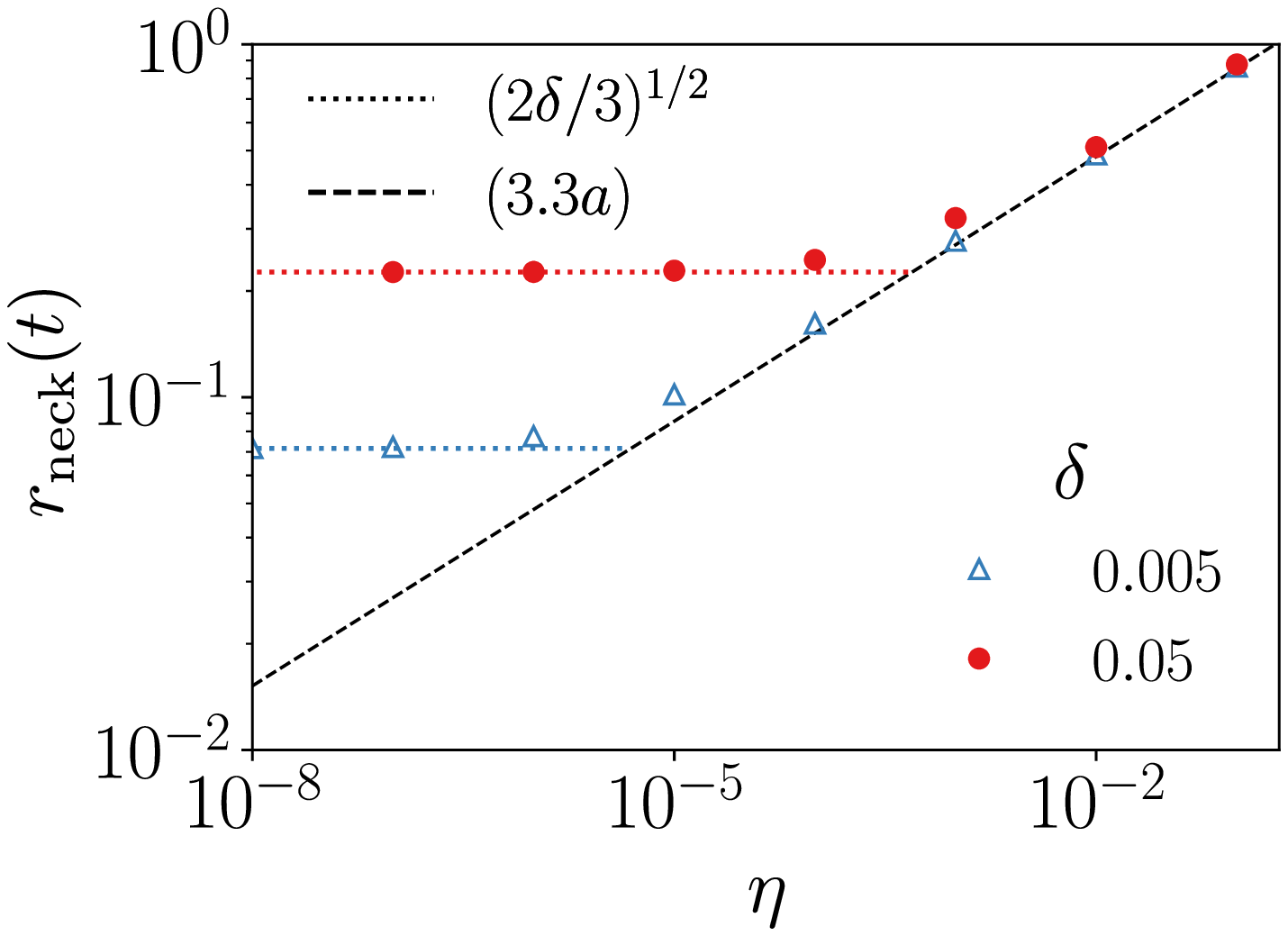}};
    \node[below right] at (fig.north west) {$(c)$};
  \end{tikzpicture}
  \begin{tikzpicture}
    \draw (0, 0) node[inner sep=0] (fig) {\includegraphics[width=0.48\textwidth]{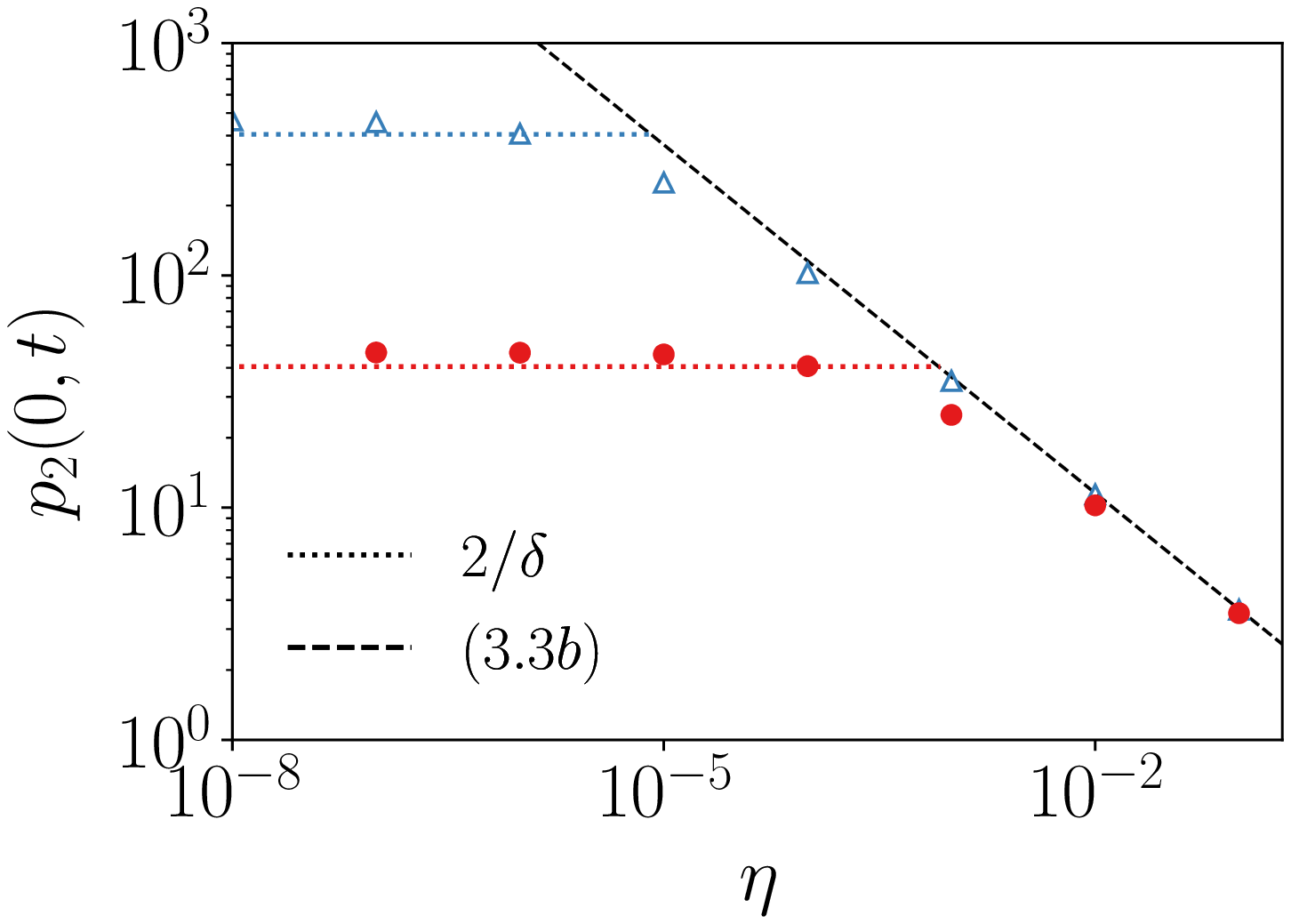}};
    \node[below right] at (fig.north west) {$(d)$};
  \end{tikzpicture}
  
	\caption{\label{fig:compressible2}
	Air layer dynamics during a droplet settling onto a thin compressible elastic layer. $(a,b)$ Time-evolution of the air layer thickness  $H$ at $r$=0 ($a$) and at its minimum value ($b$), for $\delta=0.05$ and $\eta=10^{-7},~10^{-5},~10^{-3},~10^{-1}$.
	$(c,d)$ Radius of the neck, {$r_{\rm neck}(t)={\rm argmin}_r[H(r,t)]$ ($c$)}, and pressure in the air layer at $r=0$ ($d$), both at $t=10^3$.
	}
\end{figure}

 In order to verify these predictions, we show in figure \ref{fig:compressible2} the time evolution of the air film thickness and pressure at $r=0$, of the minimum air  film thickness, and of the radial position of the neck.
 These numerical results confirm the asymptotic behavior derived herein as well as their application range, $\eta \gg \delta^2$, while the effects of elasticity are negligible when $\eta \ll \delta^2$.
 In particular, the minimum height of the air film scales as $H_{\rm min}(t)\sim t^{1/2}$ with a prefactor always larger than for a rigid wall.
 This suggests that a soft enough material would delay direct contact between the droplet and the solid.

To place these results in the context of relevant material parameters, we consider a rescaled Bond number $\delta=\mathcal{O}\left(10^{-1}\right)$, an aspect ratio $\varepsilon=\mathcal{O}\left(10^{-1}\right)$, a liquid-gas density difference $\Delta \rho=\mathcal{O}\left(10^3~{\rm kg~m}^{-3}\right)$, a surface tension coefficient $\sigma_1=\mathcal{O}(50~{\rm mN~m}^{-1})$, and a compressible layer thickness $h_s^\star\sim H_0^\star$.
The condition $\eta \gg \delta^2$ translates into the following condition for the material properties of the soft layer for compressibility effects to be dominant:  $(2G+\lambda) \ll 10^4~{\rm Pa}$.
Consequently, the analysis is primarily reserved for very soft materials, for instance substrates grafted by a layer of polymer brushes which can verify this criterion \citep{Davies2018} and for which the Winkler mode applies \citep{Gopinath2011}.

\section{Solid substrate coated with a thin viscous liquid film}

\label{sec:capillary}
Next we consider a droplet settling on a rigid substrate coated with a thin viscous liquid film (figure \ref{fig:setup}$c$) of  viscosity $\mu_3=\mu_2/\lambda $ and surface tension coefficient $\sigma_3$.
Considering the tangential stress balance between the air layer and the liquid film, $\mu_2 \partial v^\star_2/\partial z^\star = \mu_3 \partial v^\star_3/\partial z^\star$ in dimensional units, where $v_2^\star$ and $v_3^\star $ are the radial velocities in the air and liquid layer, respectively, leads to the following thin film equations for the air film thickness $H(r,t)$ and the liquid film height $h_2(r,t)$:
\begin{subequations}

\begin{align}
	\partiald{H}{t}(r,t)&= {} \frac{1}{12}\frac{1}{r}\partiald{}{r}
	\bigg[\frac{r}{H(r,t)+\lambda h_2(r,t)}\bigg( H^4(r,t)\partiald{p_2}{r}(r,t)+ \notag \\
	&\qquad  + 4\lambda H^3(r,t) h_2(r,t) \partiald{p_2}{r}(r,t) + 
	3\lambda H^ 2(r,t)h_2^2(r,t) \partiald{p_3}{r}(r,t) \bigg)\bigg],
	\label{eq:H_slip} \\
     \partiald{h_2}{t}(r,t)&= {} \frac{\lambda}{3}\frac{1}{r}\partiald{}{r}\bigg[\frac{r}{H(r,t)+\lambda h_2(r,t)}\bigg(H(r,t)h_2^3(r,t) \partiald{p_3}{r}(r,t)+ \notag \\ 
    &\qquad\qquad\qquad + \frac\lambda4 h_2^4(r,t) \partiald{p_3}{r}(r,t) + \frac34H^2(r,t)h_2^2(r,t)\partiald{p_2}{r}(r,t)\bigg)\bigg].
     \label{eq:h2_slip}
\end{align}

    \label{eq:Hh2_slip}
\end{subequations}
Equation \eqref{eq:H_slip} replaces \eqref{eq:thinfilm_H} whilst the normal stress balance \eqref{eq:pressure_h1} and force balance \eqref{eq:p_forcebalance} remain unaltered.  The pressure $p_3(r,t)$ in the viscous film is given by the normal stress balance:
\begin{align}
    \xi \left(p_3(r,t)-p_2(r,t)\right)= -\frac1r \partiald{}{r}\left(r\partiald{h_2}{r}(r,t)\right),
    \label{eq:pressureh2_capillary}
\end{align}
where $ \xi = \varepsilon^{-2} \Delta\rho ga^{\star2}/\sigma_3 $ is a rescaled Bond number defined similarly to $\delta$, $\xi=\delta \sigma_1/\sigma_3$. %, with $\rho_3$ the density of the viscous film. 
At $t=0$ and far from the droplet, the height of the layer is $h_2(r,0)=h_2(r\rightarrow\infty,t)=h_s$.
For the most common liquids, e.g. aqueous substances and common oils, we expect that surface tension coefficients remain in the same range, and hence that $\xi$ and $\delta$ have the same order of magnitude. We set $\delta=0.05$ for the numerical results presented below and vary $\xi$ from 0.01 to 0.2.

We focus on cases where the viscosity ratio $\lambda=\mu_2/\mu_3 \ll 1$, i.e., we consider the liquid film to be very viscous as compared to the air layer.
In the limit $\lambda\rightarrow0$, \eqref{eq:H_slip}
simplifies to \eqref{eq:thinfilm_H}, and we can also expect the terms in \eqref{eq:h2_slip} proportional to $\lambda^2$ and to $(\partial p_2 / \partial r)(r,t)$ to be subdominant, i.e., the deformation in the fluid layer to be primarily driven by the pressure acting on it with a negligible effect of tangential shear stresses.
Neglecting these terms is equivalent to consider no-slip in the air layer and  free-slip in the liquid layer at the air-liquid interface.
By making these assumptions, \eqref{eq:systemglobal} continues to hold and the evolution equation of the viscous film height \eqref{eq:h2_slip} simplifies to:
\begin{align}
    \partiald{h_2}{t}(r,t) = \frac{\lambda}{3}  \frac{1}{r} \partiald{}{r}\left(r h_2^3(r,t)  \partiald{p_3}{r}(r,t)\right).
    \label{eq:thinfilmviscous_h2} 
\end{align}
We discuss next what can be learned from analyzing these simplified equations at small $\lambda$ for the air layer profile and deformation of the liquid film.

\subsection{Behavior of the air layer}
\label{subsec:dimple_cap}

In the limit of very viscous films $\lambda \ll 1$, insight into the behavior of the air layer thickness $H(r,t)$ can be gained since the expression of the pressures $p_2(r,t)$ in the air layer \eqref{eq:pressure_h1}  and  $p_3(r,t)$ in the liquid film \eqref{eq:pressureh2_capillary} are similar. Indeed, substracting \eqref{eq:pressureh2_capillary} from  \eqref{eq:pressure_h1} yields: 
\begin{align}
    \left(\delta+\xi\right) p_2(r,t) - \xi p_3(r,t) = 2-\frac1r\partiald{}{r}\left(r\partiald{H}{r}(r,t)\right).
    \label{eq:pcapillarysimplified}
\end{align}
Assuming $\xi p_3(r,t) \ll \left(\delta+\xi\right)p_2(r,t)$, the system \eqref{eq:thinfilm_H}, \eqref{eq:pcapillarysimplified}, \eqref{eq:p_forcebalance} for the air layer thickness $H(r,t)$ is closed and can be solved similarly to \eqref{eq:systemglobal}.
The discussion of \S\ref{sec:solidcase} for the deposition on a rigid wall can then also be applied here, except that $\delta$ now becomes $\delta+\xi$ to account for the liquid film.
In particular the dimple radius approaches $(2(\delta+\xi)/3)^{1/2}$, the pressure in the dimple approaches $2/(\delta+\xi)$, and the results given by \eqref{eq:Yiantsios_scalings} for the height at the axis of symmetry and minimum height apply as well (with $\delta\rightarrow\delta+\xi$).
This was recognized by \citet{Yiantsios1990}  for  the  deposition of a droplet on a bath of its own fluid ($\xi=\delta$), neglecting any influence of the pressure in the bath ($p_3$ implicitly assumed to be zero), and was also used to study the approach of a droplet towards another droplet \citep{Yiantsios1991}.

\subsection{Response of a thin viscous film under constant load}
\label{sec:appendix_capillary}
Before describing the full problem we consider a simplified, generic, situation where the thin fluid film is exposed to an external load $p_e$ and with no external shear stresses.
To account for this situation we change the non-dimensional units to scale \eqref{eq:pressureh2_capillary} and \eqref{eq:thinfilmviscous_h2}
using $t=\hat{t} t_c$, $h_2=\hat{h}_2(\vect{\hat{x}},t) h_s$, $\nabla=\hat{\nabla}/x_c$, $p_e=\hat{p}_e(\vect{\hat{x}},t) p_c$; with $p_c$ the characteristic magnitude of the load which is distributed over a characteristic length $x_c$, $h_s$ the initial height of the liquid layer, and $t_c=3\xi x_c^{4}/\lambda h_s^{3}$ the characteristic time and $\vect{\hat{x}}$ is the spatial variable.
Using these scalings, we obtain the following dimensionless equation for the liquid film height:
\begin{align}
    \partiald{\hat{h}_2}{\hat{t}}(\vect{\hat{x}},\hat{t}) = \hat{\nabla} \cdot \left( \hat{h}_2^3(\vect{\hat{x}},\hat{t})
     \hat{\nabla} \left(-\hat{\nabla}^2 \hat{h}_2(\vect{\hat{x}},\hat{t}) +  \beta_{\rm cap} \hat{p}_e(\vect{\hat{x}},\hat{t}) \right) \right),
    \label{eq:thinfilmadim_cap}
\end{align}
where we temporarily consider the problem in $n$-dimensional Cartesian coordinates and with $\vect{\hat{x}}$ the spatial variable.
The parameter $\beta_{\rm cap}=p_c x_c^{2} \xi /h_s$ represents the ratio of the characteristic pressure force to the characteristic capillary force.
We consider a long and initially flat viscous film, $\hat{h}_2(\vect{\hat{x}},0)=1$, exposed to a  load uniform in time: $\hat{p}_e(\vect{\hat{x}},\hat{t})=\hat{p}_e(\vect{\hat{x}})\mathcal{H}(\hat{t})$ where $\mathcal{H}$ is the Heaviside function. By assuming small deformations, we can derive the film height profile analytically by adapting a procedure recently used to study flows induced in glassy polymer films \citep{Pedersen2021}. Indeed,  
as long as the film deformations are small, \eqref{eq:thinfilmadim_cap} can be linearized defining 
%$\hat{\mathfrak{h}}$
$\hat{\epsilon}(\vect{\hat{x}},\hat{t})$ such that $\hat{h}_2(\vect{\hat{x}},\hat{t})=1-\beta_{\rm cap}\hat{\epsilon}(\vect{\hat{x}},\hat{t})$. %; $\hat{\epsilon}$ should not be confused with the aspect ratio $\varepsilon$ introduced in \S \ref{sec:solidcase}.
By assuming $|\beta_{\rm cap}\hat{\epsilon}(\vect{\hat{x}},\hat{t})|\ll 1$ and neglecting terms that are $\mathcal{O}\left((\beta_{\rm cap}\hat{\epsilon}(\vect{\hat{x}},\hat{t}))^2\right)$,  we obtain from \eqref{eq:thinfilmadim_cap} the following linear equation:
\begin{align}
    \left(\partiald{}{\hat{t}}+\hat{\nabla}^{4}\right)\hat{\epsilon}(\vect{\hat{x}},\hat{t}) = -\hat{\nabla}^2 \hat{p}_e(\vect{\hat{x}},\hat{t}).
    \label{eq:thinfilmlinear_cap}
\end{align}
The associated Green's function in $n$-dimensional Cartesian coordinates is:
\begin{align}
    G_{\rm cap}(\vect{\hat{x}},\hat{t})=\frac{\mathcal{H}(\hat{t})}{\left(2\pi\right)^n} \int_{\mathbb{R}^n} e^{-|\vect{k}|^{4} \hat{t}} e^{i \vect{k} \cdot \vect{\hat{x}}} ~{\rm d}\vect{k},
    \label{eq:Green_cap}
\end{align}
and the solution of \eqref{eq:thinfilmlinear_cap} for an arbitrary load $\hat{p}_e(\vect{\hat{x}})$ is given by:
\begin{align}
    \hat{\epsilon}(\vect{\hat{x}},\hat{t})&=-\int_{\mathbb{R}^n} \int_0^{\hat{t}}  G_{\rm cap}(\vect{\hat{x}}-\vect{x}^\prime;\hat{t}-t^\prime)\hat{\nabla}^2\hat{p}_e(\vect{x}^\prime)~ {\rm d} t^\prime{\rm d} \vect{x}^\prime.
        \label{eq:GeneralSolution}
\end{align}
When the load is a Dirac delta function, $\hat{p}_e(\vect{\hat{x}})=\delta_{\rm Dirac}(\vect{\hat{x}})$, this simplifies to:
\begin{align}
    \hat{\epsilon}(\vect{\hat{x}},\hat{t})&=-\int_0^{\hat{t}} \hat{\nabla}^2 G_{\rm cap}(\vect{\hat{x}};t^\prime) ~{\rm d} t^\prime
    \label{eq:solutionDiracgeneral}.
\end{align}
For a confined uniform load,  $\hat{p}_e(\vect{\hat{x}})=1/\mathcal{V}_n$ if $\hat{x} \in \mathcal{B}$, $\hat{p}_e(\vect{\hat{x}})=0$ otherwise, where $\mathcal{B}=\{\vect{\hat{x}}, |\vect{\hat{x}}|\leq1\}$ is the unit $n$-ball and $\mathcal{V}_n$ its volume (in particular, $\mathcal{V}_2=\pi$), the solution \eqref{eq:GeneralSolution} becomes:
\begin{align}
    \hat{\epsilon}(\vect{\hat{x}},\hat{t})&=-\frac{1}{\pi} \int_0^{\hat{t}} \int_{\partial \mathcal{B}}\vect{\hat{\nabla}} G_{\rm cap}(\vect{\hat{x}}-\vect{x^\prime};t^\prime)\cdot\vect{n} ~{\rm d}\ell^\prime{\rm d} t^\prime,
    \label{eq:capillarysth}
\end{align}
where $\vect{n}$ is the outward normal to the boundary $\partial\mathcal{B}=\{\vect{\hat{x}}, |\vect{\hat{x}}|=1\}$.

\begin{figure}
    \centering
    \begin{tikzpicture}
    \draw (0, 0) node[inner sep=0] (fig) { \includegraphics[width=0.48\textwidth]{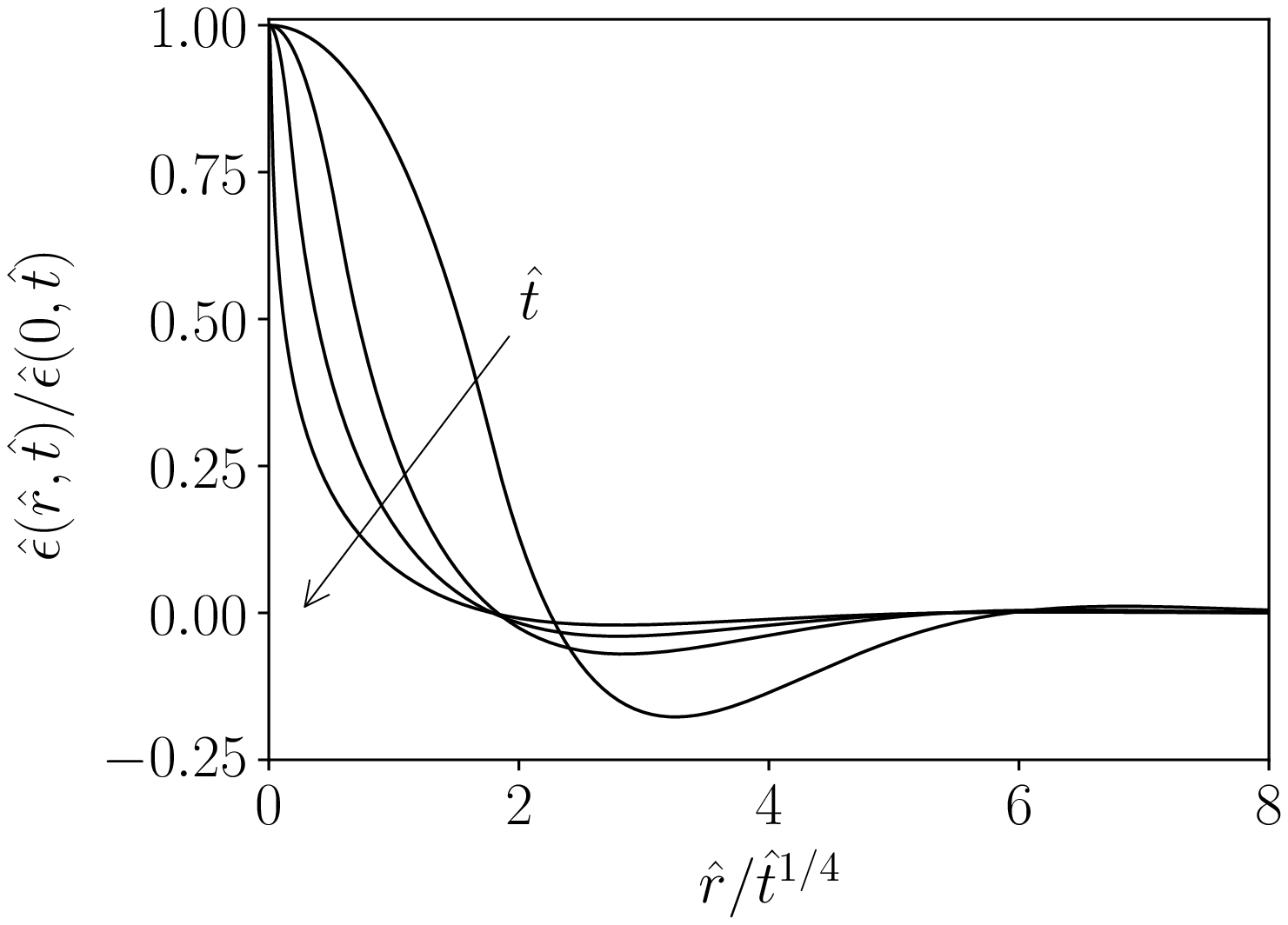}};
    \node[below right] at (fig.north west) {$(a)$};
  \end{tikzpicture} \\
  \begin{tikzpicture}
    \draw (0, 0) node[inner sep=0] (fig) {\includegraphics[width=0.48\textwidth]{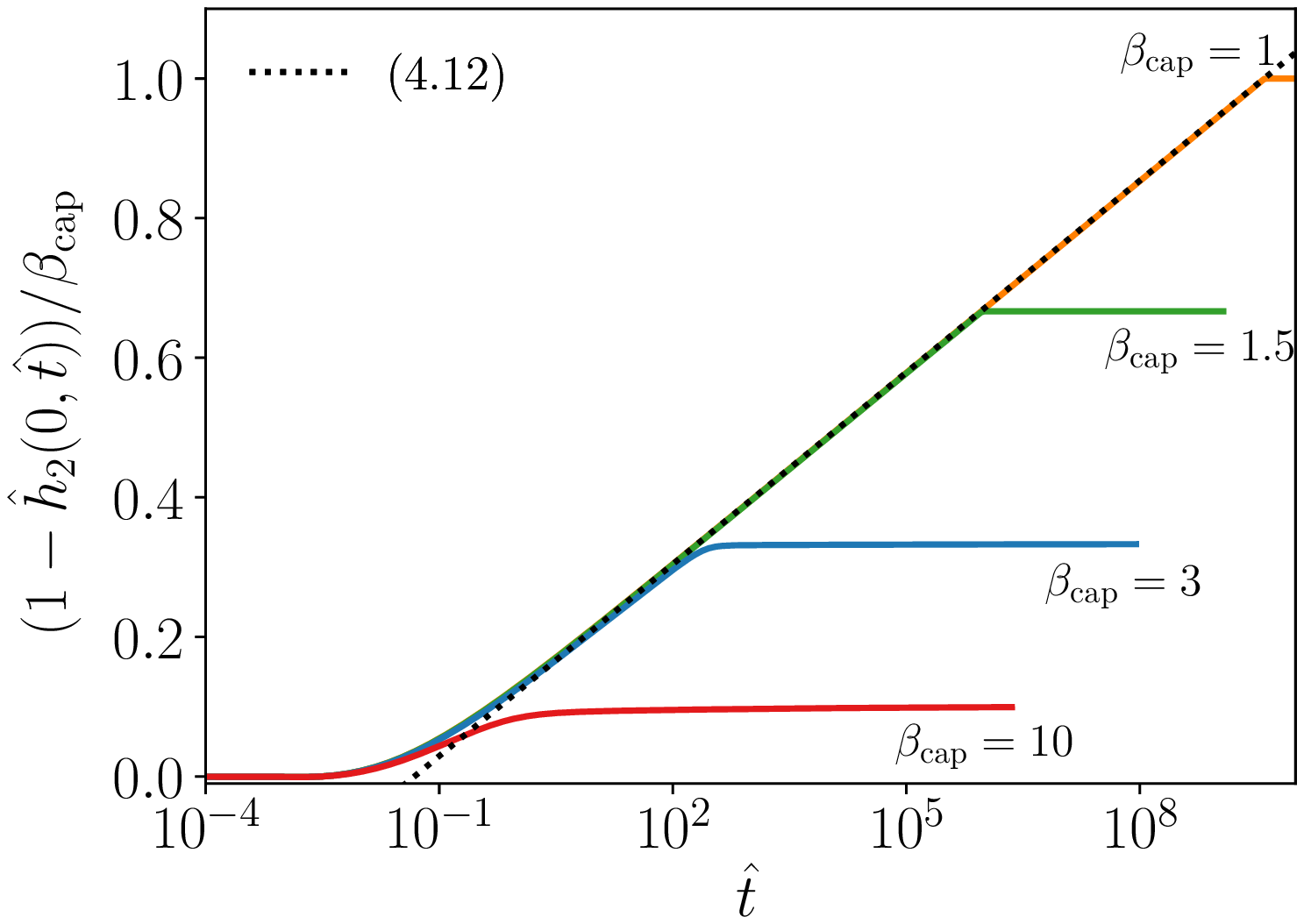}};
    \node[below right] at (fig.north west) {$(b)$};
  \end{tikzpicture}
  \begin{tikzpicture}
    \draw (0, 0) node[inner sep=0] (fig) {\includegraphics[width=0.48\textwidth]{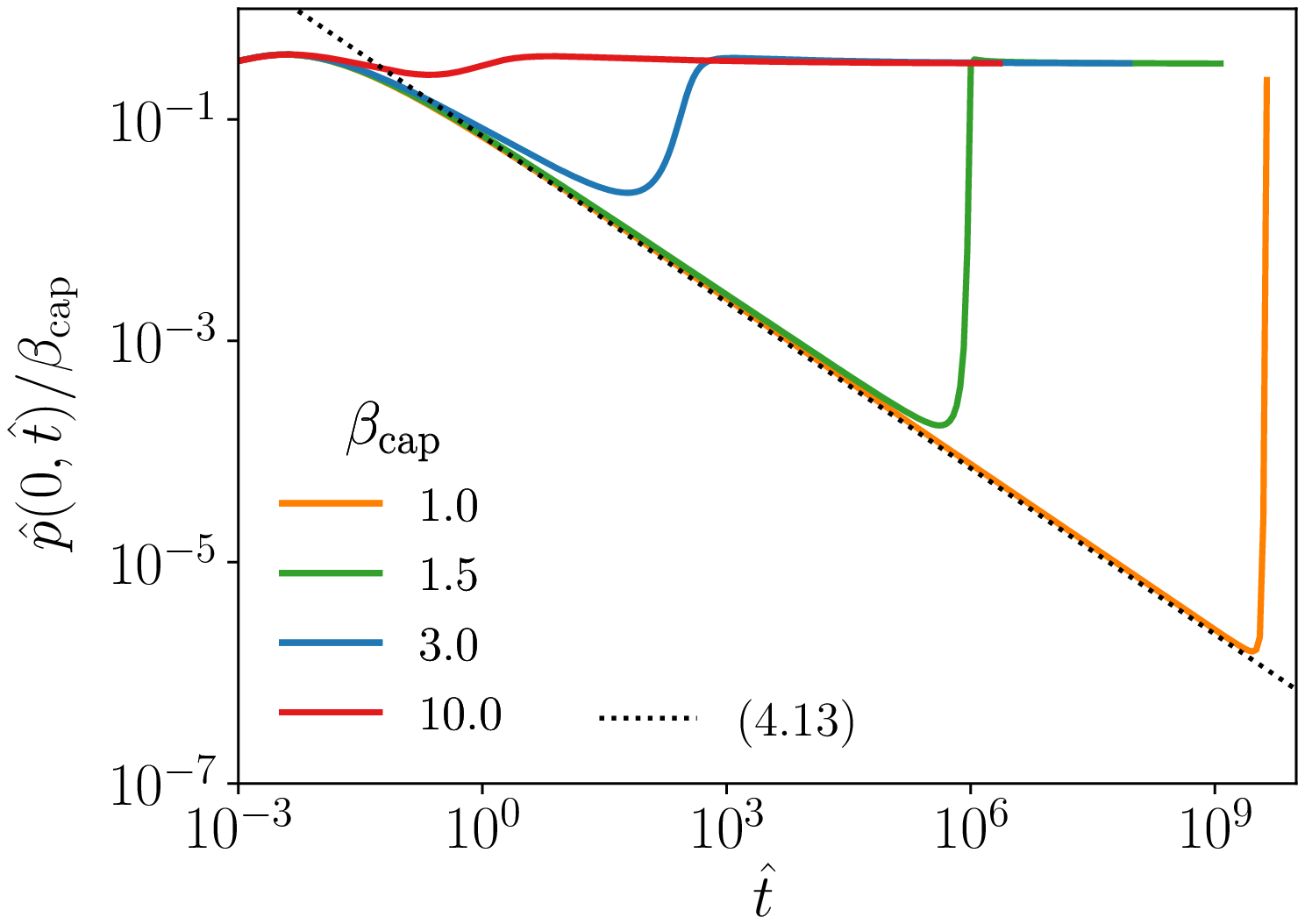}};
    \node[below right] at (fig.north west) {$(c)$};
  \end{tikzpicture}

	\caption{\label{fig:appendix_capillary}
	$(a)$ Deformation profiles of a viscous film under an external load for $\hat{t}=10^{-1},~10,~10^{3}$ and $10^{7}$ in the linear limit, using an initially flat profile and a confined uniform load defined numerically as  $\hat{p}_e(\hat{r})=p_0(1+{\rm erf}[a(1-\hat{r})])$ with $a=50$, ${\rm erf}$ the error function, $p_0$ chosen such that $\int_{\mathbb{R}^+} 2\pi p_e(\hat{r}) \hat{r} {\rm d}\hat{r}=1$.
	$(b,c)$ Evolution at the axis of symmetry ($\hat{r}=0$) of the $(b)$ height and $(c)$ pressure from simulations of the non-linear evolution equation \eqref{eq:thinfilmadim_cap}.
	}

\end{figure}

We are interested in 2-dimensional, axisymmetric solutions ($n=2$). 
Thus it is tempting to try to compute the solution for a Dirac load \eqref{eq:solutionDiracgeneral}: this generally gives the self-similar intermediate asymptotic solution \citep{Barenblatt1996} towards which solutions to any confined load would converge, similarly to the leveling scenario \citep{Benzaquen2013,Benzaquen2014} where an initial deformation is allowed to relax.
However the calculations show a singularity in time that cannot be integrated, which suggests that \eqref{eq:thinfilmlinear_cap}  does not possess a self-similar universal attractor.
Instead we compute the response to a confined uniform load, which is relevant for the droplet settling case.
By using \eqref{eq:Green_cap} and writing  spatial variables in polar coordinates, in particular letting $\vect{\hat{x}}=\left(\hat{r}\cos\theta,\hat{r}\sin\theta\right)$, and after using the identities $\int_{0}^{2\pi}\exp\left(ik\cos\theta\right){\rm d}\theta=2\pi J_0(k)$, $\int_{0}^{2\pi} \cos(\theta)\exp\left(-ik\cos\theta\right){\rm d}\theta=-2i\pi J_1(k)$ where $J_n$ is the Bessel function of the first kind of order $n$, \eqref{eq:capillarysth} simplifies to the following Hankel transform:
\begin{align}
    \hat{\epsilon}(\hat{r},\hat{t}) =  \psi\left(\frac{\hat{r}}{\hat{t}^{1/4}},\hat{t}\right),  &~~~~~~
    \psi(y,\hat{t})=\frac{\hat{t}^{1/4}}\pi \int_0^{+\infty} J_0\left(uy\right)J_1\left(\frac{u}{\hat{t}^{1/4}}\right) \frac{1-e^{-u^4}}{u^2}~{\rm d}u.
    \label{eq:cap_solution}
\end{align}
As expected, this does not converge towards a self-similar solution as illustrated in figure \ref{fig:appendix_capillary}$(a,b)$. This is a peculiar property of this problem and is at odds with  the leveling scenario mentioned above and with the elastohydrodynamic case discussed in the next section.
In particular, this means that the long-term behavior of the interface not only depends on the total weight applied on it but also on its specific distribution.

In the droplet settling dynamics, we are particularly interested in what happens at the axis of symmetry.
The integral in \eqref{eq:cap_solution} admits a closed form expression at $\hat{r}=0$ in terms of hypergeometric functions, shown in appendix \ref{sec:stuff}, from which  the following asymptotic expansion as $\hat{t}\rightarrow+\infty$ can be found:
\begin{subequations}
\begin{align}
    \hat{\epsilon}(0,\hat{t}) = \frac{\ln(\hat{t})}{8\pi} + k + \mathcal{O}\left(\hat{t}^{-1/2}\right), ~~~~&~~~~
    k = \frac{2-3\gamma_{\rm e \rm m}+4\ln(2)}{8\pi}\simeq0.121,  \\
    \hat{h}_2(0,\hat{t}) &\simeq 1-\beta_{\rm cap} k - \frac{\beta_{\rm cap}}{8\pi}\ln(\hat{t}),
\end{align}
    \label{eq:cap_linearsolution}
\end{subequations}
with $\gamma_{\rm e \rm m}\simeq0.577$ the Euler–Mascheroni constant.
It is also interesting to look at the capillary pressure at the axis of symmetry, -$\hat{\nabla}^2 \hat{h}(0,\hat{t})$, which in the linear approximation is given by $\hat{p}(0,\hat{t})=-\beta_{\rm cap} \hat{\nabla}^2 \hat{\epsilon}(0,\hat{t})$. By using \eqref{eq:cap_solution}, we can find that the asymptotic expansion of this quantity as $\hat{t}\rightarrow \infty$ is:
\begin{align}
    \hat{p}(0,\hat{t})= \frac{\beta_{\rm cap}}{8\sqrt{\pi}}\hat{t}^{-1/2} + \mathcal{O}\left(\hat{t}^{-1}\right).
    \label{eq:cap_pressure}
\end{align}

In this linear and asymptotic approximation, the time to reach $\hat{h}_2(0,t)=0$ is $\hat{\tau}=e^{8\pi\left(1/\beta_{\rm cap}-k\right)}\simeq 0.048e^{8\pi/\beta_{\rm cap}}$. 
We therefore expect \eqref{eq:cap_linearsolution} to be valid up to $\hat{t}\approx\hat{\tau}$, after which non-linear effects become important and prevent this singularity. We also expect that for this asymptotic linear solution to have enough time to develop before nonlinear effects appear we need an upper bound on $\beta_{\rm cap}$, i.e. a weak enough load.

We verify the results from this minimal model by comparing them against simulations of the non-linear evolution equation \eqref{eq:thinfilmadim_cap} in figure \ref{fig:appendix_capillary}$(b,c)$.
We indeed observe that the linear approximation \eqref{eq:thinfilmlinear_cap} and expansions \eqref{eq:cap_linearsolution} and \eqref{eq:cap_pressure} are accurate up to $\hat{t}\simeq\hat{\tau}$ when $\beta_{\rm cap} \lesssim 1.5$, while for larger values of $\beta_{\rm cap}$ the asymptotic regime is not reached before non-linearities appear.
We note that the fact that $\hat{\epsilon}(0,t)$ does not evolve as a power-law  is consistent with the lack of self-similarity of the solution.
Yet, the asymptotic pressure evolution \eqref{eq:cap_pressure} is nevertheless scale-invariant and  universal. 
We hypothesize that the logarithmic evolution given by \eqref{eq:cap_linearsolution}, $\hat{\epsilon}(0,\hat{t}) \simeq \ln(\hat{t})/8\pi + k$, is also universal but that the prefactor $k$ depends on the functional form of the load.
We verified this numerically for a few other loads, but these are not presented here.

 \begin{figure}
	    \centering
	    \begin{tikzpicture}
    \draw (0, 0) node[inner sep=0] (fig) {\includegraphics[width=0.48\textwidth]{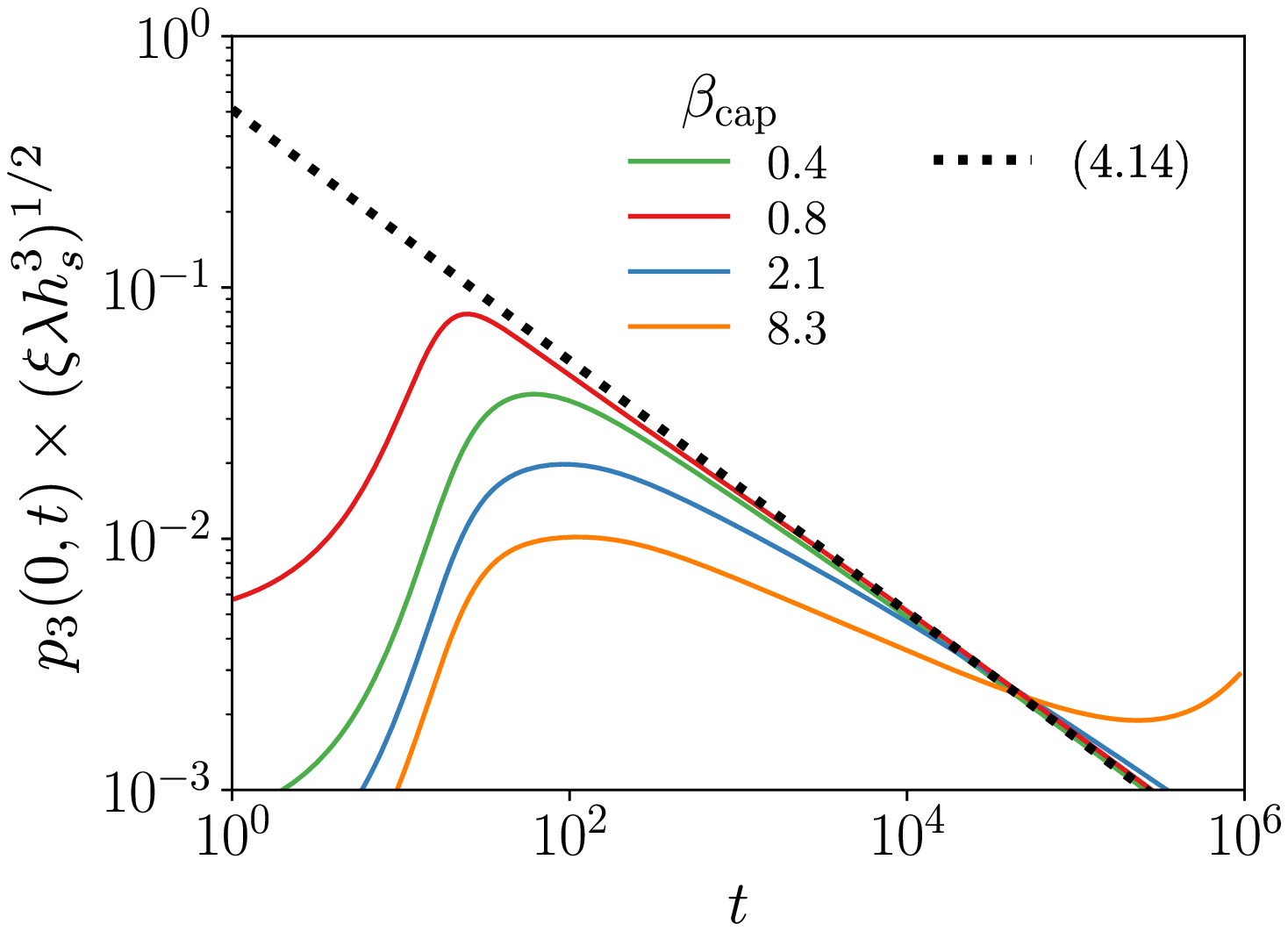}};
    \node[below right] at (fig.north west) {$(a)$};
  \end{tikzpicture}\\
  \begin{tikzpicture}
    \draw (0, 0) node[inner sep=0] (fig) {\includegraphics[width=0.48\textwidth]{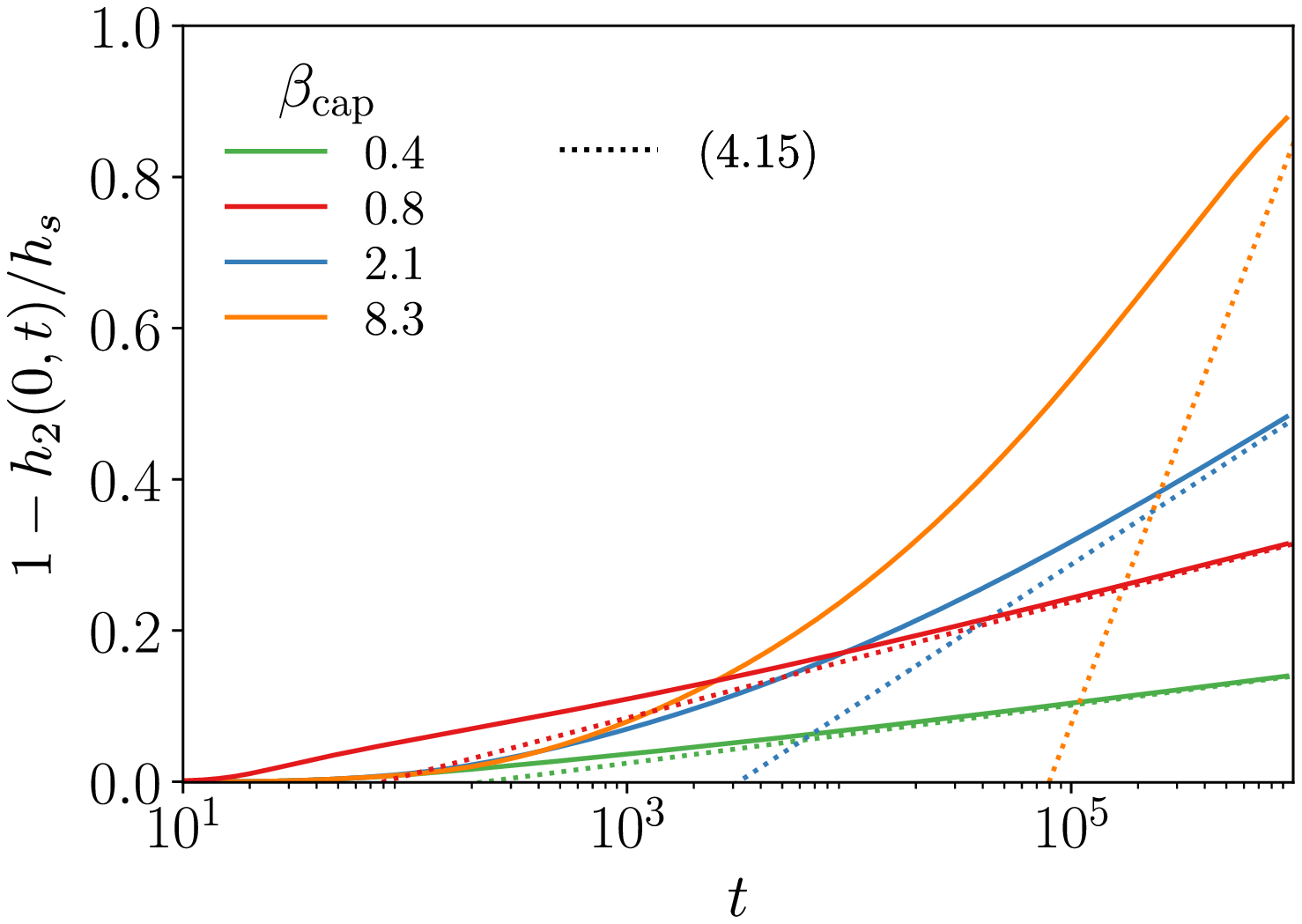}};
    \node[below right] at (fig.north west) {$(b)$};
  \end{tikzpicture}
  \begin{tikzpicture}
    \draw (0, 0) node[inner sep=0] (fig) {\includegraphics[width=0.48\textwidth]{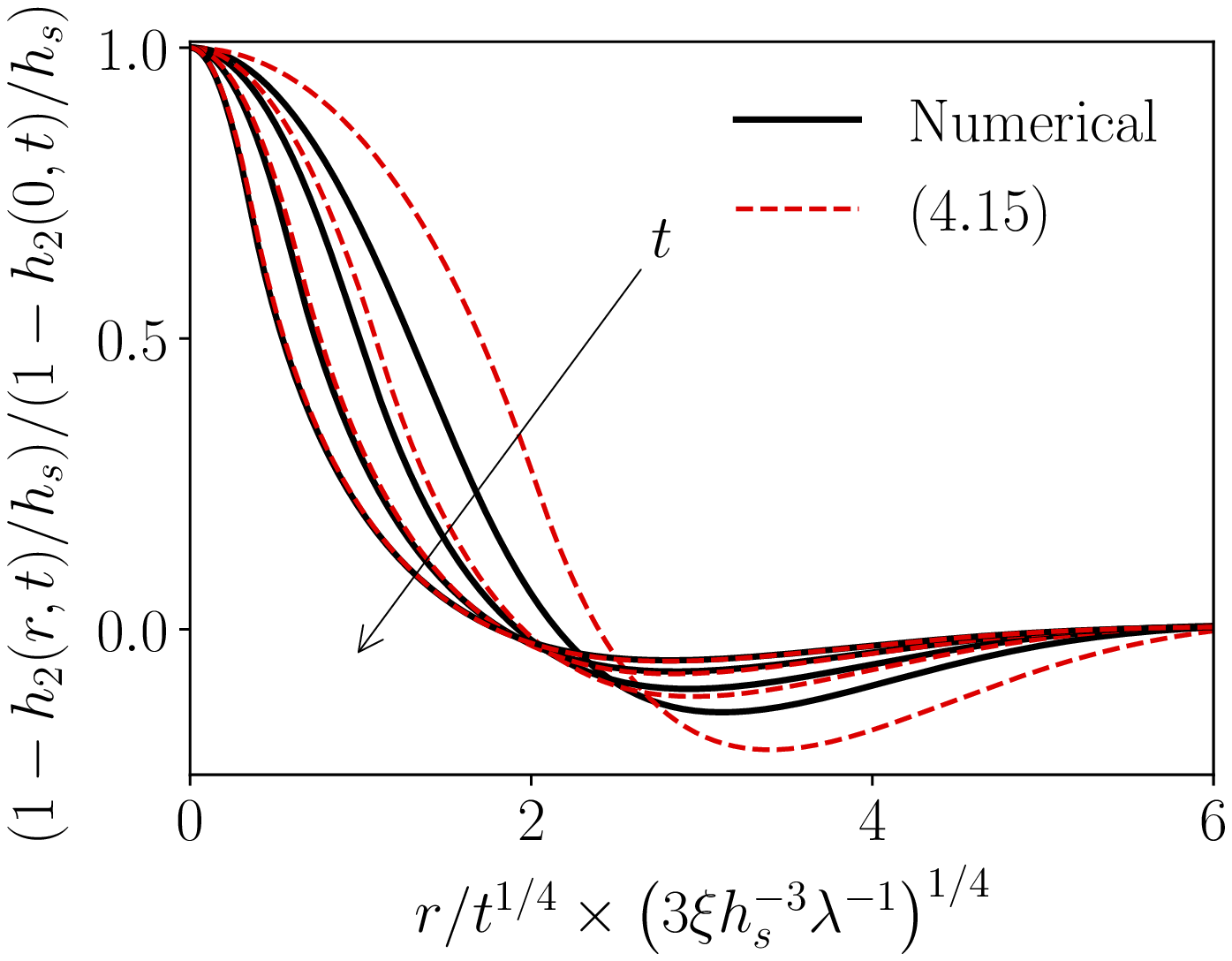}};
    \node[below right] at (fig.north west) {~$(c)$};
  \end{tikzpicture}
	\caption{\label{fig:cap_film}
	Height evolution of the coated liquid film for a droplet settling with $\delta=0.05$.
	Pressure $(a)$ and deformation $(b)$ of the liquid film at the axis of symmetry ($r=0$) for $\delta=0.05$, $\lambda=10^{-5}$ with  $h_s=0.1$, $\xi=0.2,~0.05,~0.01$ and $h_s=1$, $\xi=0.2$, corresponding to $\beta=8.4,~2.1,~0.4$ and 0.8, respectively.
	%The dotted lines represent the predictions \eqref{eq:cap_pressure2} and  \eqref{eq:cap_height}.
	$(c)$ Normalized deformation of coated the liquid film shown here for $\lambda=10^{-5}$, $h_s=0.1$, $\xi=0.2$, corresponding to $\beta=0.8$, and at times $t=10^2,~10^3,~10^4$ and $10^5$.
	%Solid lines are the numerical results and dashed lines are the prediction from \eqref{eq:cap_profile}.
	}
\end{figure}

\subsection{Response of a viscous film to a settling droplet}

We now come back to the complete droplet settling case, where we anticipated from the discussion of \S \ref{subsec:dimple_cap} that the dimple radius evolves towards a constant, $(2(\delta+\xi)/3)^{1/2}$, and that the pressure in the dimple is also constant, equal to $2/(\delta+\xi)$.
Assuming that this is indeed the case, we expect the results derived in \S\ref{sec:appendix_capillary} to apply in the case of droplet settling as well.
Accordingly, we define  $r_c=(2(\delta+\xi)/3)^{1/2}$ and $p_c=2\pi/(\delta+\xi)$; this gives $t_c=4\xi(\delta+\xi)^2/3h_s^3\lambda$, $\beta_{\rm cap}=4\pi\xi/3h_s$ and the rescaled height $\hat{h}_2(\hat{r},\hat{t})$ is governed by \eqref{eq:thinfilmadim_cap} where the external pressure $\hat{p}_e$ is now that in the air layer, $\hat{p}_2(\hat{r},\hat{t})$.

We first need to verify that \eqref{eq:pcapillarysimplified} applies, i.e. that $\xi p_3(r,t) \ll \left(\delta+\xi\right)p_2(r,t)$.
Assuming $\beta_{\rm cap} \lesssim 1$ for an asymptotic regime to be reached and $\hat{t}\lesssim \hat{\tau} \simeq 0.05 e^{8\pi/\beta_{\rm cap}}$, i.e. ${t\lesssim 0.1 \xi(\delta+\xi)^2 h_s^{-3} \lambda^{-1} e^{6h_s/\xi}}$ for the deformations  of the liquid film interface to remain small, the full film profile can be derived as shown in \S\ref{sec:appendix_capillary}. 
In particular, \eqref{eq:cap_pressure} gives the pressure at $r=0$ in the liquid film as:
\begin{align}
    p_3(0,t) = \frac{\beta_{\rm cap}}{8\sqrt{\pi}} \frac{h_s}{\xi r_c^2} \left(\frac{t}{t_c}\right)^{-1/2} = \frac{\sqrt{\pi}}{2\sqrt{3}} \left(\frac{\xi}{h_s^3 \lambda t}\right)^{1/2}
    \label{eq:cap_pressure2}.
\end{align}
In this regime ($\beta_{\rm cap} \lesssim 1$, $\hat{t}\lesssim\hat{\tau}$), we therefore expect the pressure in the liquid to continuously decrease and the condition $\xi p_3(r,t) \ll (\delta+\xi) p_2(r,t)$ will eventually be satisfied: this justifies a posteriori the discussion of \S\ref{subsec:dimple_cap} and the scaling we have chosen for $r_c$ and $p_c$.
We can now expect from \eqref{eq:cap_solution} and \eqref{eq:cap_linearsolution} that the sheet profile $h_2(r,t)$ evolves according to: 
\begin{subequations}
\begin{align}
    1-\frac{h_2(r,t)}{h_s}=\beta_{\rm cap}\psi\left(\frac{\hat{r}}{\hat{t}^{1/4}},\hat{t}\right)=\frac{4\pi\xi}{3h_s}
    \psi\left(\frac{r}{\left(h_s^3\lambda t\right)^{1/4}},\frac{3h_s^3\lambda}{4\xi(\delta+\xi)^2} t\right),
    \label{eq:cap_profile} \\
    1-\frac{h_2(0,t)}{h_s} = \beta_{\rm cap} \left(\frac{\ln(t/t_c)}{8\pi}+k\right) \simeq \frac{\xi}{6 h_s}\left[
    \ln\left(\frac{h_s^3\lambda}{\xi(\delta+\xi)^2}t\right) + 2.75
    \right].
    \label{eq:cap_height}
\end{align}
\label{eq:cap_all}
\end{subequations}

\subsection{Numerical results}
\label{subsec:cap_num}
 \begin{figure}
  \centering
  \begin{tikzpicture}
    \draw (0, 0) node[inner sep=0] (fig) {\includegraphics[width=0.48\textwidth]{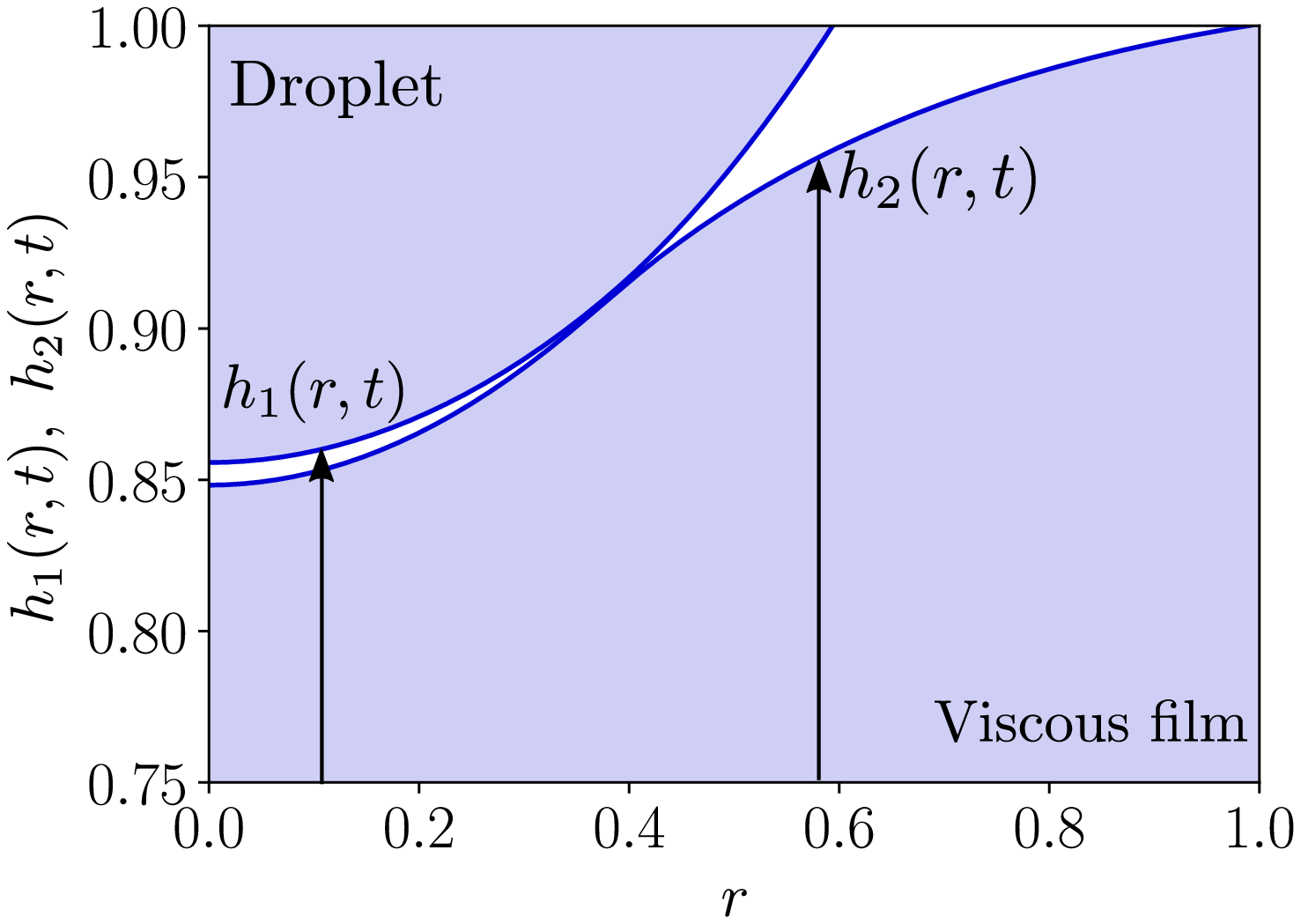}};
    \node[below right] at (fig.north west) {$(a)$};
  \end{tikzpicture}
  \begin{tikzpicture}
    \draw (0, 0) node[inner sep=0] (fig) {\includegraphics[width=0.48\textwidth]{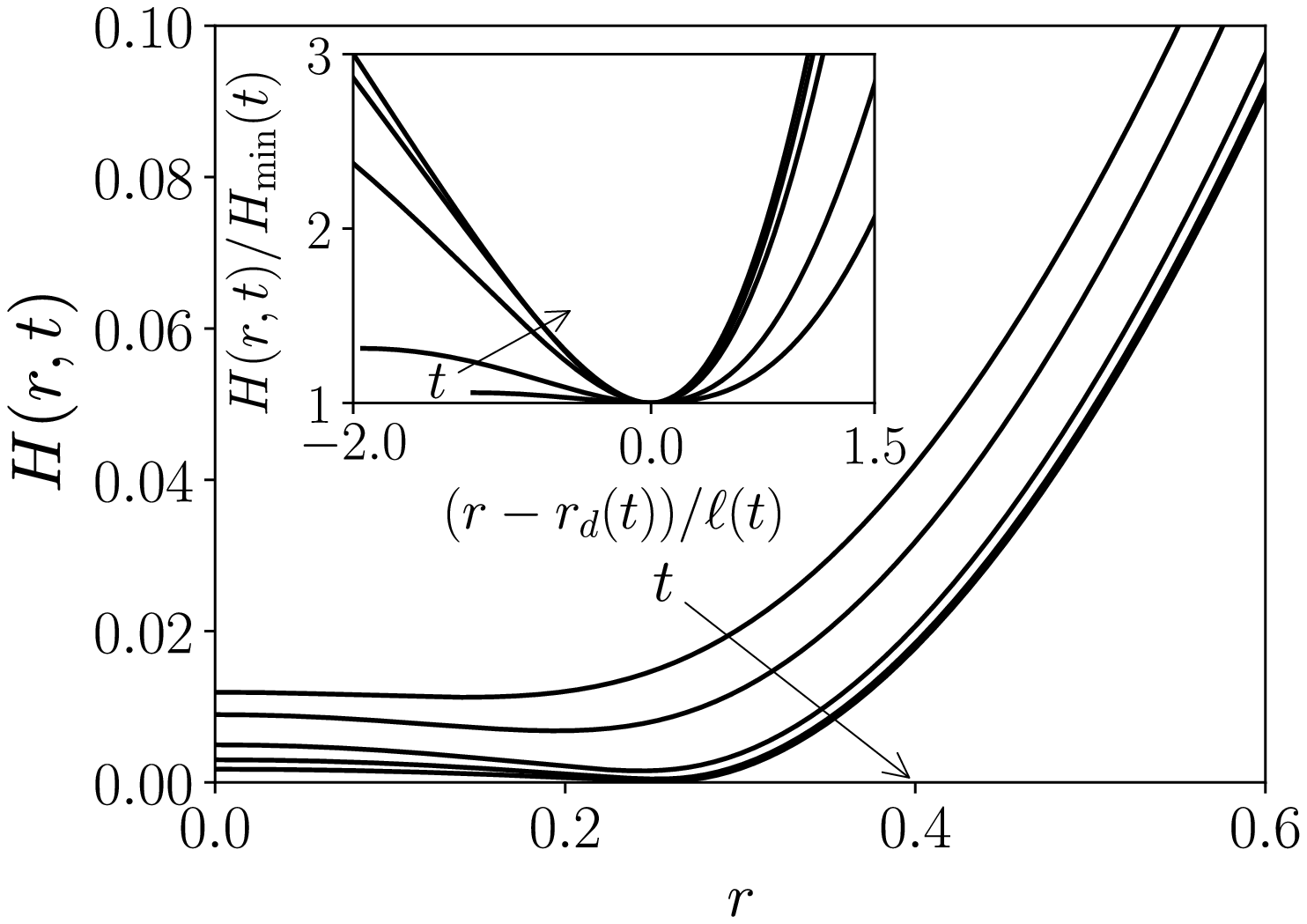}};
    \node[below right] at (fig.north west) {$(b)$};
  \end{tikzpicture}
  \begin{tikzpicture}
    \draw (0, 0) node[inner sep=0] (fig) {\includegraphics[width=0.48\textwidth]{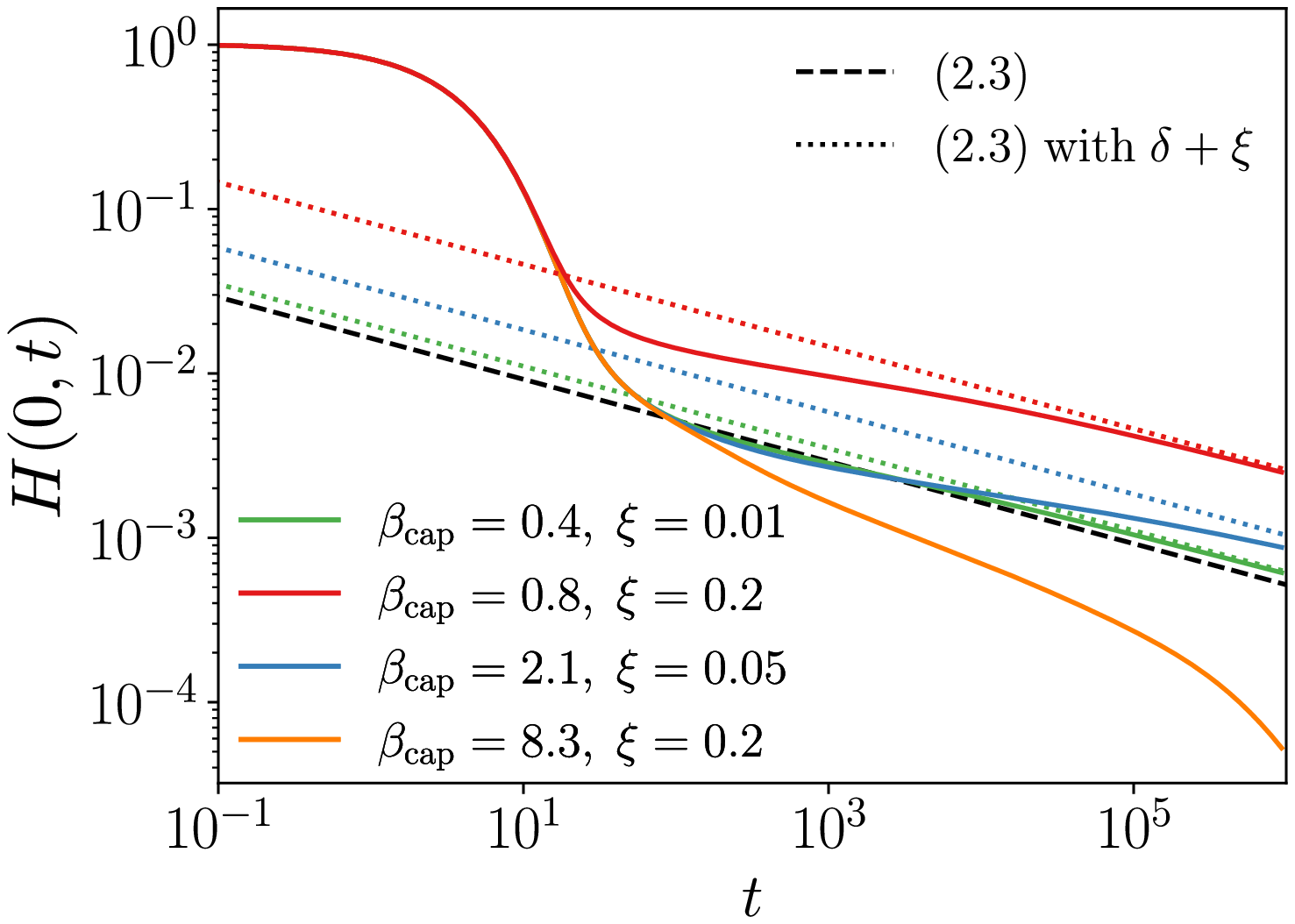}};
    \node[below right] at (fig.north west) {$(c)$};
  \end{tikzpicture}
  \begin{tikzpicture}
    \draw (0, 0) node[inner sep=0] (fig) {\includegraphics[width=0.48\textwidth]{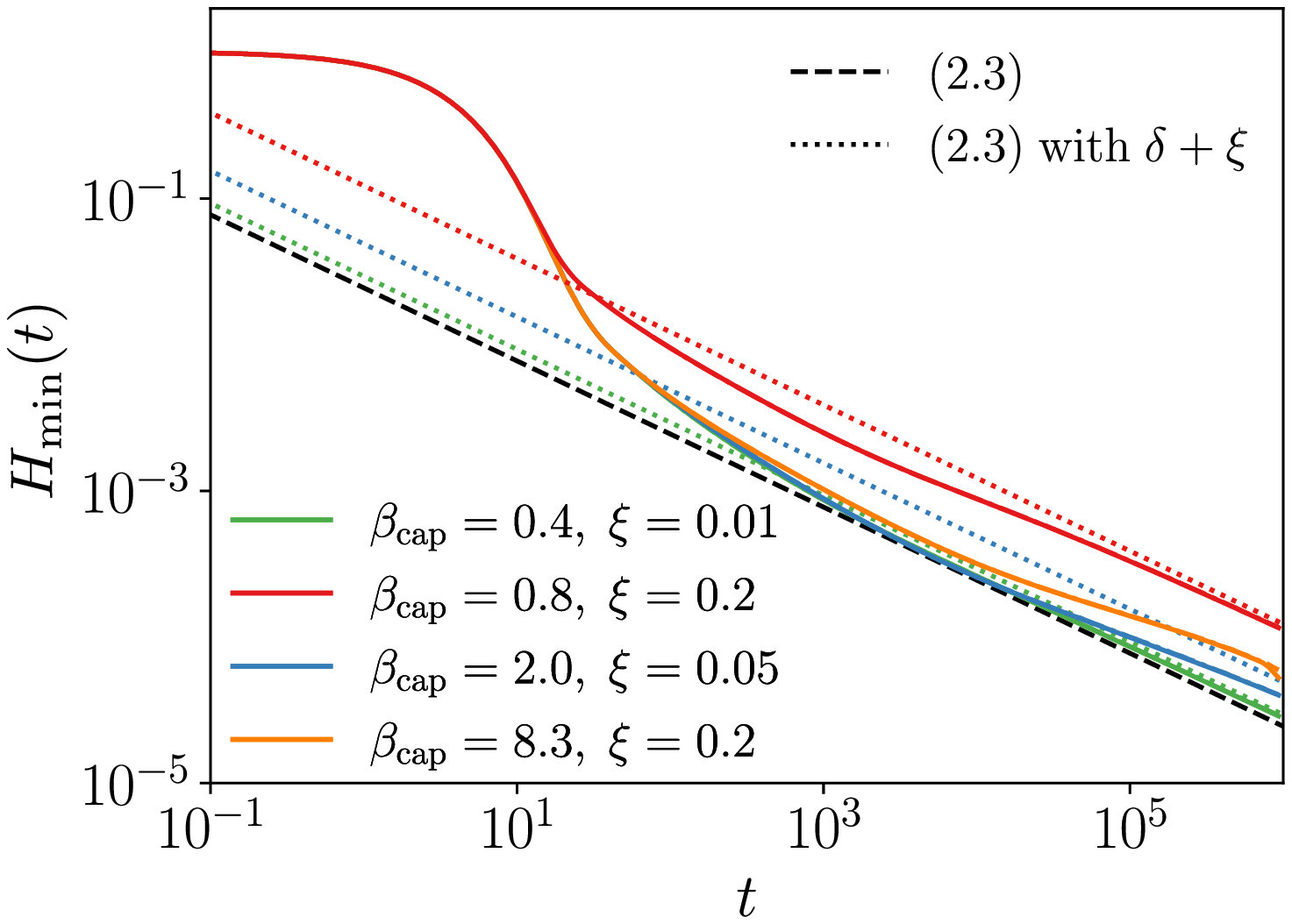}};
    \node[below right] at (fig.north west) {$(d)$};
  \end{tikzpicture}
	\caption{\label{fig:cap_gap} Evolution of the air layer for the settling of a droplet with $\delta=0.05$ on a viscous liquid film.
	$(a)$ Profiles for $\xi=0.2$, $h_s=1$, $\lambda=10^{-5}$ of $(a)$ the droplet  $h_1(r,t)$ and the liquid film $h_2(r,t)$ at $t=5000$.
	Air film profile $H(r,t)=h_1(r,t)-h_2(r,t)$ for $\xi=0.05$, $h_s=1$, $\lambda=10^{-5}$ at $t=50,~10^2$, $10^3$, $10^4$ and $10^5$. The neck is predicted to be located at  $r=(2(\delta+\xi)/3)^{1/2}\simeq0.26$.
	The inset of $(b)$ shows the rescaled neck structure near $r=r_d(t)$ with $\ell(t)=(\delta+\xi)^{1/2} t^{-1/4}$.
	$(c,d)$  Thickness of the air film at $r=0$ $(c)$ and at its minimum ($d$) for $\delta=0.05$, $\lambda=10^{-5}$ with  $h_s=0.1$, $\xi=0.2,~0.05,~0.01$ and $h_s=1$, $\xi=0.2$, corresponding to $\beta_{\rm cap}=8.4,~2.1,~0.4$ and 0.8, respectively.
	}
\end{figure}

In figure \ref{fig:cap_film} we show the time evolution of the pressure in the liquid film  and its height from numerical solutions of the complete non-linear equations, i.e. considering the normal stress balances \eqref{eq:pressure_h1} and \eqref{eq:cap_pressure}, the force balance \eqref{eq:p_forcebalance}, and the governing equations \eqref{eq:Hh2_slip} accounting for the full shear stress balance.
We present numerical results for $\lambda=10^{-5}$, and show in appendix \S \ref{sec:appendixSLIP} that this viscosity ratio is small enough for the effects of shear stresses to be negligible.
The results are in close agreement with the theoretical expectations  \eqref{eq:cap_pressure2} and \eqref{eq:cap_all} for $\beta\lesssim 1$, a condition required for the asymptotic regime to be reached.

The results presented in figure \ref{fig:cap_gap} for the air layer dynamics show that the discussion of \S \ref{subsec:dimple_cap} indeed applies.
The long-term asymptotics of the air film thickness $H(r,t)$ is similar to that of the solid case, with $\delta\rightarrow\delta+\xi$.
We note that the air layer thickness profile has a dimple structure while both the droplet and the liquid film profiles are monotonically increasing near the axis of symmetry and that the neck of the dimple is located at the inflection point of the film profile.
As discussed by \citet{Duchemin2020} when a droplet settles on a solid surface, we expect that this neck approaches a self-similar form, $H(r,t)=H_{\rm min}(t) F\left(\frac{r-r_d(t)}{\ell(t)}\right)$,  where $F$ is linear for $r<r_d$ and quadratic for $r>r_d$, and where $\ell(t)$ is the radial extent of the neck region. This is verified in figure \ref{fig:cap_gap}$(b)$, where we used $\ell(t)=(\delta+\xi)^{1/2} t^{-1/4}$ following the scaling analysis presented in \S\ref{subsec:pheno}.

\begin{figure}
  \centering
  \begin{tikzpicture}
    \draw (0, 0) node[inner sep=0] (fig) {\includegraphics[width=0.48\textwidth]{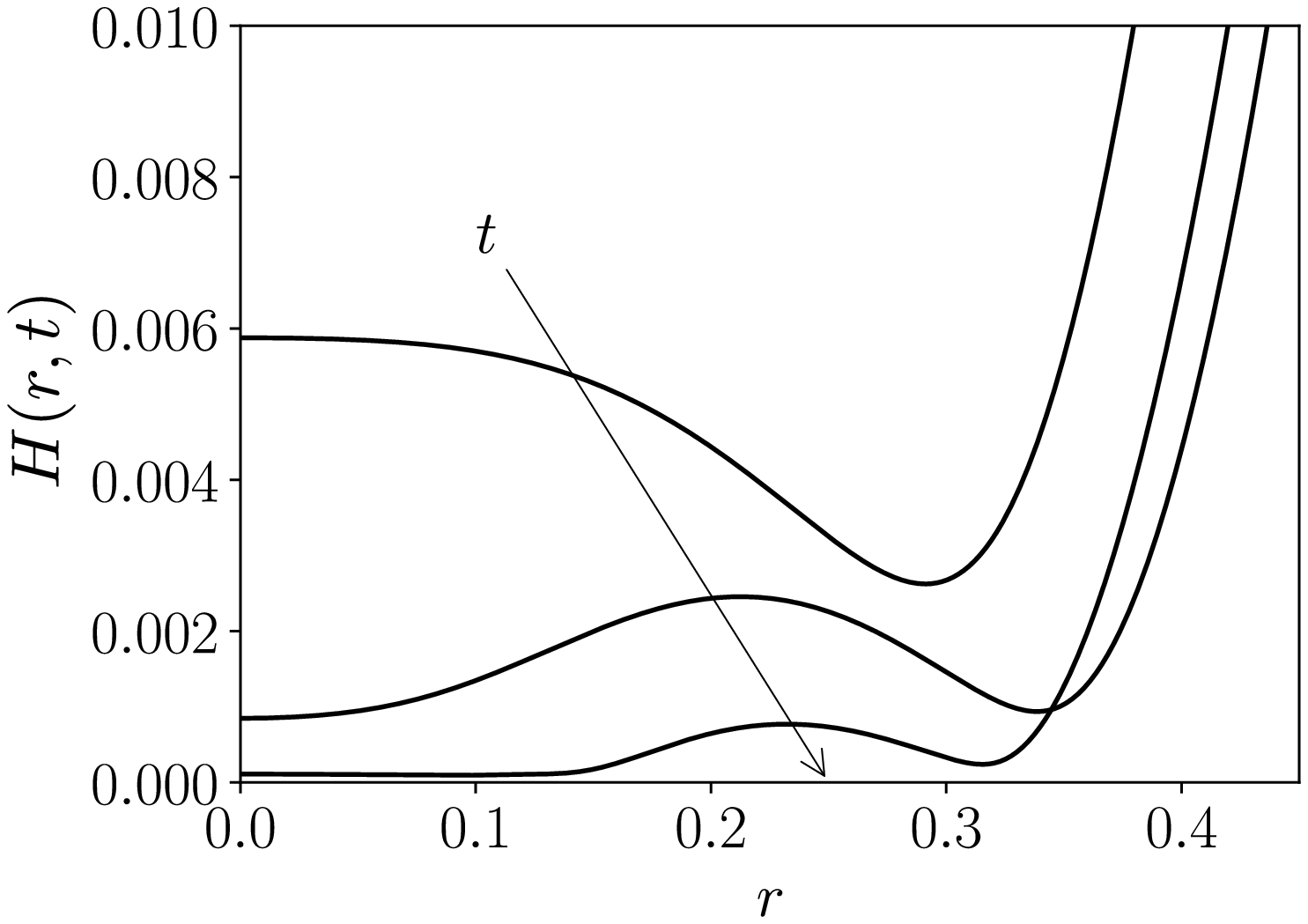}};
    \node[below right] at (fig.north west) {$(a)$};
  \end{tikzpicture}
  \begin{tikzpicture}
    \draw (0, 0) node[inner sep=0] (fig) {\includegraphics[width=0.48\textwidth]{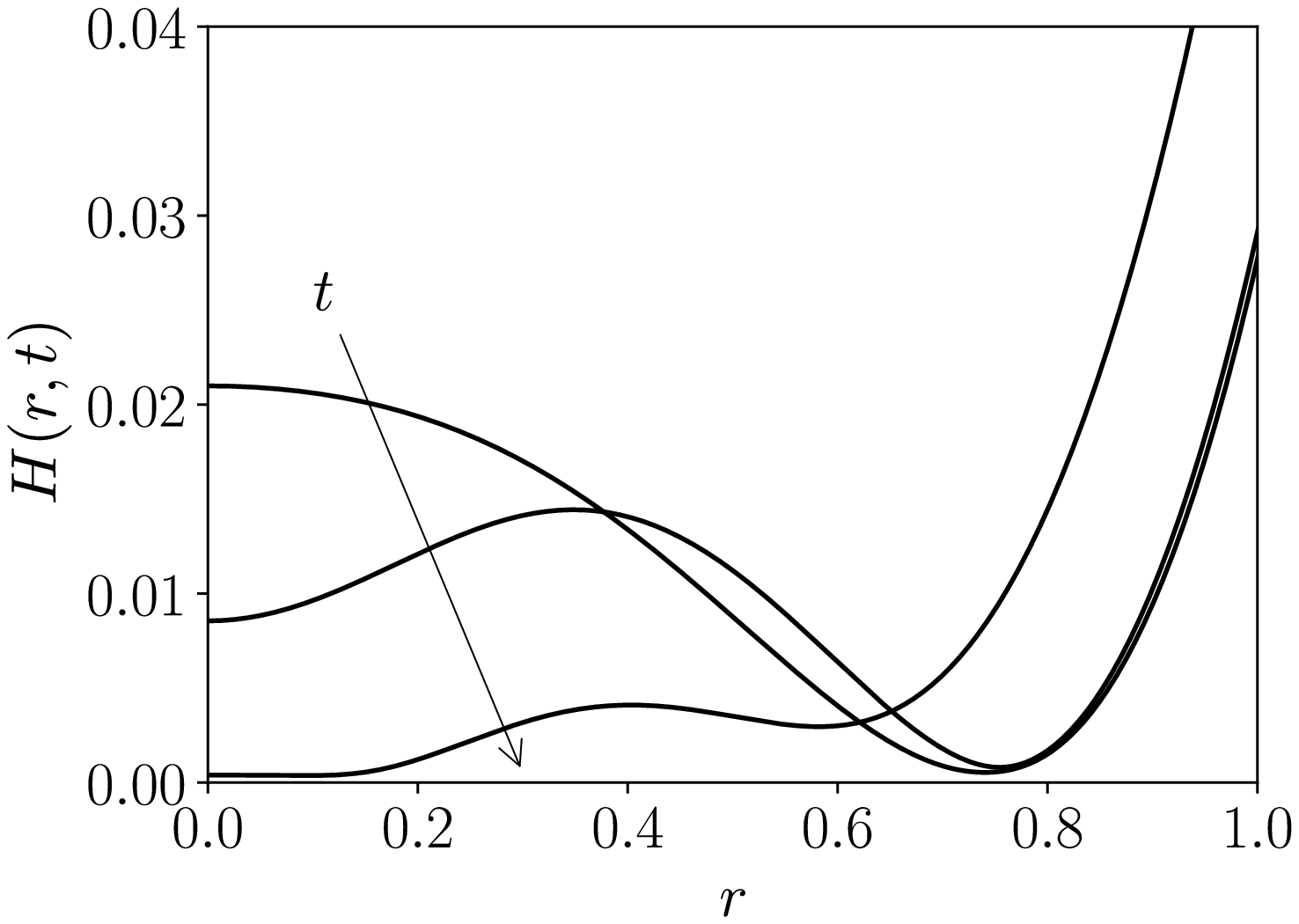}};
    \node[below right] at (fig.north west) {$(b)$};
  \end{tikzpicture}

	\caption{\label{fig:cap_wrimple}
	$(a)$ Interface profile of the air layer thickness for a droplet settling on a viscous film  at $t=500,$ $10^4$ and $10^5$, for $\xi=0.2$, $h_s=0.1$, $\lambda=10^{-3}$. 
	$(b)$
	Interface profile of the air layer thickness for a droplet settling on an elastic sheet supported by a thin viscous layer  at $t=10.000,~20.000$ and $40.000$, for $\alpha=10$, $h_s=1$, $\lambda=10^{-3}$.
	}
\end{figure}

When the deformations of the liquid film become large, the pressure $p_3(r,t)$ increases and can no longer be neglected in \eqref{eq:pcapillarysimplified}.
This leads the air layer into another regime.
For $\beta_{\rm cap} \lesssim 1$, the time at which deformations become large can be estimated from \S \ref{sec:appendix_capillary}
as $\tau \approx 0.05 t_0 e^{8\pi/\beta_{\rm cap}}\simeq 0.1 \xi(\delta+\xi)^2 h_s^{-3}\lambda^{-1} e^{6h_s/\xi}$.
The effect of large deformations on the liquid film evolution can be seen in figure \ref{fig:cap_film}$(a,b)$ for the case $\beta_{\rm cap}=8.3$ at time $t\gtrsim 10^4$. 
As shown in figure \ref{fig:cap_gap}$(c)$, the resulting effect on the air film is characterized by a sudden decrease of its thickness at the axis of symmetry.
In fact, figure \ref{fig:cap_wrimple}$(a)$ shows that when non-linearities of the liquid film are present the profile evolves towards a more complicated shape presenting a local maximum not located at the axis of symmetry and a minimum not located at the neck.
Such a shape is not uncommon in thin film drainage and has been observed, both experimentally and numerically, for droplets and bubbles approaching a rigid substrate \citep{Clasohm2005,Ajaev2007,Ajaev2008}, and has been referred to as rippled deformation, or as a wimple \citep{Chan2011}.
This wimple occurs due to a competition of two effects in the lubrication pressure: the capillary-driven deformations of the drop/bubble, along with an additional physical effect.
In prior work this additional effect originated from surface forces, such as van der Waals interaction or electrostatic effects.
Here we do not consider such effects, but it is the non-linear deformations of the liquid film towards which the droplet settles that lead to a wimple.
Such a rippled shape also seems to develop in the numerical simulations of \citet{Duchemin2020} during the late stage of a large drop settling on a liquid film.
The results we presented above are valid before the appearance of a wimple; an understanding of this shape and the associated drainage dynamic would be an important step towards an understanding of droplet settling when the parameter $\beta_{\rm cap}=4\pi\xi/3h_s$ becomes large, i.e. on very thin films.

\section{Solid substrate coated with a thin viscous film and an elastic sheet atop}
\label{sec:EHD} 
 We now consider the case where the interface of the thin viscous film is subsituted by an elastic sheet (figure \ref{fig:setup}$d$), so that its surface tension becomes irrelevant.
 We neglect any tension in the sheet and only consider its bending, an assumption valid when the deformation of the sheet remains small compared to its thickness \citep{LandauBook}.
 This  leads to a thin film equation for $h_2(r,t)$ similar to \eqref{eq:thinfilm_H} and \eqref{eq:thinfilmviscous_h2}, where the pressure enters the equation as $p_3^\star(r^\star,t^\star)=p_2^\star + B \nabla^{\star4}h_2^\star(r^\star,t^\star)$ (e.g. \citet{Hosoi2004}).
 The operator $\nabla^4(.)=\frac{1}{r}\partiald{}{r}\left(r\partiald{}{r}\left(\frac1r\partiald{}{r}\left(r\partiald{(.)}{r}\right)\right)\right)$ is the axisymmetric bi-Laplacian and $B=Ed^{\star3}/12(1-\nu^2)$ is the bending stiffness of the sheet, with $d^\star,~E$ and $\nu$ its thickness, Young's modulus and Poisson ratio, respectively.
 Using the scaling defined in \S \ref{sec:solidcase}, the non-dimensional governing equation for the height of the film reads:
\begin{subequations}
\begin{align}
    \partiald{h_2}{t}(r,t) &= \frac{\lambda}{12}  \frac{1}{r} \partiald{}{r}\left(r h_2^3(r,t)  \partiald{p_3}{r}(r,t)\right),
    \label{eq:thinfilm_h2} \\
    \alpha \left(p_3(r,t)-p_2(r,t)\right) &=\nabla^4 h_2 (r,t)
    ,
    \label{eq:thinfilm_h2_pressurebend}
\end{align}
\label{eq:EHD}
\end{subequations}
with $\lambda=\mu_2/\mu_3$,  $\mu_3$ the viscosity of the fluid supporting the elastic sheet. The effect of bending is quantified  by the parameter $\alpha=\Delta \rho g a^{\star4}/B=(a^\star/\ell_g^\star)^4$, analoguous to an elastic Bond number, which compares the droplet radius to the elastogravity length $\ell_g^\star=(B/ g \Delta\rho)^{1/4}$.
At the initial time $t=0$ and far from the droplet, the height of the sheet is $h_2(r,0)=h_2(r\rightarrow\infty,t)=h_s$.
We set $\delta=0.05$ for the numerical results presented below.

We note that for a typical density ratio between gas and liquids $\Delta \rho=10^3~{\rm kg~m}^{-3}$, and considering very soft elastic sheets, e.g. $\nu \simeq 0.5$, $E=\mathcal{O}(10^9~{\rm Pa})$,  $d=\mathcal{O}(100~{\rm nm})$, $\ell_g$ is approximately 50 $\mu$m, which is also the typical upper bound we expect for droplet size in order for the lubrication scalings presented in \S\ref{subsec:governing} to be valid.
We are therefore mostly interested in the cases where $\alpha\leq1.$

 \subsection{Response of an elastic sheet supported by a viscous film under constant load}
 \label{sec:appendix_EHD}

  \begin{figure}
    \centering
    \begin{tikzpicture}
    \draw (0, 0) node[inner sep=0] (fig) {\includegraphics[width=0.5\textwidth]{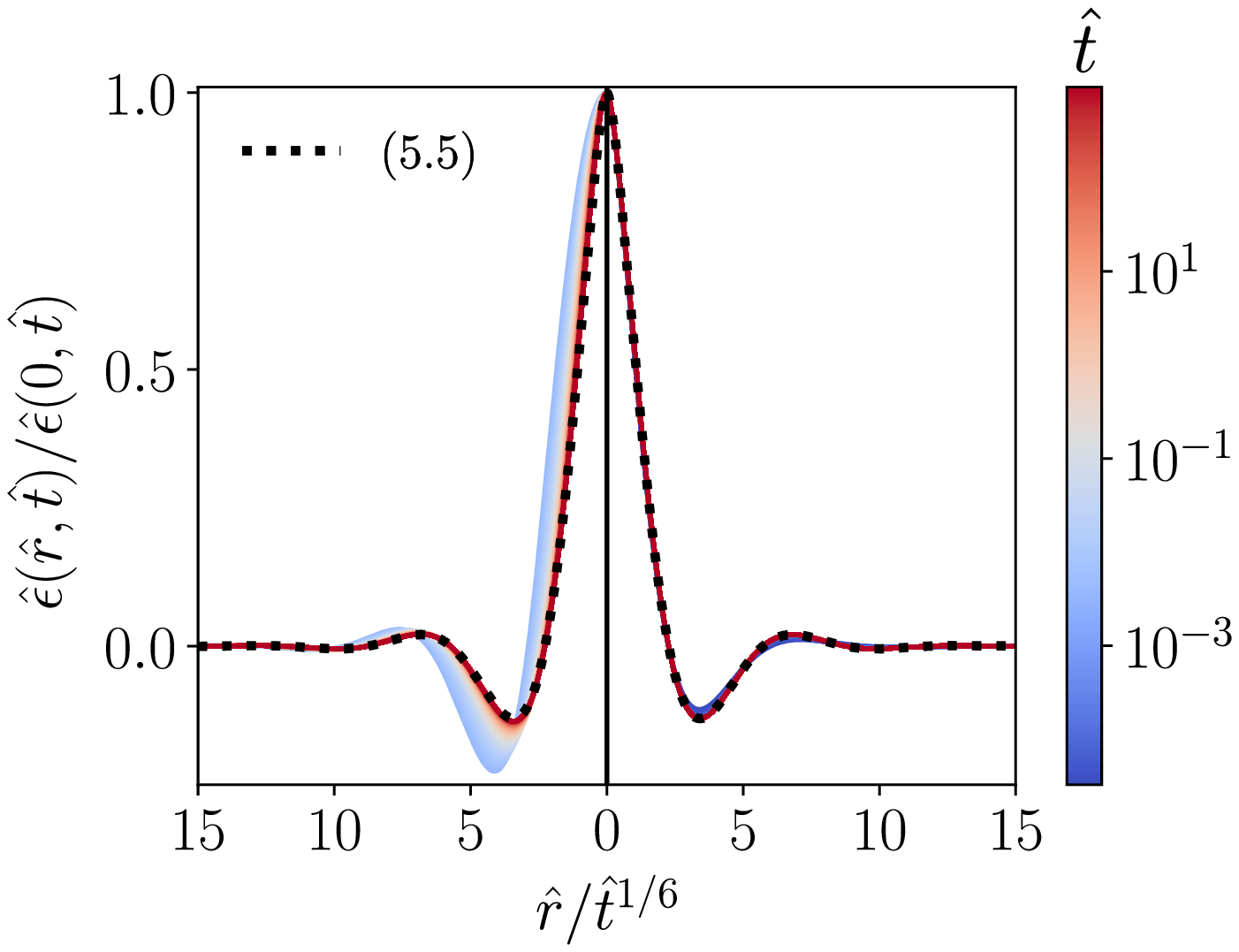}};
    \node[below right] at (fig.north west) {$(a)$};
  \end{tikzpicture}
  \begin{tikzpicture}
    \draw (0, 0) node[inner sep=0] (fig) {\includegraphics[width=0.48\textwidth]{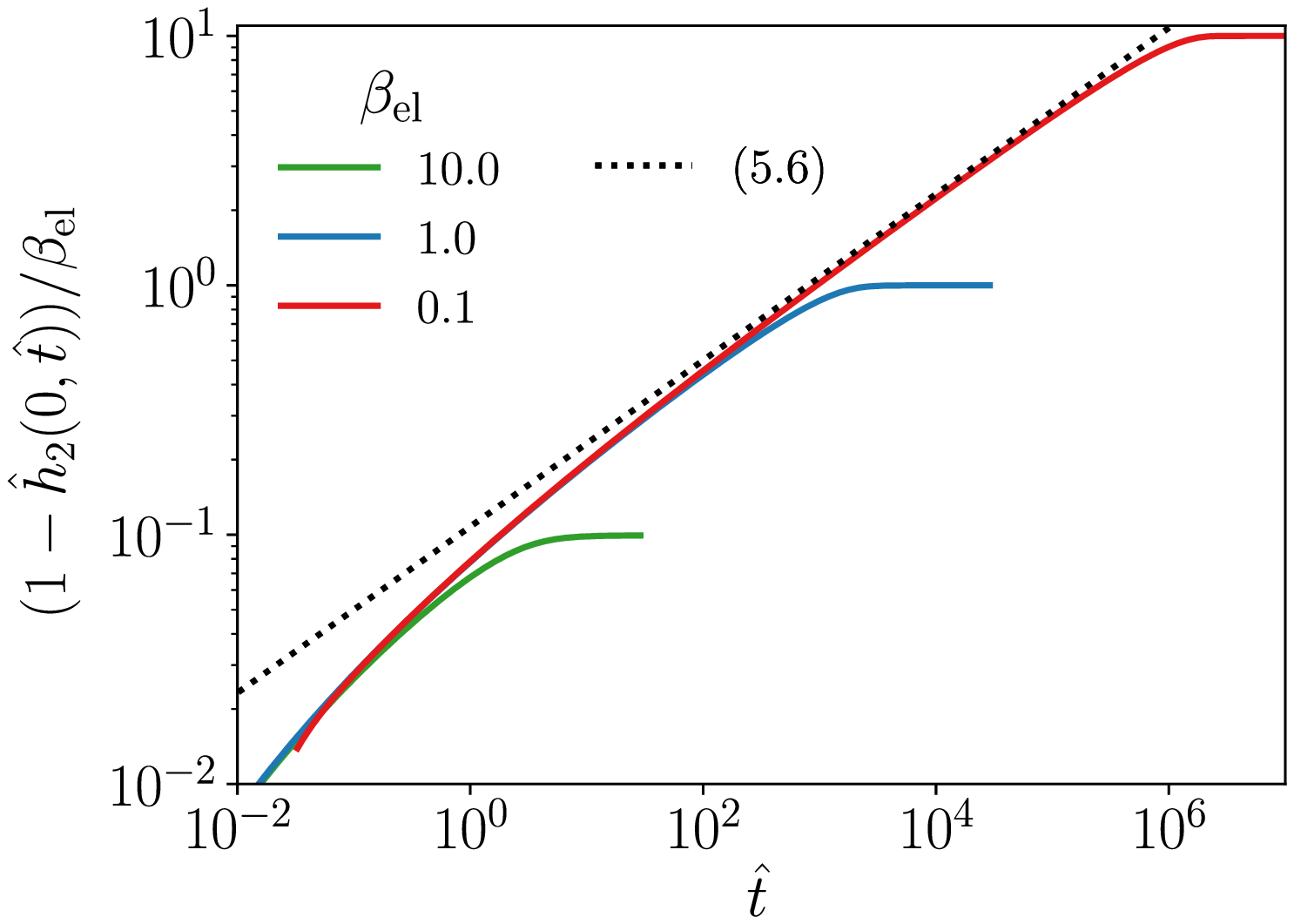}};
    \node[below right] at (fig.north west) {$(b)$};
  \end{tikzpicture}
	\caption{\label{fig:bendingtheory}
	Deformation profile of an elastic sheet supported by viscous fluid and exposed to an external load.
	$(a)$
	Solution of the linear evolution equation \eqref{eq:thinfilmlinear} in cylindrical coordinates for (right panel) a narrow load approaching a Dirac distribution, $p_e(\hat{r})=p_0{\rm exp}[-(\hat{r}/a)^2]/\hat{r}$ with $a=1/200$ and (left panel) a confined uniform load, $p_e(\hat{r})=p_0(1+{\rm erf}[a(1-\hat{r})])$ with $a=50$, ${\rm erf}$ the error function.
	In both cases $p_0$ is chosen such that $\int_{\mathbb{R}^+} 2\pi p_e(\hat{r}) \hat{r} {\rm d}\hat{r}=1$.
	The rescaled solution converges to the self-similar solution \eqref{eq:dirac2D_EHD}, and faster for the narrow load (right) than for the distributed one (left).
	$(b)$ Evolution of the deformation at the axis of symmetry ($\hat{r}=0$) from the non-linear evolution equation \eqref{eq:thinfilmadim} and for a confined  uniform load.
	}
\end{figure}
 
We start by looking at the general case of an external load $p_e(\vect{x})$ exerted over an elastic sheet supported by a thin viscous film, before coming back to the complete droplet settling situation.  The rescaled governing equation of the sheet height for a load $p_e(\vect{\hat{x}})$ reads:
\begin{align}
    \partiald{\hat{h}_2}{\hat{t}}(\vect{\hat{x}},\hat{t}) = \hat{\nabla} \cdot \left( \hat{h}_2^3(\vect{\hat{x}},\hat{t})\hat{\nabla} \left(\hat{\nabla}^4 \hat{h}_2(\vect{\hat{x}},\hat{t}) + \beta_{\rm el}  \hat{p}_e(\vect{\hat{x}}) \right) \right),{}
    \label{eq:thinfilmadim}
\end{align}
with $t=\hat{t} t_c$, $h_2=\hat{h}_2(\vect{\hat{x}},t) h_s$, $\nabla=\hat{\nabla}/x_c$, $p_e=\hat{p}_e(\vect{\hat{x}}) p_c$; with $p_c$ the characteristic magnitude of the load which is distributed over a characteristic length $x_c$, $h_s$ the initial height of the elastic sheet, and $t_c=12\alpha x_c^6/h_s^3 \lambda$  the characteristic time.
The parameter $\beta_{\rm el}=\alpha x_c^4 p_c / h_s$ characterizes the ratio of the force due to the applied pressure to the characteristic bending force.
For small deformations of the elastic sheet, \eqref{eq:thinfilmadim} can be linearized by defining $\hat{\epsilon}(\vect{\hat{x}},\hat{t})$ such that $\hat{h}_2(\vect{\hat{x}},\hat{t})=1-\beta_{\rm el}\hat{\epsilon}(\vect{\hat{x}},\hat{t})$ and considering $\vert\beta_{\rm el} \hat{\epsilon}(\vect{\hat{x}},\hat{t})\vert\ll 1$. This leads to the following equation when neglecting terms that are $\mathcal{O}\left((\beta_{\rm el} \hat{\epsilon}(\vect{\hat{x}},\hat{t}))^2\right)$ in \eqref{eq:thinfilmadim}:
\begin{align}
    \left(\partiald{}{\hat{t}}-\hat{\nabla}^{6}\right)\hat{\epsilon}(\vect{\hat{x}},\hat{t}) = -\hat{\nabla}^2 \hat{p}_e(\vect{\hat{x}}).
    \label{eq:thinfilmlinear}
\end{align}
The Green's function of this equation in $n$-dimensional Cartesian coordinates is:
\begin{align}
    G_{\rm el}(\vect{\hat{x}},\hat{t})=\frac{\mathcal{H}(\hat{t})}{\left(2\pi\right)^n} \int_{\mathbb{R}^n} e^{-|\vect{k}|^{6} \hat{t}} e^{i \vect{k} \cdot \vect{\hat{x}}} ~{\rm d}\vect{k},
    \label{eq:Green}
\end{align}
and the solutions to a Dirac load and to a confined uniform load are given by \eqref{eq:solutionDiracgeneral} and \eqref{eq:capillarysth}, respectively, provided $G_{\rm el}$ be replaced with $G_{\rm cap}$.
In particular for $n=2$ the solution to a Dirac load, $\hat{p}_e(\vect{\hat{x}})=\delta_{\rm Dirac}(\vect{\hat{x}})$, exists and simplifies to the following Hankel transform:
\begin{align}
    \hat{\epsilon}(\hat{r},\hat{t}) = \hat{t}^{1/3} \phi\left(\frac{\hat{r}}{\hat{t}^{1/6}}\right),
    &\qquad \phi(y)=\frac{1}{2\pi}\int_0^{+\infty} J_0(uy)\frac{1-e^{-u^6}}{u^3}{\rm d}u,
    \label{eq:dirac2D_EHD}
\end{align}
where $\hat{r}=\left\lVert\vect{x}\right\rVert$. At the axis of symmetry $\hat{r}=0$ this gives:
\begin{subequations}
\begin{align}
    \hat{\epsilon}(0,\hat{t})&=\frac{\Gamma(2/3)}{4\pi}\hat{t}^{1/3}, \\
    \hat{h}(0,\hat{t})=1-\left(\frac{\hat{t}}{\hat{\tau}}\right)^{1/3}, ~~&
~~~\hat{\tau} = \left(\frac{4\pi}{\Gamma(2/3)\beta_{\rm el}}\right)^3,
\end{align}
\label{eq:dirac2D_EHD_origin}
\end{subequations}
with $\hat{\tau} \simeq 799/\beta_{\rm el}^3$ the characteristic time scale of the process and $\Gamma$ Euler's Gamma function ($\Gamma(1/3)\simeq 2.68$, $\Gamma(2/3)\simeq1.35$).
The bending pressure $\hat{p}(\hat{r},\hat{t})=\beta_{\rm el}\hat{\nabla}^4\hat{\epsilon}(\hat{r},\hat{t})$ at the axis of symmetry is given by:
\begin{align}
    \hat{p}(0,\hat{t})=\beta_{\rm el} \frac{\Gamma(1/3)}{12\pi}  \hat{t}^{-1/3}.
    \label{eq:p_cap_appendix}
\end{align}

The self-similar solution \eqref{eq:dirac2D_EHD} is only valid when the load is a Dirac distribution. However it can be expected that it is also an intermediate asymptotic solution \citep{Barenblatt1996} towards which solutions to other loads converge.
We verify this in figure \ref{fig:bendingtheory}$(a)$ for a uniform and confined load: $\hat{p}_e(\hat{r}<1)=1/\pi$ and $\hat{p}_e(\hat{r}>1)=0$. This is a property also present for the levelling scenario for elastic \citep{Pedersen2020} %citep{Rubinstein}
and capillary \citep{Benzaquen2013, Benzaquen2014} interfaces, including polymer melts \citep{Mcgraw2013}, when an initial local deformation is allowed to relax, as well as for gravity currents \citep{Ball2019} governed by a similar equation.  However under load and contrary to the aforementioned cases, the deformation grows with time and the linear approximation always eventually breaks down so that  \eqref{eq:dirac2D_EHD} is only valid in a hypothetical intermediate regime. 
In figure \ref{fig:bendingtheory}$(b)$ we show by solving numerically the non-linear equation \eqref{eq:thinfilmadim} in the case of a confined uniform load that the the self-similar regime is indeed reached for  $\beta_{\rm el} < 1$.
The solution to the linear equation predicts that the height at the axis of symmetry becomes zero in finite time: this singularity is prevented by non-linear effects that become significant from time $\hat{t}\simeq\hat{\tau}/10$. For larger loads, $\beta_{\rm el} > 1$, the asymptotic self-similar regime does not have enough time to be reached before non-linear effects appear.

 \begin{figure}
  \centering
    \begin{tikzpicture}
    \draw (0, 0) node[inner sep=0] (fig) {\includegraphics[width=0.51\textwidth]{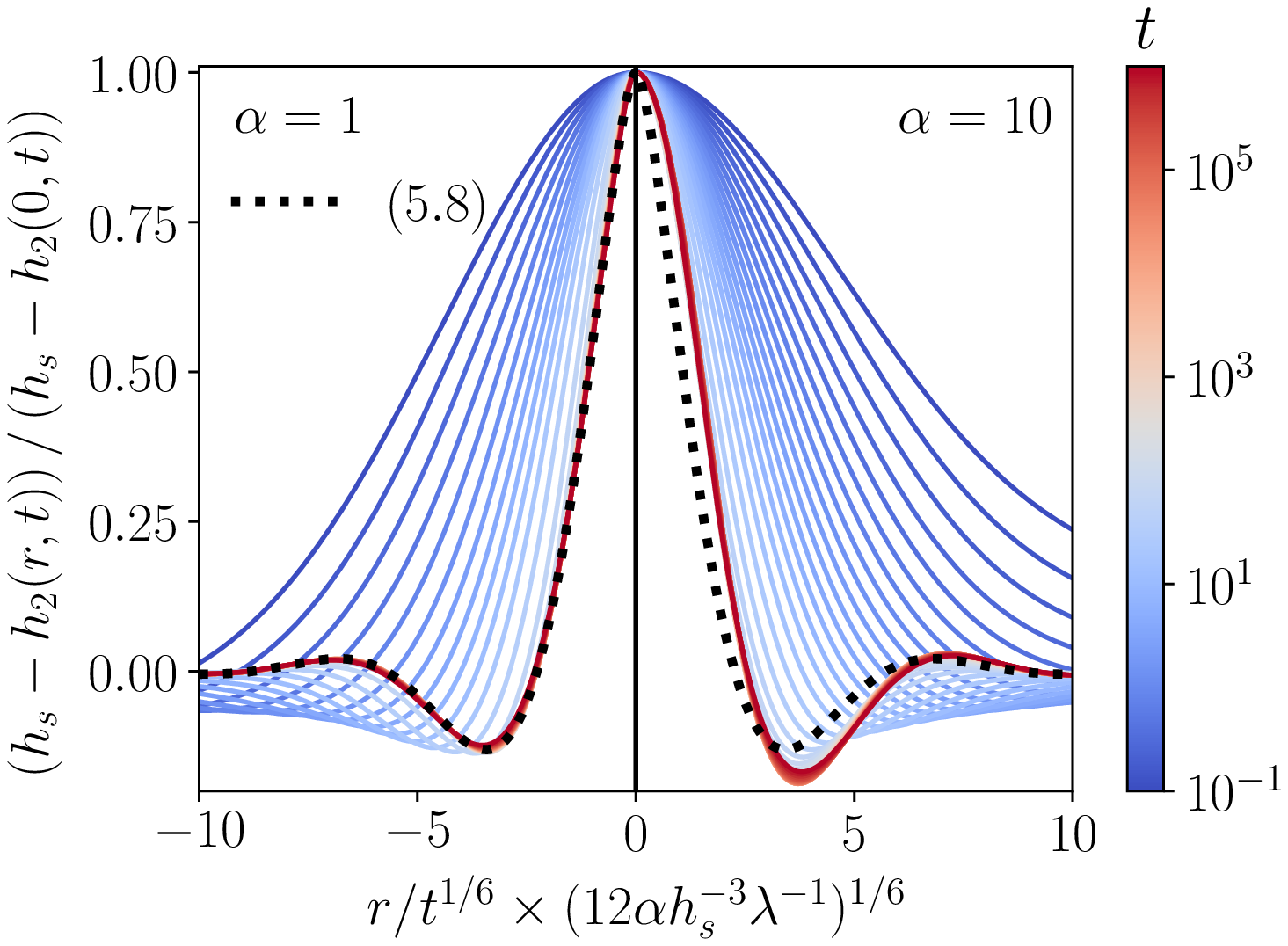}};
    \node[below right] at (fig.north west) {$(a)$};
  \end{tikzpicture}
  \begin{tikzpicture}
    \draw (0, 0) node[inner sep=0] (fig) {\includegraphics[width=0.48\textwidth]{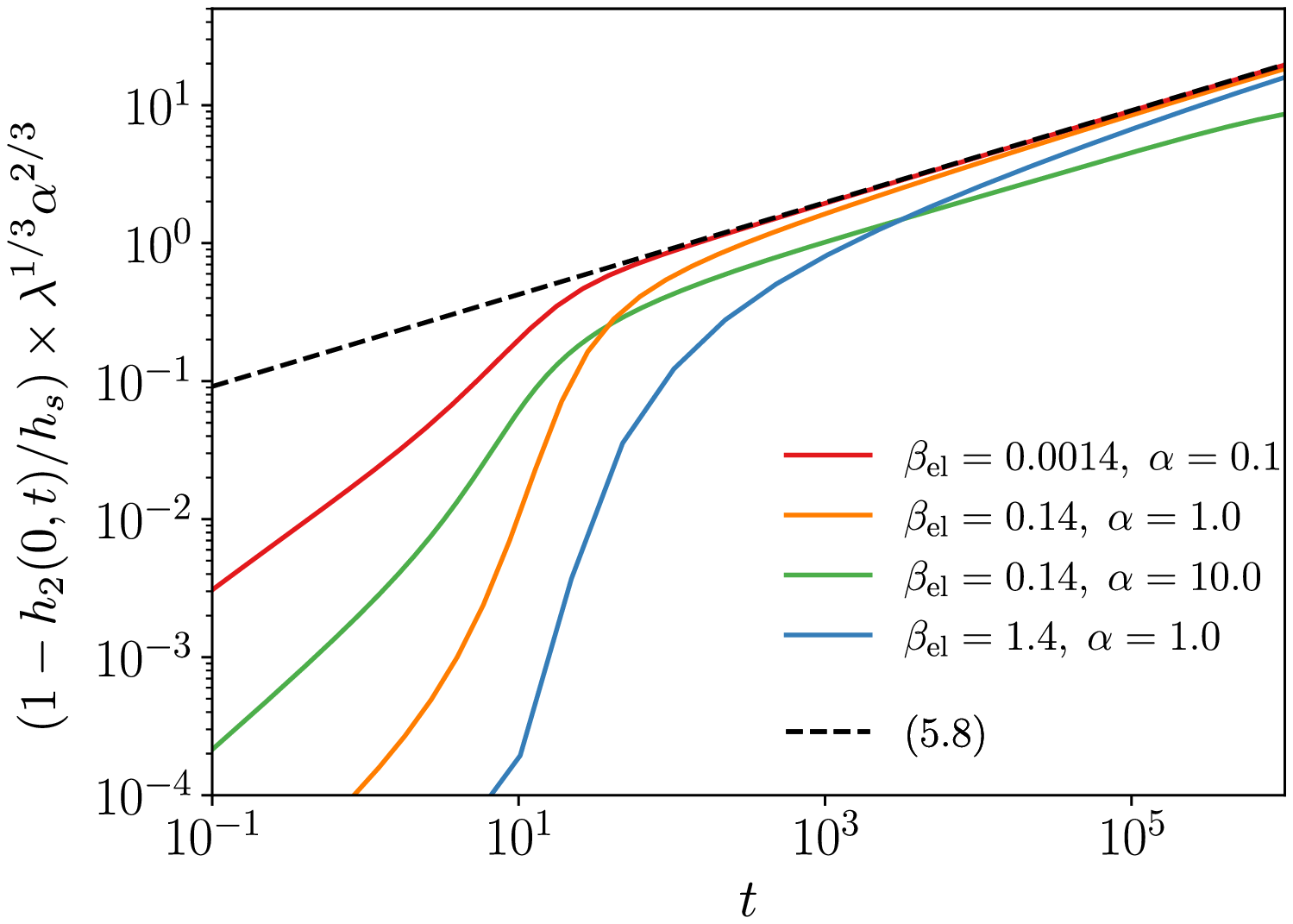}};
    \node[below right] at (fig.north west) {$(b)$};
  \end{tikzpicture}
	\caption{\label{fig:EHD_sheet}
	Deformations of an elastic sheet supported by a viscous layer for a droplet settling with $\delta=0.05$.
	$(a)$ Normalized deformation of the elastic sheet, shown here for $\lambda=10^{-5}$, $h_s=10$, and (left panel) $\alpha=1$, (right panel) $\alpha=10$.
	$(b)$ Deformation of the elastic sheet at the axis of symmetry ($r=0$) for $\lambda=10^{-5}$ and various values of $\alpha$ and $h_s$, leading to different values of $\beta_{\rm el}$.
	}
\end{figure}

 \subsection{Response of an elastic sheet to a settling droplet}
We return to the case of a settling droplet where the weight of the droplet acts on the sheet, set by \eqref{eq:p_forcebalance}, so that we can expect to be able to characterize its self-similar asymptotic response  without a priori knowledge of the structure of the air layer.
We let $r_c=(2\delta/3)^{1/2}$ and $p_c=2\pi/\delta$,  the values obtained in the rigid case. Then % such that $\hat{p_2}$ is approximately  1 for $0\leq \hat{r}<1$.
$\beta_{\rm el}=8\pi\alpha\delta/9 h_s$, $t_c=32\alpha\delta^3/9h_s^3\lambda$, 
and the rescaled height $\hat{h}_2(\hat{r},\hat{t})$ is governed by \eqref{eq:thinfilmadim}, with $p_2(\hat{r},\hat{t})$ in lieu of $p_e(r)$.
For $\beta_{\rm el} \lesssim 1$ the sheet reaches a self-similar evolution, until non-linear effects appear at time $\hat{\tau}\simeq 800/\beta_{\rm el}^3  \simeq 37 (h_s/\alpha\delta)^3$, when $h_2(0,t) \ll h_s$.
For $\beta_{\rm el} \gtrsim 1$, the sheet does not enters the self-similar regime before non-linear effects play a role.

For $\beta_{\rm el} \lesssim 1$ and $t<t_c\hat{\tau}$, we then expect from \eqref{eq:dirac2D_EHD} and \eqref{eq:dirac2D_EHD_origin} the following self-similar evolution of the elastic sheet:
\begin{subequations}
\begin{align}
    \frac{h_s-h_2(r,t)}{h_s-h_2(0,t)}&=\phi\left(\frac{\hat{r}}{\hat{t}^{1/6}}\right)  =\phi\left(\frac{r}{t^{1/6}} \times \left[\frac{12\alpha}{h_s^3\lambda}\right]^{1/6} \right)
    \label{eq:EHD_profile},\\
    1-\frac{h_2(0,t)}{h_s} &= \left(\frac{\hat{t}}{\hat{\tau}}\right)^{1/3} \simeq 0.2 \times \left(\frac{t}{\lambda \alpha^2}\right)^{1/3},
    %0.2 is Gamma(2/3)/(324^(1/3)),
    \label{eq:EHD_height}
\end{align}
\label{eq:EHD_height_all}
\end{subequations}
while the pressure in the viscous film beneath the sheet at $r=0$ is found from \eqref{eq:p_cap_appendix}:
\begin{align}
    p_3(0,t) = \frac{\Gamma(1/3)}{12\pi} \frac{\beta_{\rm el} h_s}{\alpha r_c^4} \left(\frac{t}{t_0}\right)^{1/3} \simeq 0.68 \left(\frac{\alpha}{\lambda h_s^3 t}\right)^{1/3}.
    \label{eq:p3elas}
\end{align}
Figure \ref{fig:EHD_sheet} shows that if $\beta_{\rm el} \lesssim 1$ to ensure that the self-similar dynamics is reached, and for $t \lesssim t_0 \tau \simeq 130\lambda^{-1}\alpha^{-2}$ to ensure that deformations remain small, \eqref{eq:EHD_height_all} can be verified. 
The results also show a third necessary condition, $\alpha \lesssim 1$, which is linked to the breakdown of the assumption of a localized load that we discuss next.

 \begin{figure}
  \centering
  
  \begin{tikzpicture}
    \draw (0, 0) node[inner sep=0] (fig) {\includegraphics[width=0.48\textwidth]{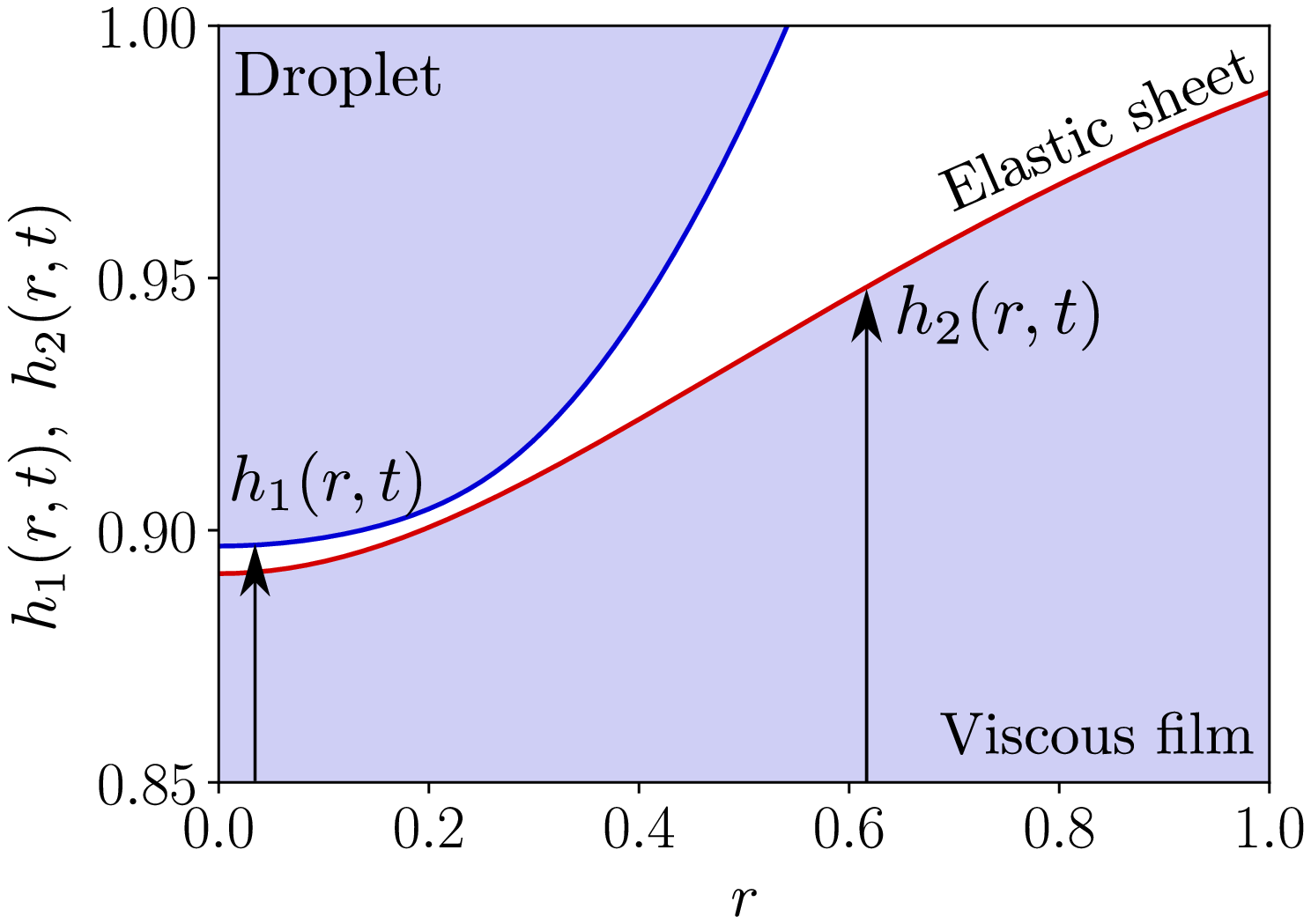}};
    \node[below right] at (fig.north west) {$(a)$};
  \end{tikzpicture}
  \begin{tikzpicture}
    \draw (0, 0) node[inner sep=0] (fig) {\includegraphics[width=0.48\textwidth]{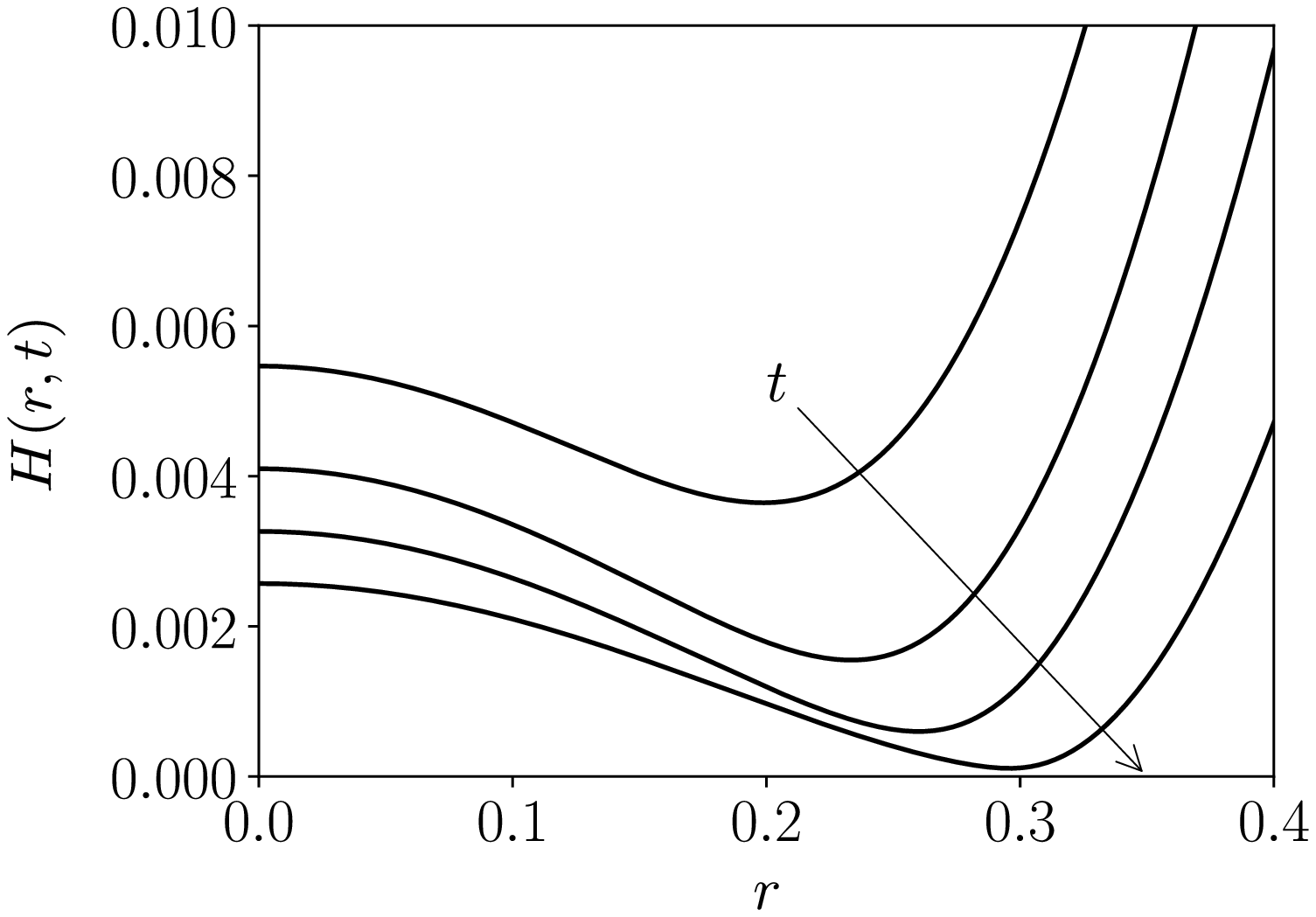}};
    \node[below right] at (fig.north west) {$(b)$};
  \end{tikzpicture}
  
  \begin{tikzpicture}
    \draw (0, 0) node[inner sep=0] (fig) {	\includegraphics[width=0.32\textwidth]{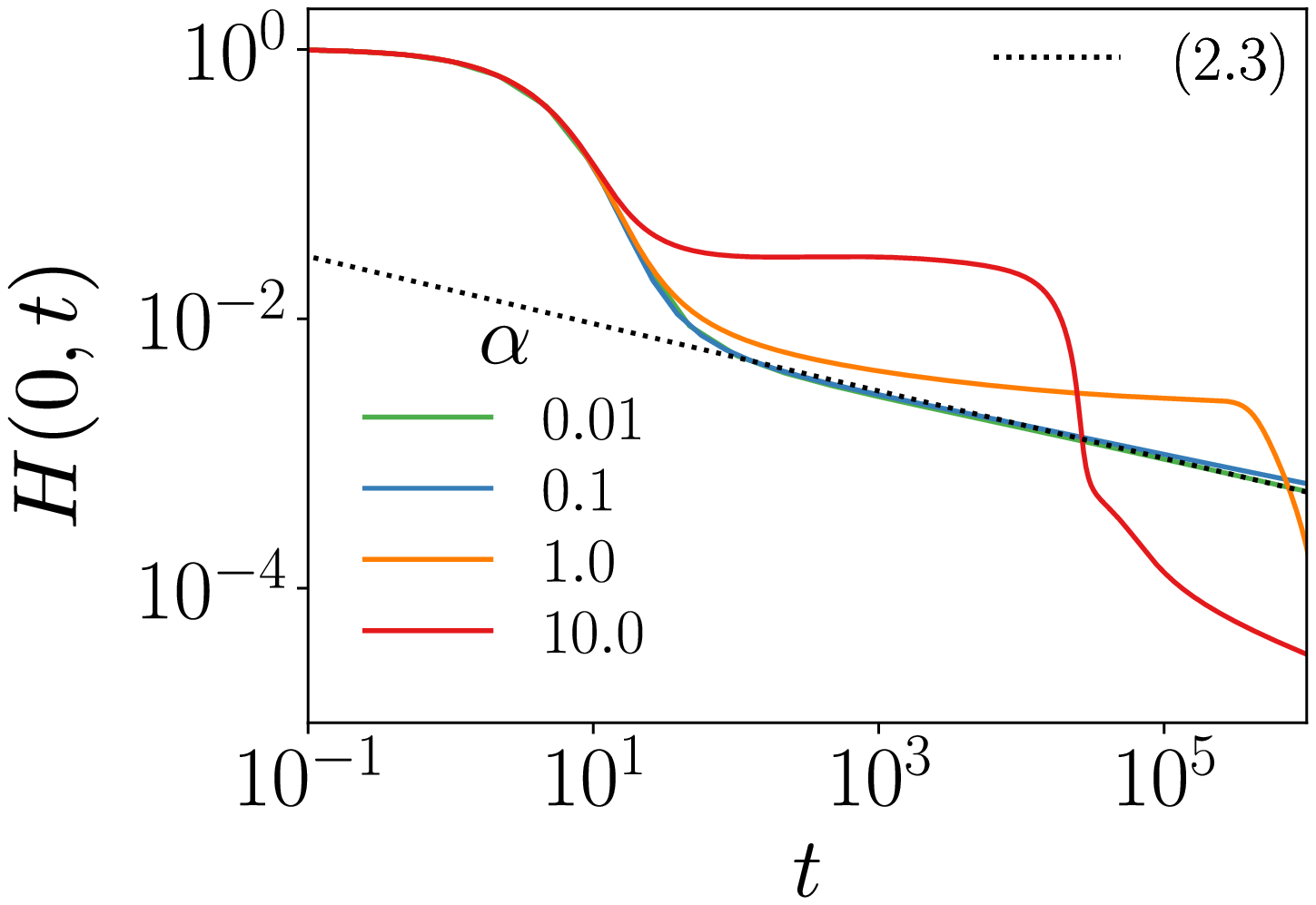}};
    \node[below right] at (fig.north west) {$(c)$};
  \end{tikzpicture}
  \begin{tikzpicture}
    \draw (0, 0) node[inner sep=0] (fig) {\includegraphics[width=0.32\textwidth]{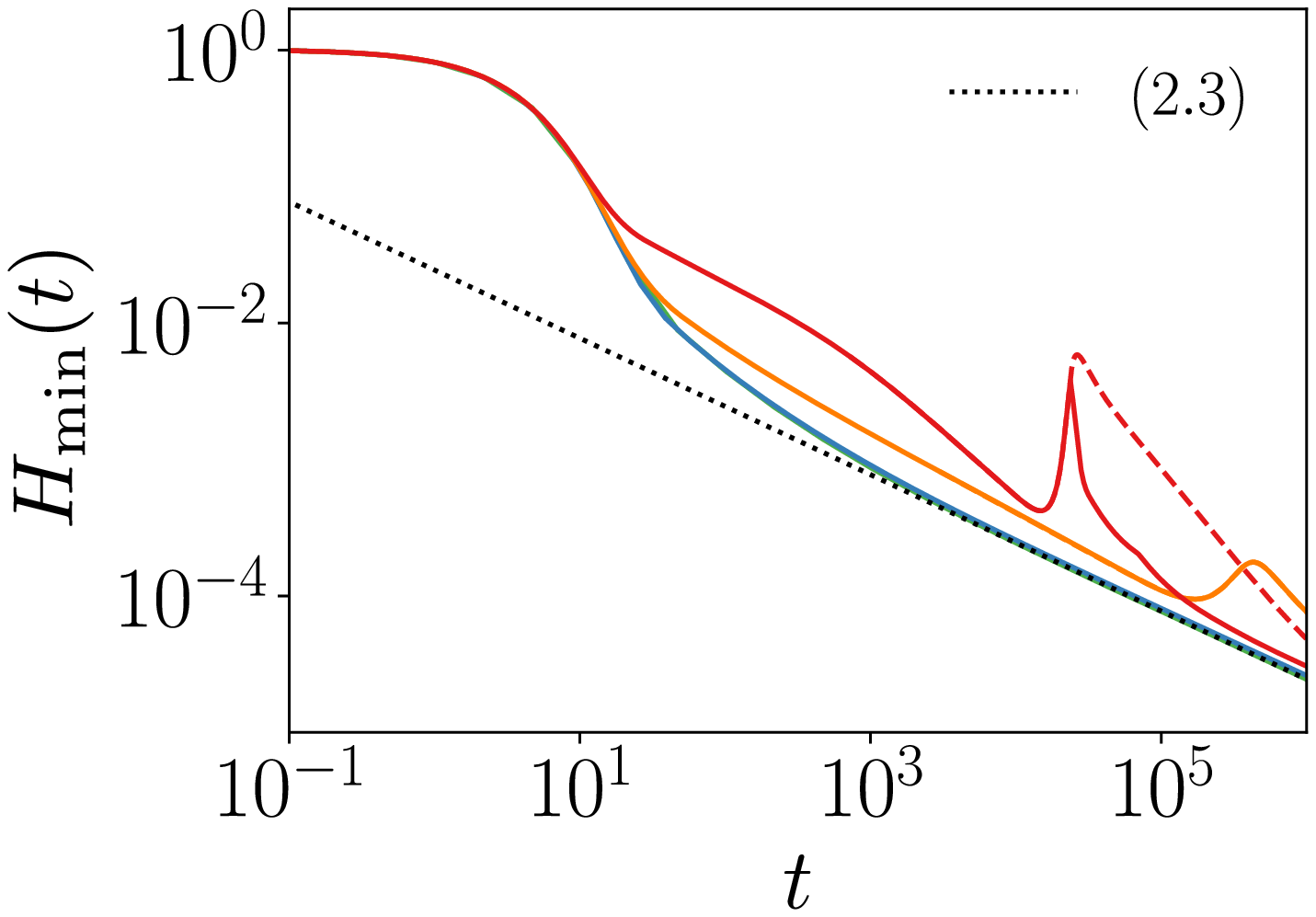}};
    \node[below right] at (fig.north west) {$(d)$};
  \end{tikzpicture}
  \begin{tikzpicture}
    \draw (0, 0) node[inner sep=0] (fig) {\includegraphics[width=0.32\textwidth]{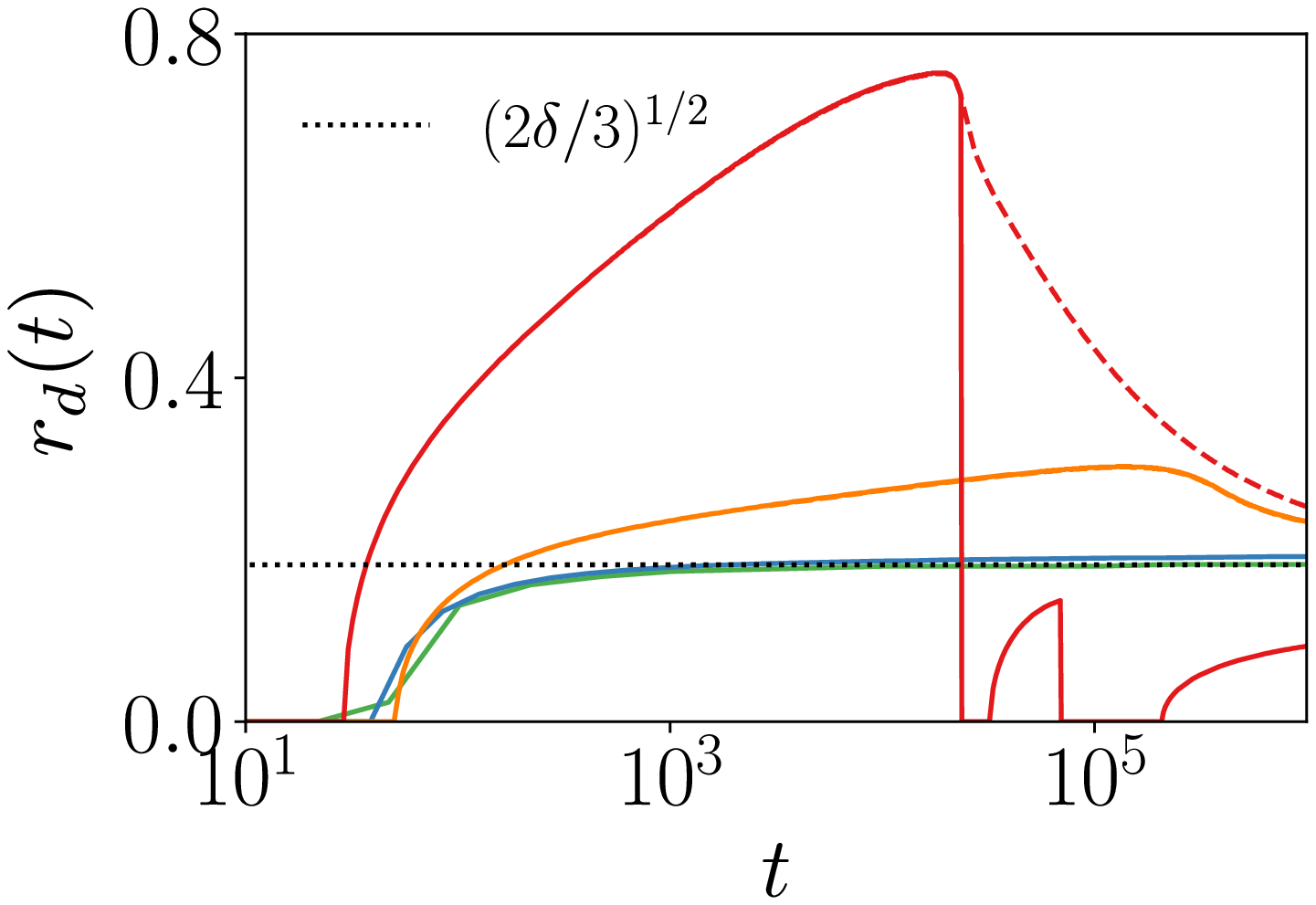}};
    \node[below right] at (fig.north west) {$(e)$};
  \end{tikzpicture}
  
  \begin{tikzpicture}
    \draw (0, 0) node[inner sep=0] (fig) {\includegraphics[width=0.32\textwidth]{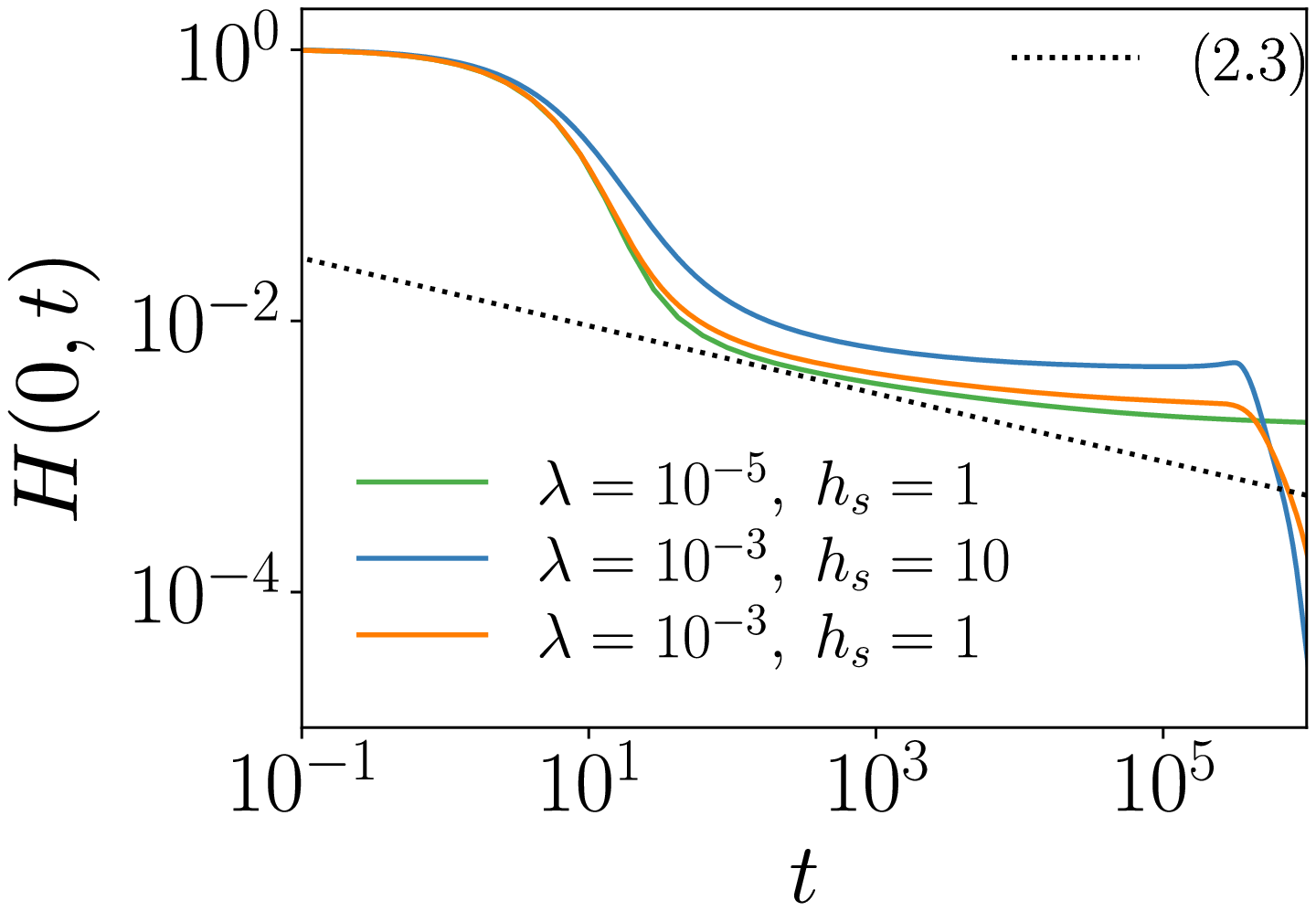}};
    \node[below right] at (fig.north west) {$(f)$};
  \end{tikzpicture}
  \begin{tikzpicture}
    \draw (0, 0) node[inner sep=0] (fig) {\includegraphics[width=0.32\textwidth]{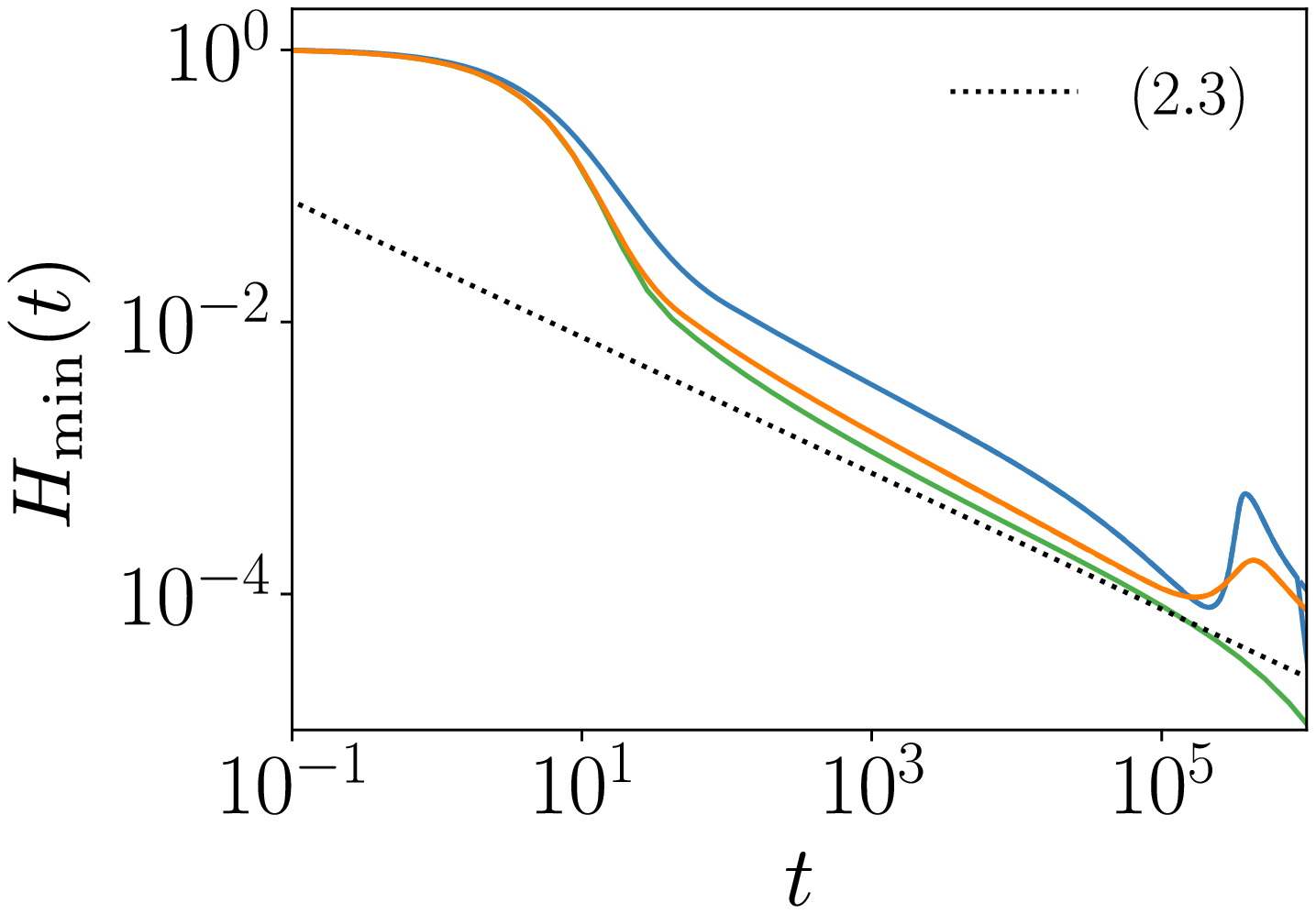}};
    \node[below right] at (fig.north west) {$(g)$};
  \end{tikzpicture}
  \begin{tikzpicture}
    \draw (0, 0) node[inner sep=0] (fig) {	\includegraphics[width=0.32\textwidth]{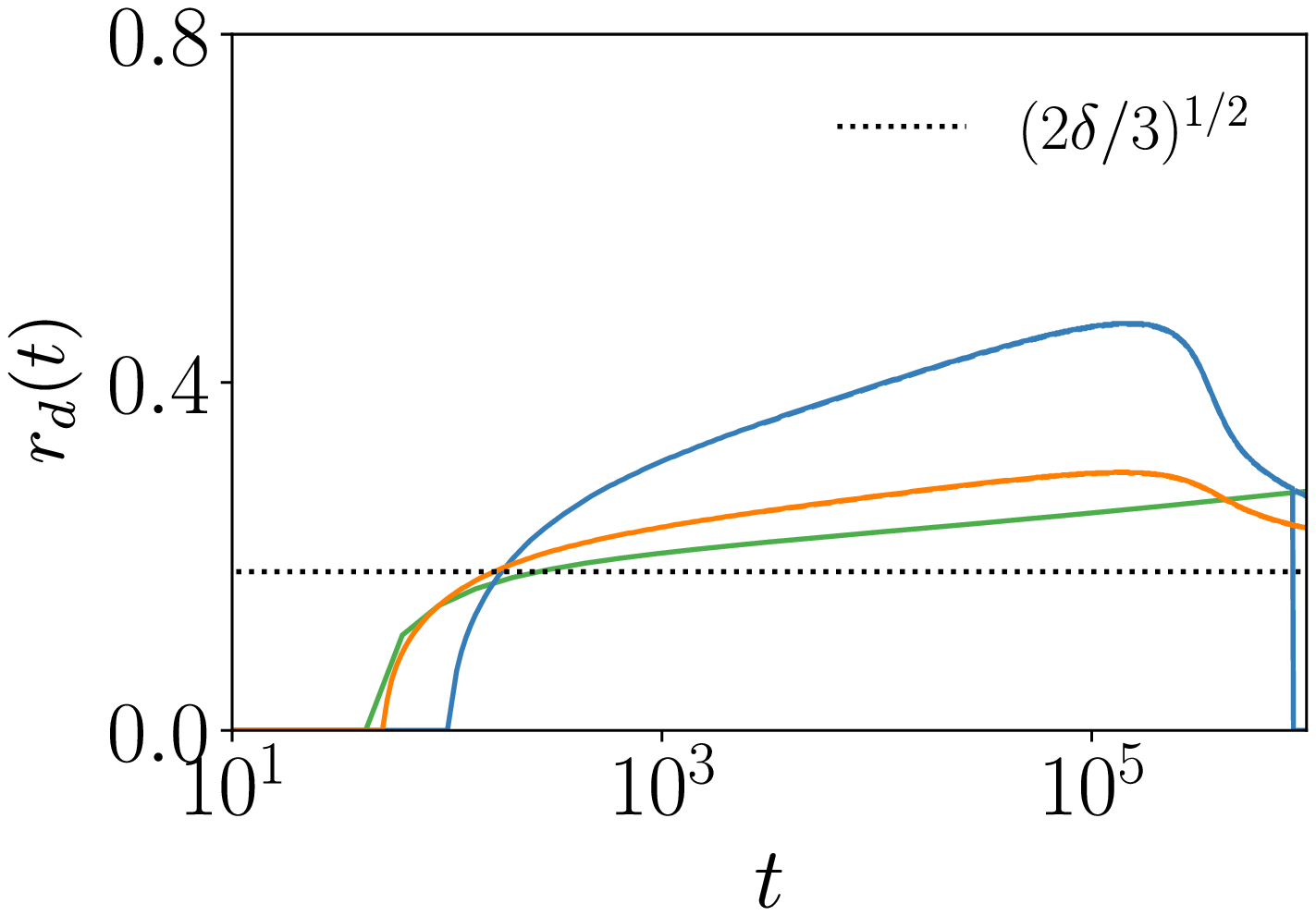}};
    \node[below right] at (fig.north west) {$(h)$};
  \end{tikzpicture}
  
    \caption{
    \label{fig:EHD_gap}
    Evolution of the air layer thickness for a droplet settling on an elastic sheet with $\delta=0.05$.
	$(a)$ Profiles for $\alpha=1$, $h_s=1$, $\lambda=10^{-3}$ of the droplet $h_1(r,t)$ and the viscous film $h_2(r,t)$ at $t=250$.
	We note that the thickness of the elastic sheet is not represented to scale for clarity.
	$(b)$ Profile of the air layer $H(r,t)=h_1(r,t)-h_2(r,t)$ for $\alpha=1$, $h_s=1$, $\lambda=10^{-3}$ and  $t=250,$ 1000, 5000, and $10^5$. 
	$(c-e)$ Evolution of the air layer thickness at the axis of symmetry ($r=0$), at its minimum value, and of the dimple radius for $\lambda=10^{-3}$, $h_s=1$ and $\alpha=0.01,~0.1,~1,~10$.
	$(f-h)$ Likewise for $\alpha=1$ and various values of $\lambda$ and $h_s$
	 The dashed lines in $(d)$ and $(e)$ show the the height and radius at the local minimum corresponding to the dimple neck, for  $\alpha=10$ and $t\simeq 10^4$ the minimum radius is not located anymore at the neck.
	}
\end{figure}

 \subsection{Numerical results}
 In figure \ref{fig:EHD_gap} we show the evolution of the air layer thickness for a droplet with $\delta=0.05$ from numerical simulations of \eqref{eq:systemglobal} and \eqref{eq:EHD}.
For $\alpha < 1$, i.e. for droplets smaller than the elastogravity length, we observe the same behavior of the air layer as in the solid case discussed in \S\ref{subsec:pheno}.
This gives a criterion for when elasticity effects can be neglected in the droplet settling process.
In this scenario, we derived analytically and verified above the self-similar response of the elastic sheet.

For $\alpha \gtrsim 1$, we observe a significant deviation in the dynamics of the air layer as shown in figure \ref{fig:EHD_gap}$(a-e)$.
It continues to present a dimple, but the height at the axis of symmetry  saturates and evolves towards a constant value. The minimum height, at the neck, continues to decrease while remaining larger than for the rigid case.
Contrary to the rigid case, the radius of the neck of the dimple also monotonically increases with time.
We note that this dimple is present while both the droplet and the sheet heights are increasing near the axis of symmetry, and its neck is approximately located at the inflection point of the second spatial derivative of $h_2$.
Contrary to the case of a rigid solid coated by a thin viscous film, without an elastic sheet atop, this new behavior not only depends on the property of the sheet characterized by $\alpha$, but also on the properties on the supporting viscous film height and viscosity through $h_s$ and $\lambda$ as show in figure \ref{fig:EHD_gap}$(f-h)$. Namely, the thickness and dimple radius increase with increasing initial height and decreasing viscosity.

The continuous increase of the radius of the neck of the dimple helps to explain why the analytical results for the sheet height presented in  the previous section fail when $\alpha \gtrsim 1$, even when $\beta_{\rm el} \lesssim 1$ and $\hat{t}<\hat{\tau}$.
The analysis is based on the results derived in \S\ref{sec:appendix_EHD}, which formally only applies if the load that the droplet puts on the sheet were concentrated into a single point in space.
The results continue to apply asymptotically when the load is localized to a small region as compared to the lateral extent of the deformations of the sheet, which evolves as $\hat{r}_{\rm sheet}(\hat{t})\sim 3.4 \hat{t}^{1/6}$, i.e. $r_{\rm sheet}(t)\sim  \alpha^{-1/6} h_s^{1/2} \lambda^{1/6} t^{1/6} $ (we define $r_{\rm sheet}(t)$ as the radial position of the maximum of $h_2(r,t)$, see figure \ref{fig:EHD_sheet}$a$).
For large enough $\alpha$ the dimple radius increases as fast as the radial extent of the sheet deformations and the load exerted on the sheet cannot be assumed to be localized to a very small region anymore (figure \ref{fig:EHD_gap}$e$). 

At long times  $t\gtrsim\tau\simeq 37 (h_s/\alpha\delta)^3$ and regardless of the value of $\alpha$, non-linear effects of the thin film become important and the behavior completely changes. Specifically, the air film no longer forms a dimple but instead evolves into a rippled deformation similar to the capillary case (\S \ref{subsec:cap_num}), as illustrated in figures \ref{fig:cap_wrimple}$(b)$ and \ref{fig:EHD_gap}$(d,e)$, where it can be seen that the minimum thickness is not located anymore at the neck of the dimple.

\section{Discussion}
\label{sec:conclusion}

We have studied the settling of a droplet onto three  classes of thin soft layers coating a rigid solid: a compressible elastic material, a viscous fluid, or an elastic sheet supported by a viscous film.
First, when the soft layer is made of a compressible elastic material (\S\ref{sec:compressible_elastic}), we found that depending on its softness the air layer separating the droplet and the elastic solid  transitions from the dimple characterizing droplet settling onto a rigid wall to a layer with near-uniform thickness.
This thickness is always larger than the minimum thickness of the rigid case.
Second, we studied in \S\ref{sec:capillary} the settling towards a thin viscous film, generalizing the analysis of \citet{Yiantsios1990} for an infinite bath of fluid. We derived the long-term asymptotics of both the film and of the air layer separating the film and the droplet in the limit of small film deformations. %, when its evolution can be characterized by a linear evolution equation.
The dimple shows the same structure as for the case of a rigid substrate, but its thickness is asymptotically larger by a factor $1+\xi/\delta = 1 + \sigma_1/\sigma_3 $ with $\sigma_1$ and $\sigma_3$ the surface tension of the droplet and of the liquid film, respectively.
When the film deformation becomes large, we observed rippled deformations which modify the drainage dynamics and in particular lead to a sudden decrease of the minimum air layer thickness.
Finally, we described in \S \ref{sec:EHD}  a droplet  settling towards an elastic sheet supported by a thin layer of viscous fluid.
We found that when the droplet is smaller than the elastogravity length the dimple profile of the air layer is similar to the the solid case,
and derived analytically the response of the elastic sheet. 
For droplets of size similar to or larger than the elastogravity length, we observed numerically an increase of both the minimum thickness of the dimple and of its radius.
In the three cases studied, the deformation of the surface always lead to an increase of the minimum thickness of the dimple, provided that these deformations are small. This suggests that surface deformations upon droplet settling can delay impact or coalescence of the droplet with the substrate.

We observed a shift in dynamics for large deformations of the viscous film coating a solid substrate, with or without an elastic sheet on top.
The dimple evolves then towards a more complex structure which strongly affects the drainage process and, to our knowledge, had not been described before for interactions between droplets and compliant surfaces. This so-called rippled dimple occurs because of the non-linear response of viscous films to large deformations.
A better understanding of this regime could also have applications beyond settling droplets settling and impacting drops.
Indeed the response of a capillary interface under load appears for instance in the deformation of glassy materials by  nanobubbles  \citep{Ren2020} and is relevant for lubricant-impregnated surfaces \citep{Pack2017} as well as other industrial applications \citep{Carou2009,Lunz2018}, while an elastic sheet supported by a viscous fluid can be found in various situations, from technological \citep{Rogers2010} and biological \citep{Bongrand2018} applications up to the geophysical scale \citep{Michaut2011}. 

In our study of the settling of a droplet onto a thin liquid film, we focused on very viscous films and verified that our choice of parameters justifies the assumptions required for no-slip of air at the interfaces of both the droplet and film.
While this assumption only alters the time scale of the dynamic on solid surfaces, where the no-slip condition continues to apply at one of the interfaces, both numerical \citep{Yiantsios1990,Duchemin2020}  and experimental \citep{Lo2017} studies have shown that slip can significantly alter the dynamic of droplet deposition on a liquid film.
The analysis presented in \S\ref{sec:capillary} henceforth only applies in a regime where the air layer is wide enough and the viscosity ratios large enough.
%$\mu_2/\mu_1 \ll \epsilon H_{\rm min}$, for the no-slip condition to apply but eventually fails as $H_{\rm min}$ decreases.
It also applies  when considering the  presence of surfactants, which can alter the boundary conditions at the interfaces, in either the droplet or the film. %\citep{Manikantan2020,Liu2020}.
It has long been known that a drop falling onto a dirty interface takes much more time to coalesce than on a clean one \citep{Reynolds1881,Rayleigh1882}, and indeed small amounts of surfactant have been shown to considerably delay the coalescence process \citep{Amarouchene2001}.

For a droplet settling on an elastic sheet, we rationalized the dynamics for droplets smaller than the elastogravity length and only considered bending stresses in the description of the sheet.
When the deformations of the sheet become significant compared to its thickness, additional effects from stretching could appear.
Including stretching in the elastic model requires a more complete description, e.g. the full Föppl-von Kármán equations \citep{LandauBook}.

Finally, in all the cases we have studied the minimum thickness of the air layer continuously decreases with time and there would eventually be direct contact between the droplet and the substrate. This usually occurs due to the influence of surface forces such as van der Walls or electrostatic interactions \citep{Oron1997,Israelachvili2011}, rarefied gas effects \citep{Duchemin2012},  nano-roughness \citep{Kolinski2014a,Li2015} or other small scale phenomena which can otherwise be neglected for most of the drainage dynamic. We did not consider them in this study but isolated the coupling between the deformations of the droplet and of the substrate. They should be incorporated when the minimum thickness of the air layer reaches the order of a hundred nanometers or less as they lead to the rupture of the air film then and to contact \citep{Couder2005,Chan2011}.
In the case of settling onto a viscous film, this supporting layer could also rupture due to surface forces \citep{Zhang1999,Carlson2016}.
Contact can occur in an axisymmetric manner \citep{Chan2011}, but \citet{Lo2017} showed experimentally that when a large drop settles on a rigid surface, symmetry eventually breaks and contacts occur earlier than for an axisymmetric situation. Axisymmetry could be preserved for deposition on thin liquid films in their experiments.
It would be interesting to understand when the deposition process can become non-axisymmetric and, more generally, to investigate to which extent the observations made and conclusions drawn in this work hold for drops with larger Bond number.

This work also opens the question of possible effects of non-uniform substrates on the drop settling dynamics.
Directional transport of droplets above a substrate can for instance be reached through a Leidenfrost dynamics above a textured solid \citep{Lagubeau2011} or a Marangoni dynamics of a drop above a liquid film with a temperature gradient \citep{Davanlou2015}.
Gradients of substrate stiffness \citep{Style2013} or bending rigidity \citep{Bradley2019} can also lead to the transport of a droplet. Our work gives a first minimal description of how gravitational settling dynamics of droplets can be affected by a thin compliant layer, but there are many natural extensions such as how non-linear elastic effects, gradients in substrate properties, and adhesive contact affect the flow.\\

\section*{Declaration of Interests}
 The authors report no conflict of interest.

\section*{Acknowledgements}
We acknowledge the financial support of the Research Council of Norway through the program NANO2021, project number 301138, and thank Miroslav Kuchta and Christian Pedersen for stimulating discussions about the numerical methods.

\appendix
%!TEX root = version_JFM.tex

\section{Expression of $\psi(0,t)$}
\label{sec:stuff}

The height at the axis of symmetry $r=0$ of a viscous film under an axisymmetrical, confined, and uniform load can be expressed from \eqref{eq:cap_solution} using the following identity:
\begin{align}
    \psi(0,t)=\frac{1}{\pi}\int_0^{+\infty} J_1(\rho)\frac{1-e^{-\rho^4 t}}{\rho^2}~{\rm d \rho} &=  \frac{\ln t + 8 \left(2-3\gamma_{\rm e \rm m}+4\ln 2\right)}{\pi} \\
    &+
    \frac{t}{64\sqrt{\pi}}  {}_1F_3\left[
    \left\{\frac12\right\};\left\{\frac32,\frac32,2\right\};\frac{1}{256t}
    \right] \nonumber \\
    &- \frac{1}{1536\pi t}
    {}_2F_4\left[
    \left\{1,1\right\};\left\{\frac32,2,2,\frac52\right\};\frac{1}{256t}
    \right], \nonumber
\end{align}
where $\gamma_{\rm e \rm m}$ is the Euler-Mascheroni constant,
$J_1$ is the first order Bessel function of the first kind, and ${}_pF_q$ is the $(p,q)$ hypergeometric function defined as:
\begin{align}
    {}_pF_q\left[
    \left\{a_1, \dots, a_p\right\};\left\{b_1, \dots b_q\right\};x
    \right] = \sum_{n=0}^{+\infty} \frac{(a_1)_n \dots (a_p)_n}{(b_1)_n \dots (b_q)_n} \frac{x^n}{n!},
\end{align}
with $(a)_n=\prod_{k=1}^n(a+k-1)$ the rising factorial of $a$.

\section{Shear stresses and droplet settling onto a solid substrate coated with a thin viscous liquid film}
\label{sec:appendixSLIP}
\begin{figure}
  \centering
  
  \begin{tikzpicture}
    \draw (0, 0) node[inner sep=0] (fig) {\includegraphics[width=0.32\textwidth]{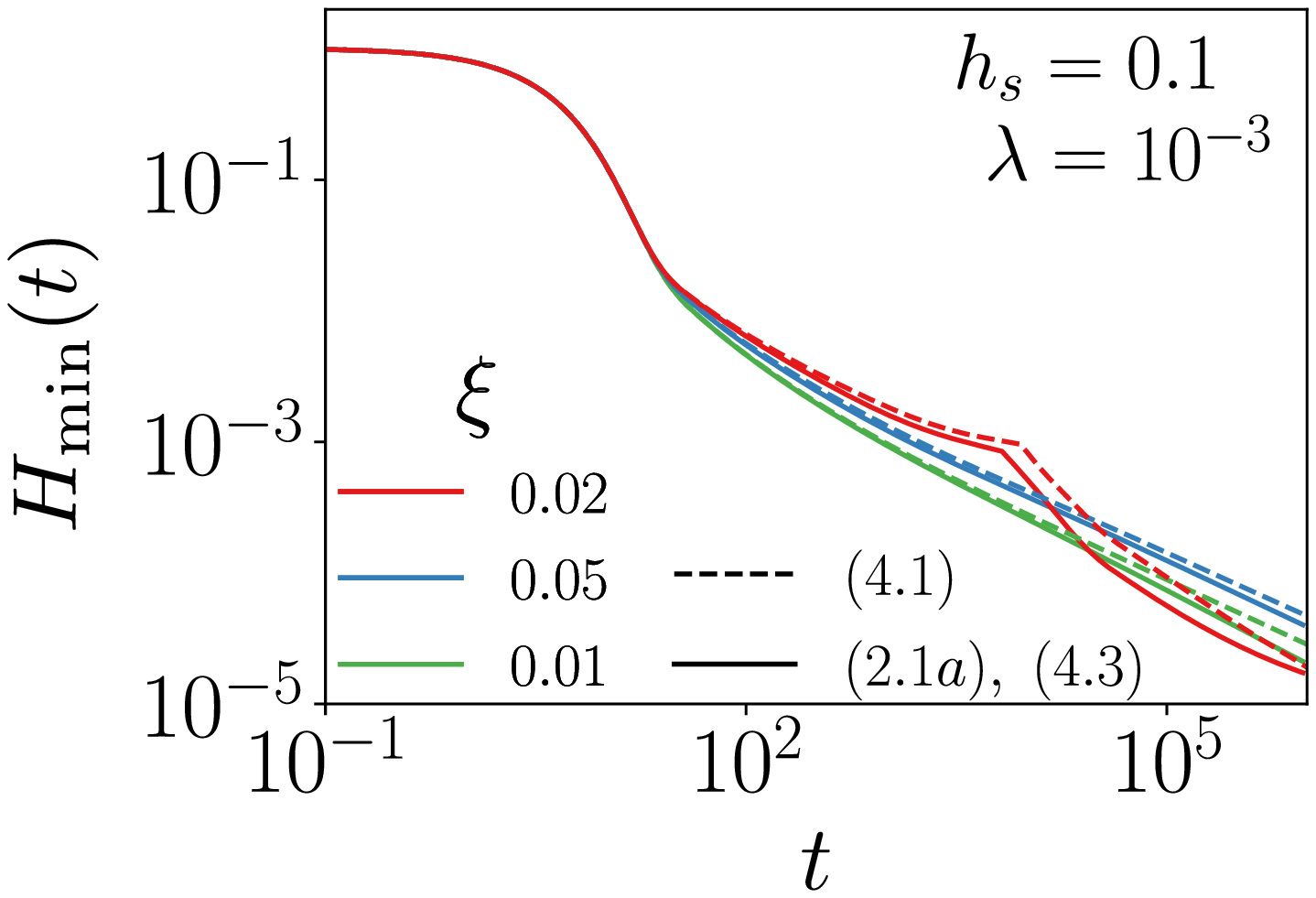}};
    \node[below right] at (fig.north west) {$(a)$};
  \end{tikzpicture}
  \begin{tikzpicture}
    \draw (0, 0) node[inner sep=0] (fig) {\includegraphics[width=0.32\textwidth]{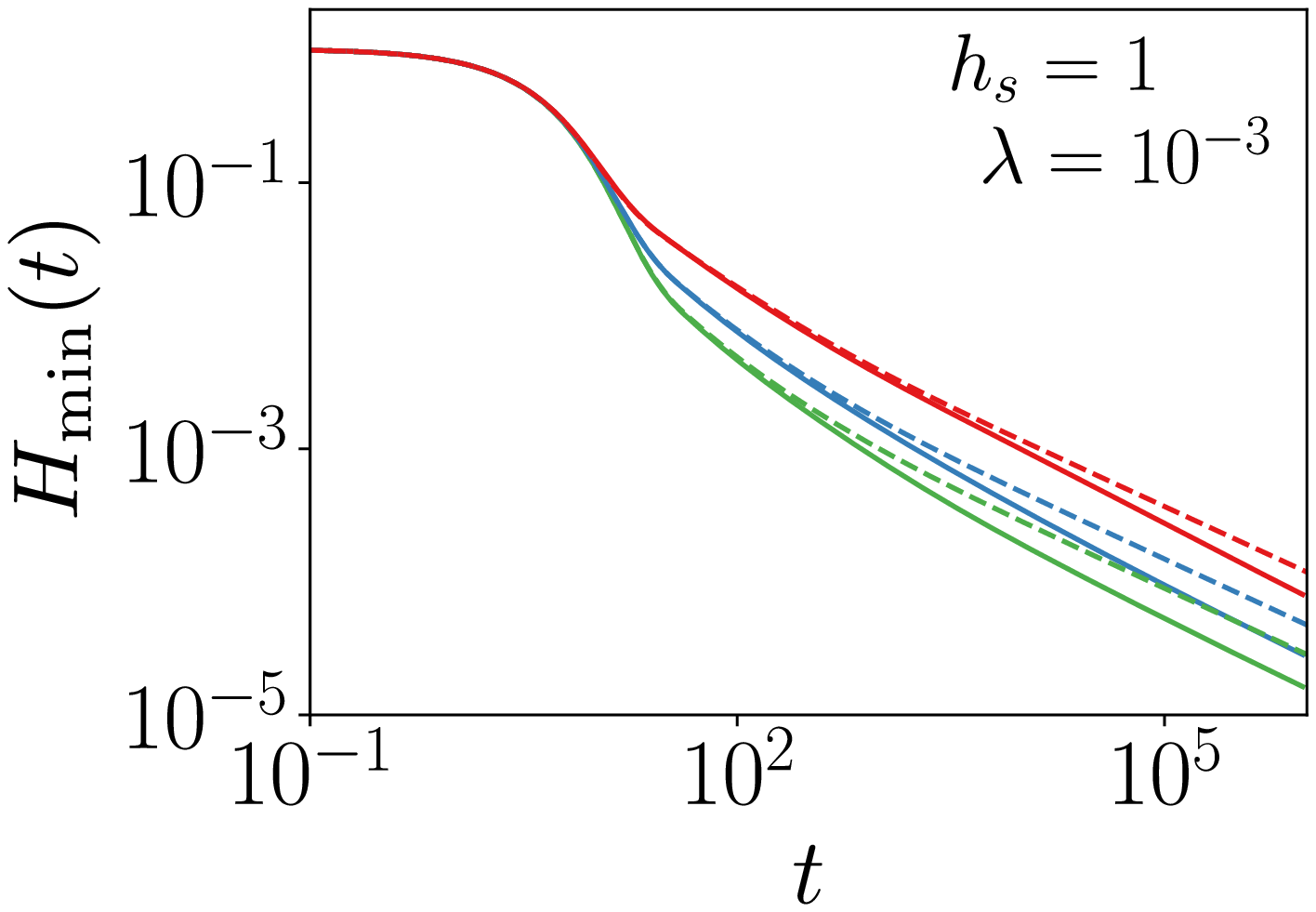}};
    \node[below right] at (fig.north west) {$(b)$};
  \end{tikzpicture}
  \begin{tikzpicture}
    \draw (0, 0) node[inner sep=0] (fig) {	\includegraphics[width=0.32\textwidth]{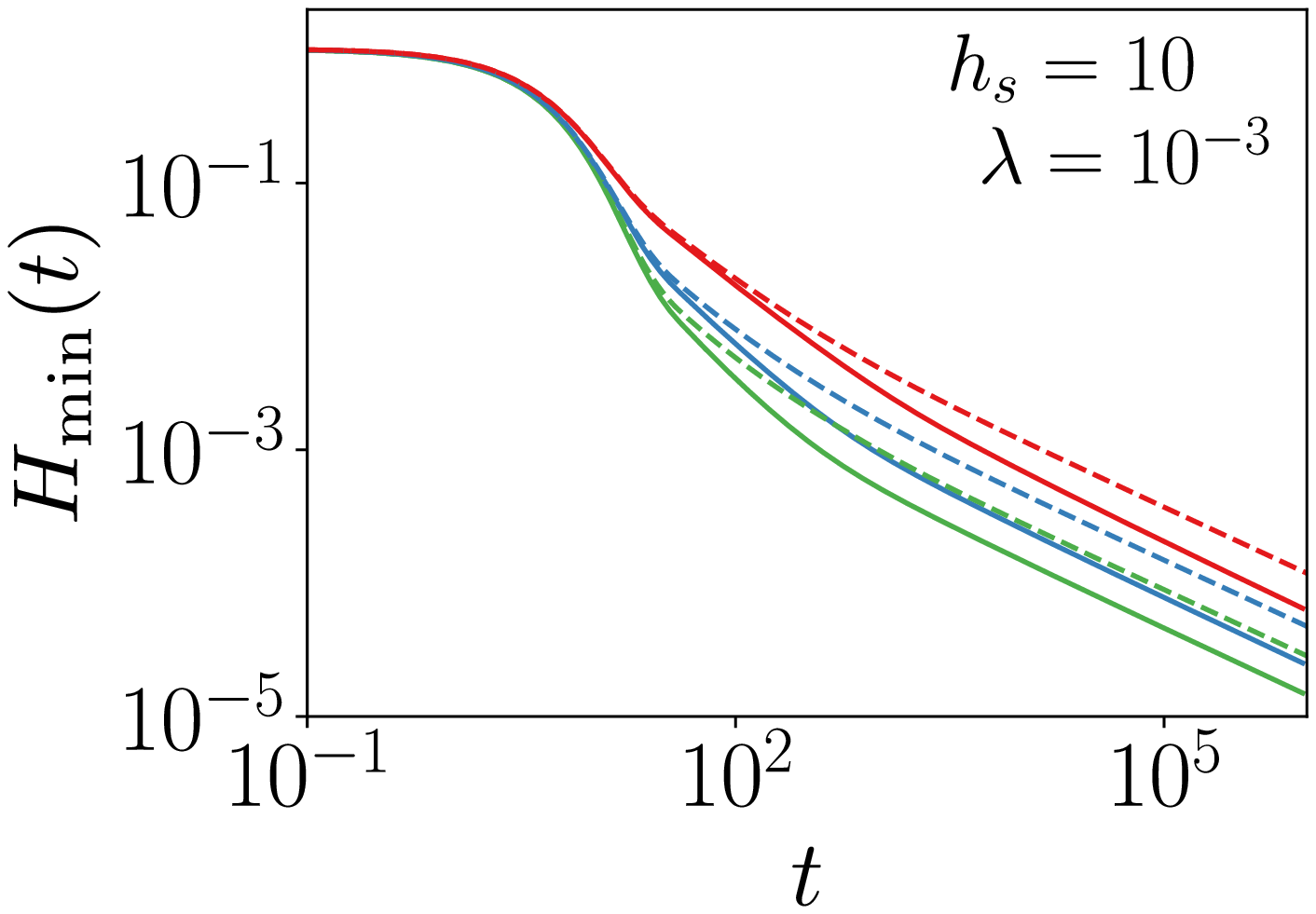}};
    \node[below right] at (fig.north west) {$(c)$};
  \end{tikzpicture}
  
  \begin{tikzpicture}
    \draw (0, 0) node[inner sep=0] (fig) {\includegraphics[width=0.32\textwidth]{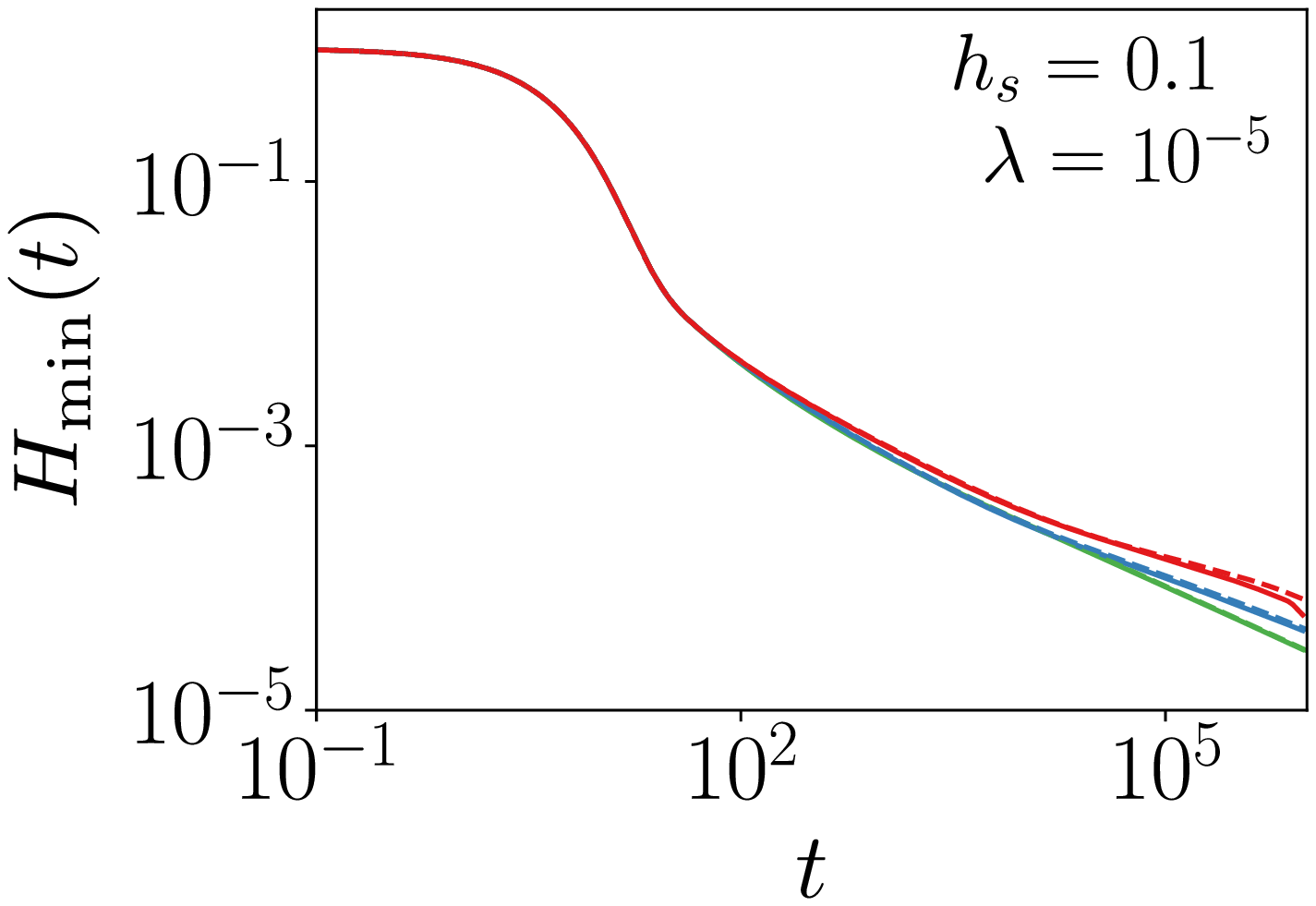}};
    \node[below right] at (fig.north west) {$(d)$};
  \end{tikzpicture}
  \begin{tikzpicture}
    \draw (0, 0) node[inner sep=0] (fig) {\includegraphics[width=0.32\textwidth]{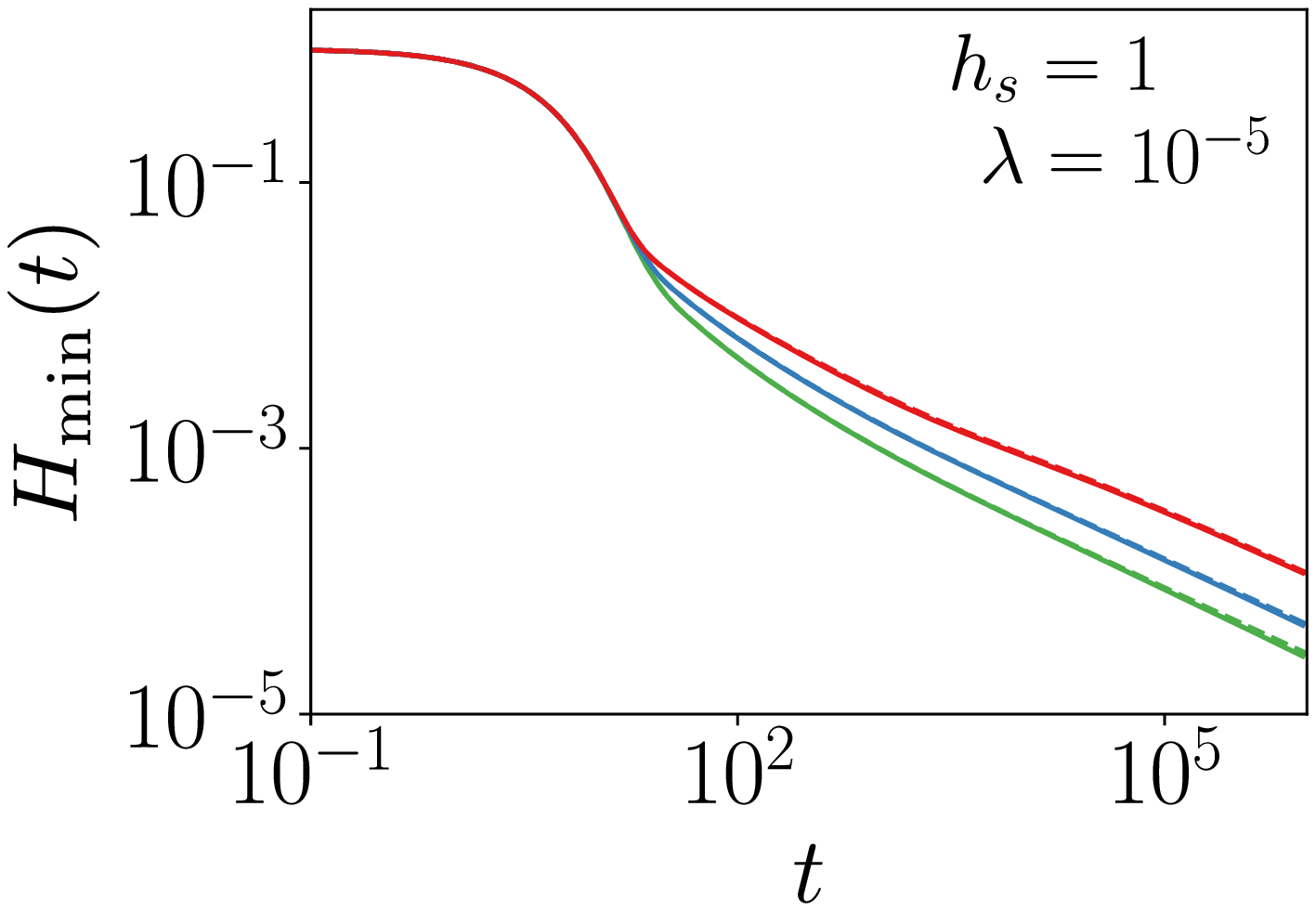}};
    \node[below right] at (fig.north west) {$(e)$};
  \end{tikzpicture}
  \begin{tikzpicture}
    \draw (0, 0) node[inner sep=0] (fig) {\includegraphics[width=0.32\textwidth]{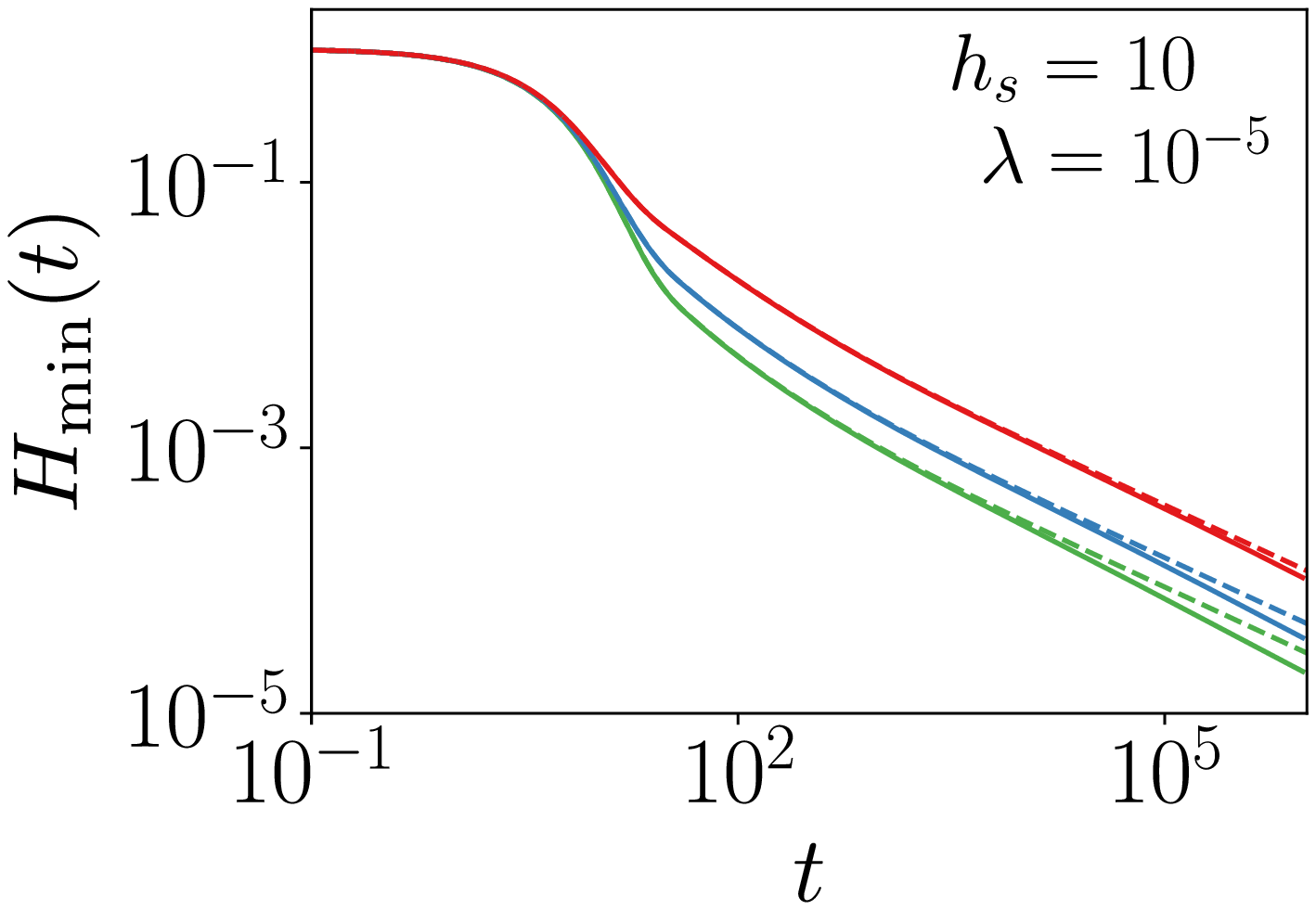}};
    \node[below right] at (fig.north west) {$(f)$};
  \end{tikzpicture}
  
    \caption{
    \label{fig:supp2_1}Evolution of the minimum air layer thickness  for droplet deposition on a viscous film and various values of $\xi$, $\lambda$ and $h_s$. The parameter $\delta$ is set to 0.05.
    The system of equations solved consists of  \eqref{eq:pressure_h1}. \eqref{eq:cap_pressure}, \eqref{eq:p_forcebalance}, and either (solid lines) accounting for the full stress balance using \eqref{eq:H_slip} and \eqref{eq:h2_slip}, or (dashed lines) using the simplified equations  \eqref{eq:thinfilm_H} and \eqref{eq:thinfilmviscous_h2}.
	}
\end{figure}

\begin{figure}
  \centering
  
  \begin{tikzpicture}
    \draw (0, 0) node[inner sep=0] (fig) {\includegraphics[width=0.32\textwidth]{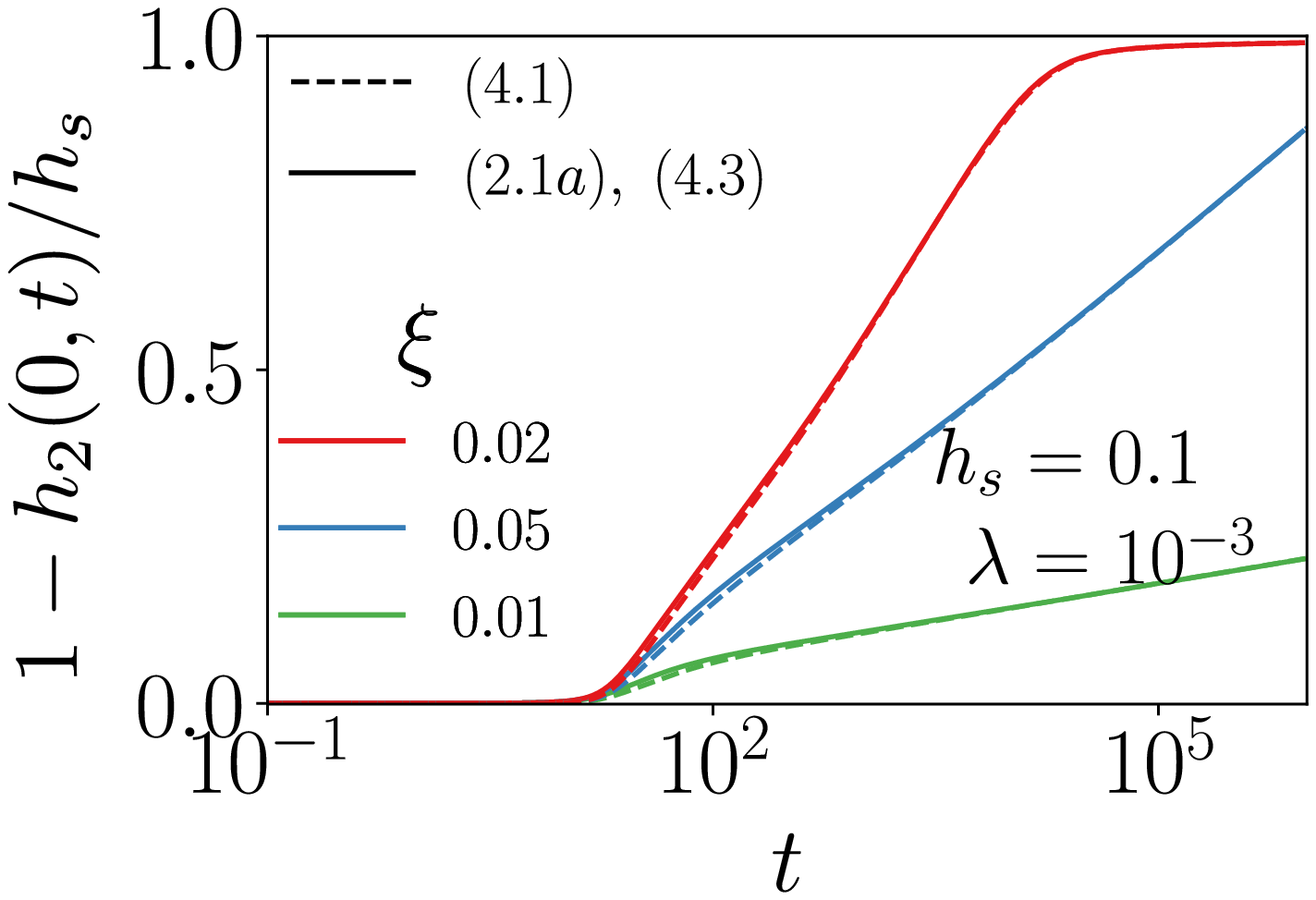}};
    \node[below right] at (fig.north west) {$(a)$};
  \end{tikzpicture}
  \begin{tikzpicture}
    \draw (0, 0) node[inner sep=0] (fig) {\includegraphics[width=0.32\textwidth]{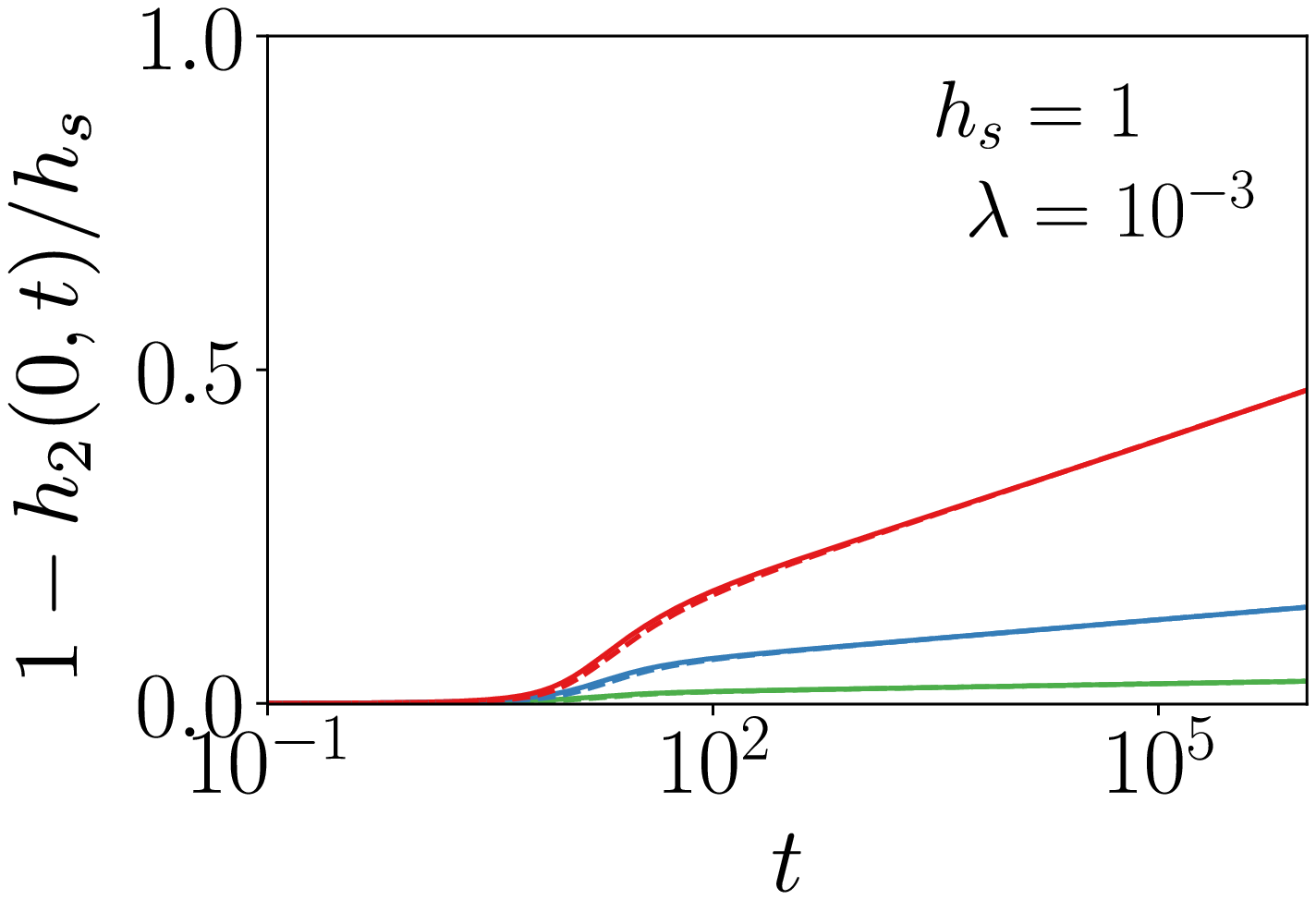}};
    \node[below right] at (fig.north west) {$(b)$};
  \end{tikzpicture}
  \begin{tikzpicture}
    \draw (0, 0) node[inner sep=0] (fig) {	\includegraphics[width=0.32\textwidth]{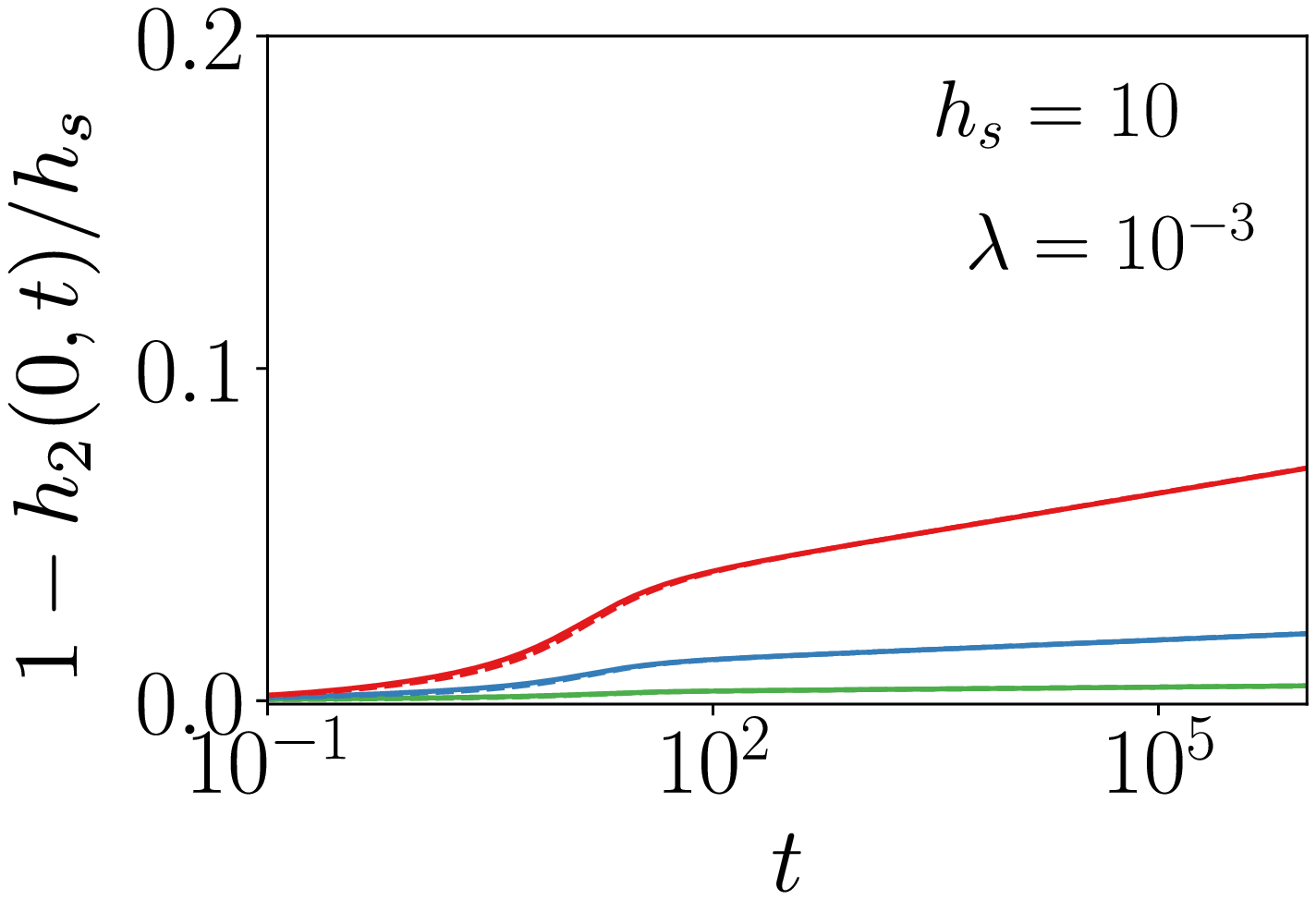}};
    \node[below right] at (fig.north west) {$(c)$};
  \end{tikzpicture}
  
  \begin{tikzpicture}
    \draw (0, 0) node[inner sep=0] (fig) {\includegraphics[width=0.32\textwidth]{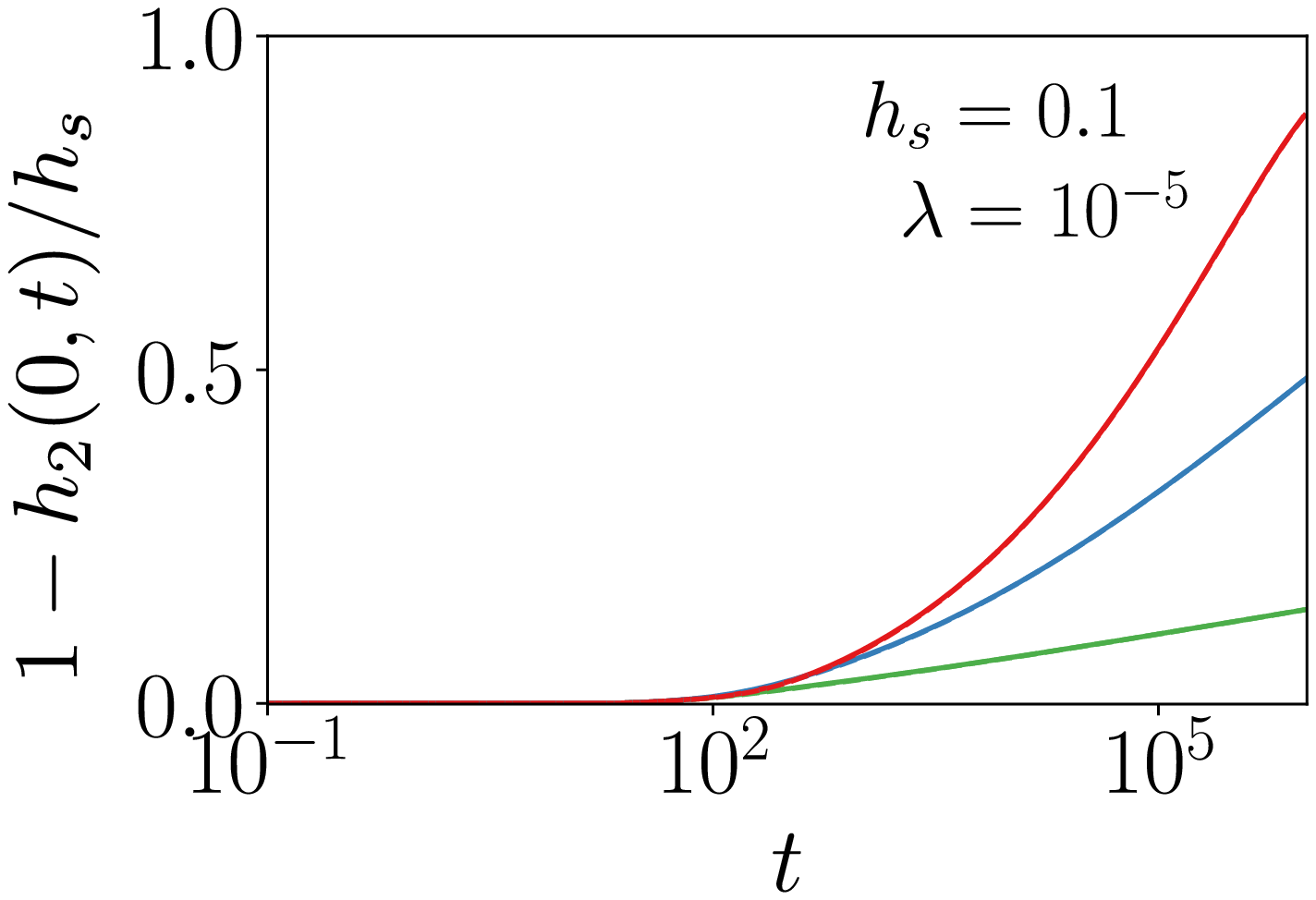}};
    \node[below right] at (fig.north west) {$(d)$};
  \end{tikzpicture}
  \begin{tikzpicture}
    \draw (0, 0) node[inner sep=0] (fig) {\includegraphics[width=0.32\textwidth]{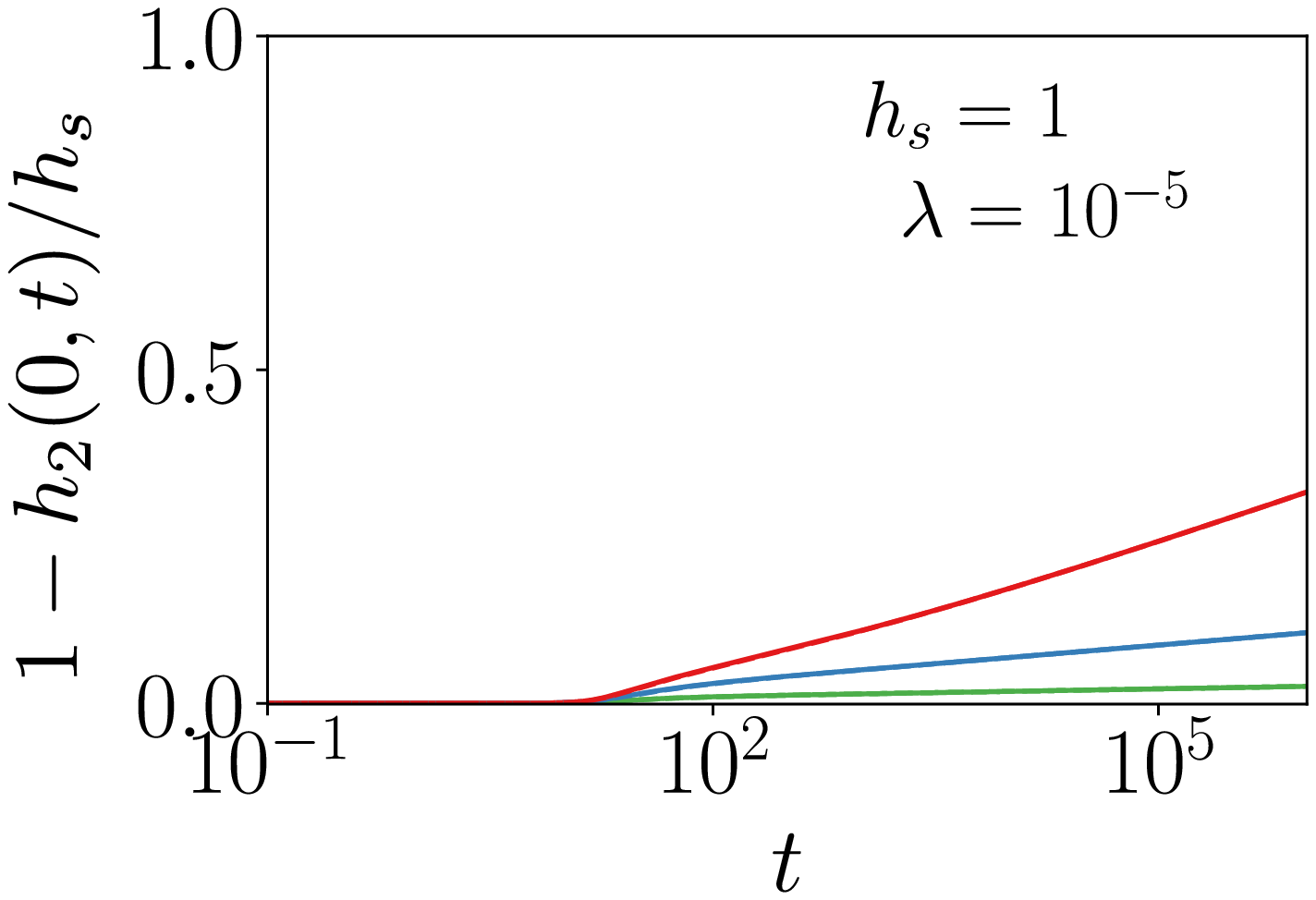}};
    \node[below right] at (fig.north west) {$(e)$};
  \end{tikzpicture}
  \begin{tikzpicture}
    \draw (0, 0) node[inner sep=0] (fig) {\includegraphics[width=0.32\textwidth]{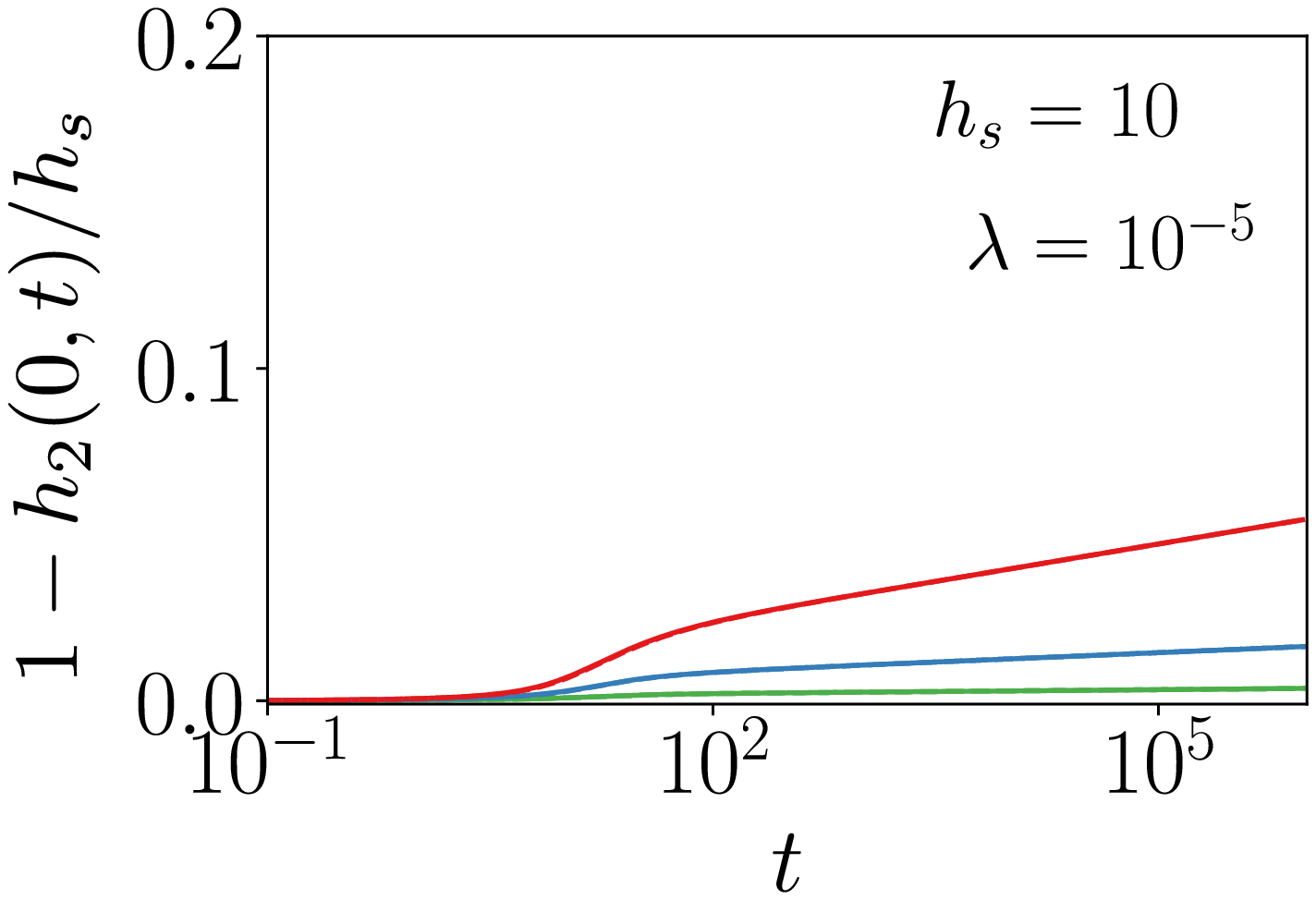}};
    \node[below right] at (fig.north west) {$(f)$};
  \end{tikzpicture}
  
    \caption{
    \label{fig:supp2_2}Evolution of the liquid film height at the axis of symmetry ($r=0$) for droplet deposition on a viscous film and various values of $\xi$, $\lambda$ and $h_s$. The parameter $\delta$ is set to 0.05.
    The system of equations solved consists of  \eqref{eq:pressure_h1}. \eqref{eq:cap_pressure}, \eqref{eq:p_forcebalance}, and either (solid lines) accounting for the full stress balance using \eqref{eq:H_slip} and \eqref{eq:h2_slip}, or (dashed lines) using the simplified equations  \eqref{eq:thinfilm_H} and \eqref{eq:thinfilmviscous_h2}. 
	}
\end{figure}

For a droplet settling onto a viscous film discussed in \S \ref{sec:capillary}, considering the tangential stress balance leads to the governing equations \eqref{eq:H_slip} and \eqref{eq:h2_slip} for the thickness of the air layer separating the drop and the film and for the height of the film, respectively.
These two equations are expected to simplify to \eqref{eq:thinfilm_H} and \eqref{eq:thinfilmviscous_h2} as the viscosity ration $\lambda$ between the air and the liquid goes to zero, which allows to find analytical results in that limit for the evolution of both the gap and the viscous film.
We solved numerically the system of equations constituting of the normal stress balances \eqref{eq:pressure_h1} and \eqref{eq:cap_pressure}, the force balance \eqref{eq:p_forcebalance}, and the governing equations for the air film thickness and liquid film height as either (i) accouting for the full stress balance using \eqref{eq:H_slip} and \eqref{eq:h2_slip}, or (ii) using the simplified equations  \eqref{eq:thinfilm_H} and \eqref{eq:thinfilmviscous_h2} obtained when considering no slip in the air and free slip in the viscous layer. 
Figure \ref{fig:supp2_1} shows the resulting evolution of the minimum gap thickness $H_{\rm min}(t)$ and that indeed, for $\lambda \ll 1$, the two formulations are almost equivalent. 
The evolution of the viscous film is even less affected than that of gap as shown in figure \ref{fig:supp2_2}.

\end{document}